\documentclass[journal]{IEEEtran}

\usepackage{amssymb}
\usepackage[cmex10]{amsmath}
\usepackage{amsfonts}
\usepackage{lineno}
\modulolinenumbers[5]
\usepackage{textgreek}
\usepackage{tikz}
\usetikzlibrary{arrows, arrows.meta, decorations.pathreplacing}
\usepackage{graphicx}
\usepackage{grffile}
\graphicspath{{./fig/}}
\DeclareGraphicsExtensions{.pdf,.jpeg,.jpg,.png}
\usepackage{xcolor}
\usepackage{array}
\usepackage{multirow}
\usepackage{tabularx}
\usepackage{tabulary}
\usepackage{booktabs}
\usepackage{setspace}

\newcommand{\isot}[2]{\textsuperscript{#2}#1}

\DeclareMathOperator*{\argmax}{arg\, max}

\usepackage{soul}
\usepackage{etoolbox}
\makeatletter
\patchcmd{\SOUL@ulunderline}{\dimen@}{\SOUL@dimen}{}{}
\patchcmd{\SOUL@ulunderline}{\dimen@}{\SOUL@dimen}{}{}
\patchcmd{\SOUL@ulunderline}{\dimen@}{\SOUL@dimen}{}{}
\newdimen\SOUL@dimen
\makeatother

\usepackage{url}
\usepackage[plain]{fancyref}
\usepackage{hyperref}

\newcolumntype{Y}{>{\centering\arraybackslash}X}

\begin{document}
\IEEEpubid{\begin{minipage}{0.85\textwidth}\ \\[12pt]\\ \\ \\ \centering
  \textcopyright 2021 IEEE.  Personal use of this material is permitted. Permission from IEEE must be obtained for all other uses, in any current or future media, including reprinting/republishing this material for advertising or promotional purposes, creating new collective works, for resale or redistribution to servers or lists, or reuse of any copyrighted component of this work in other works.
\end{minipage}}

\title{Correlations between Panoramic Imagery and Gamma-Ray Background in an Urban Area}

\author{M.~S.~Bandstra,
        B.~J.~Quiter,
        M.~Salathe,
        K.~J.~Bilton,
        J.~C.~Curtis,
        S.~Goldenberg,
        and T.~H.~Y.~Joshi

\thanks{This work was performed under the auspices of the U.S.~Department of Energy by Lawrence Berkeley National Laboratory under Contract DE-AC02-05CH11231. The project was funded by the U.S.~Department of Energy, National Nuclear Security Administration, Office of Defense Nuclear Nonproliferation Research and Development (DNN R\&D). This research used resources of the National Energy Research Scientific Computing Center (NERSC), a U.S.~Department of Energy Office of Science User Facility operated under Contract No.~DE-AC02-05CH11231.}
\thanks{M.~S.~Bandstra, B.~J.~Quiter, M.~Salathe, J.~C.~Curtis, and T.~H.~Y.~Joshi are with the Applied Nuclear Physics Program at Lawrence Berkeley National Laboratory, Berkeley, CA 94720 USA (e-mail: msbandstra@lbl.gov).}%
\thanks{K.~J.~Bilton is with the Department of Nuclear Engineering, University of California, Berkeley, CA 94720 USA}%
\thanks{S.~Goldenberg is with the Computer Science Department, The College of William and Mary, Williamsburg, VA 23185 USA}%
}


\maketitle

\begin{abstract}
When searching for radiological sources in an urban area, a vehicle-borne detector system will often measure complex, varying backgrounds primarily from natural gamma-ray sources.
Much work has been focused on developing spectral algorithms that retain sensitivity and minimize the false positive rate even in the presence of such spectral and temporal variability.
However, information about the environment surrounding the detector system might also provide useful clues about the expected background, which if incorporated into an algorithm, could improve performance.
Recent work has focused on extensive measuring and modeling of urban areas with the goal of understanding how these complex backgrounds arise.
This work presents an analysis of panoramic video images and gamma-ray background data collected in Oakland, California by the Radiological Multi-sensor Analysis Platform (RadMAP) vehicle.
Features were extracted from the panoramic images by semantically labeling the images and then convolving the labeled regions with the detector response.
A linear model was used to relate the image-derived features to gamma-ray spectral features obtained using Non-negative Matrix Factorization (NMF) under different regularizations.
We find some gamma-ray background features correlate strongly with image-derived features that measure the response-adjusted solid angle subtended by sky and buildings, and we discuss the implications for the development of future, contextually-aware detection algorithms.
\end{abstract}

\IEEEpeerreviewmaketitle%



\section{Introduction}
\IEEEPARstart{V}{ehicle}-borne gamma-ray detection systems have been developed and deployed for many years and play a key role in radiological and nuclear security missions, especially the search for sources outside of regulatory control~\cite{hjerpe_statistical_2001, aage_search_2003, ziock_large_2004, mitchell_mobile_2009, penny_dual-sided_2011, zelakiewicz_sorisstandoff_2011, curtis_simulation_2020}.
These systems can be rapidly deployed and carry large volume detectors, giving them advantages in efficiency and the ability to cover wide areas relative to human-portable systems.
However, with their higher efficiencies, these systems also suffer from the complex natural radiological backgrounds often found in urban areas~\cite{aucott_routine_2013, archer_systematic_2015}, whose spatial and temporal complexities can be exacerbated by the added mobility of vehicle-borne systems and limit the sensitivity of detection algorithms~\cite{ziock_large_2004, jarman_comparison_2008, aucott_effects_2014}.

The natural backgrounds encountered by a vehicle-borne system include the three main terrestrial ``KUT'' sources (\isot{K}{40}, the \isot{U}{238} decay series, and the \isot{Th}{232} decay series), which are found in some quantity in most minerals, soils, asphalt, and building materials.
The background also contains the progeny of \isot{Rn}{222}, which, while a major portion of the U-238 series, can also be suspended in the air; and cosmic emission, which takes the form of a power-law continuum and 511-keV line emission.
(The decay products of \isot{Rn}{220} from the \isot{Th}{232} series can also escape from the soil and be suspended in air, but due to the very short half-life of \isot{Rn}{220} compared to \isot{Rn}{222}, this decay chain is a negligible contribution to the background.)
The reader is directed to~\cite{runkle_photon_2009} for a review of background sources and~\cite{sandness_accurate_2009} for thorough measurements and modeling of the backgrounds encountered by a ground-based gamma-ray detector system.
Because the compositions of KUT in building materials can vary by orders of magnitude~\cite{trevisi_natural_2012} and the sizes of buildings and other structures can also vary widely in urban areas, urban radiological backgrounds can vary significantly, even over distances as small as several meters~\cite{aucott_routine_2013}.

In order to improve their sensitivity to sources of interest, many recent detection algorithms focus on capturing the background complexity through analysis of the full gamma-ray spectrum instead of only a portion of the spectrum~\cite{cosofret_utilization_2014, pfund_improvements_2016, tandon_detection_2016, miller_gamma-ray_2018, bilton_non-negative_2019}.
However, these algorithms are not yet close to the Poisson statistical limit~\cite{bilton_non-negative_2019}, presumably due to the temporal variability of the background, which is difficult to compensate for.
One way to potentially improve the performance of detection algorithms in urban settings may be to include some non-radiological contextual information.
The most extreme hypothetical example would be an algorithm that is able to perfectly predict the Poisson mean of the current background spectrum through the use of contextual information.
Spectroscopic algorithms in this case should therefore be able to achieve the Poisson limit of detection sensitivity.
However, such a contextual algorithm does not exist, and instead we assert that finding correlations between contextual features and spectroscopic features, even if they are weak, could provide useful information to algorithms that could allow them to improve their performance, e.g., by adapting their detection thresholds to the current background environment.

Some radiation portal monitor research has focused on the use of contextual information to provide cues for detection algorithms.
The advantage that portal monitors have is that they are stationary, so any rapid changes in background must be due to shielding by the vehicles being monitored~\cite{lo_presti_baseline_2006} or radon washout.
Sensors to detect the presence of vehicles are typically used to estimate background suppression profiles as vehicles pass through the sensors~\cite{lo_presti_baseline_2006, burr_alarm_2007}.
In some cases, cameras for monitoring the location of vehicles have been used, although to correlate particular vehicles with sources and not to identify background variations~\cite{karnowski_design_2010, ziock_performance_2013}.
Rainfall sensors have been used with RPMs to estimate the increases in background due to radon washout~\cite{livesay_rain-induced_2014}.

For mobile detector systems the problem is more complex; both the movement of the detector system itself and the movement of objects in the scene around the system can change the radiological conditions surrounding the detector.
Previous research has shown that by being aware of the city~\cite{mitchell_gamma-ray_2015} or the region within a city~\cite{aucott_routine_2013} that a mobile system is in can give some idea about the distribution of backgrounds encountered.
Other research has focused on ``clutter,'' i.e., vehicles and people near the detector system, which temporarily shield some of the background emissions and depress measured count rates.
Using contextual sensors such as cameras and LiDAR to detect nearby clutter can be used to identify when an algorithm threshold should be increased so as not to alarm on the changing rates~\cite{stewart_understanding_2018}.
One recent attempt was made to train a deep neural network on panoramic images to predict the measured spectrum, with some promising results~\cite{kaffine_background_2017}.

This work is a part of the Modeling Urban Scenarios and Experiments (MUSE) collaboration~\cite{nicholson_multiagency_2017, archer_modeling_2017}, where the RadMAP vehicle~\cite{bandstra_radmap:_2016} was used to explore the connections between gamma-ray backgrounds and various contextual sensors.
Previous work using RadMAP within MUSE consisted of analysis of panoramic imagery at a small mock urban area (the Military Operations in Urban Terrain or MOUT facility in Fort Indiantown Gap, Pennsylvania, or FtIG)~\cite{bandstra_attribution_2020}, for which numerous gamma-ray ground truth measurements had been made~\cite{swinney_methodology_2018}.
This work investigates applying methods developed for analyzing data collected at the MOUT facility to data from a dense urban area for which no ground truth data exists but which offers realistic complexity and covers a much larger survey area.
An earlier version of this analysis was presented in~\cite{bandstra_correlations_2019}, and this version expands upon it by examining more spectral features, quantitatively comparing the results of the correlations between imagery and spectral features, discussing the results in more detail, and discussing future prospects for this type of analysis.

This paper will discuss the preparation of the dataset (\Fref{sec:data}), the linear model used to guide the analysis (\Fref{sec:methods}), the extraction of gamma-ray background features (\Fref{sec:spectra}), the extraction of features from panoramic imagery (\Fref{sec:panos}), and then a search for correlations between the two feature sets (\Fref{sec:analysis}).
Finally, the implications for detection algorithms and the development of improved models will be discussed in \Fref{sec:discuss}.


\section{The RadMAP dataset}\label{sec:data}
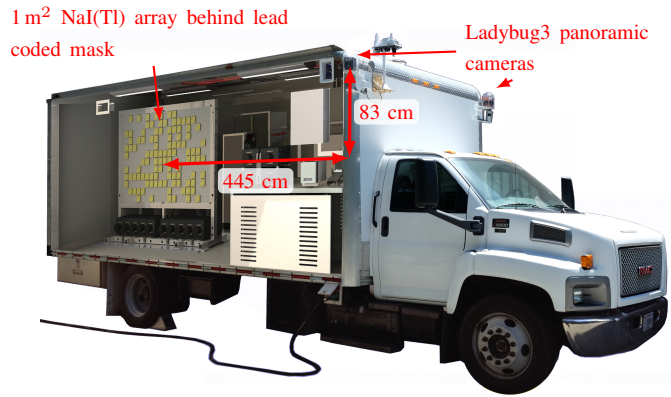
\begin{figure}[t!]
\centering
\begin{tikzpicture}
    \node[anchor=south west, inner sep=0] (image) at (0,0) {\includegraphics[width=0.95\columnwidth, clip]{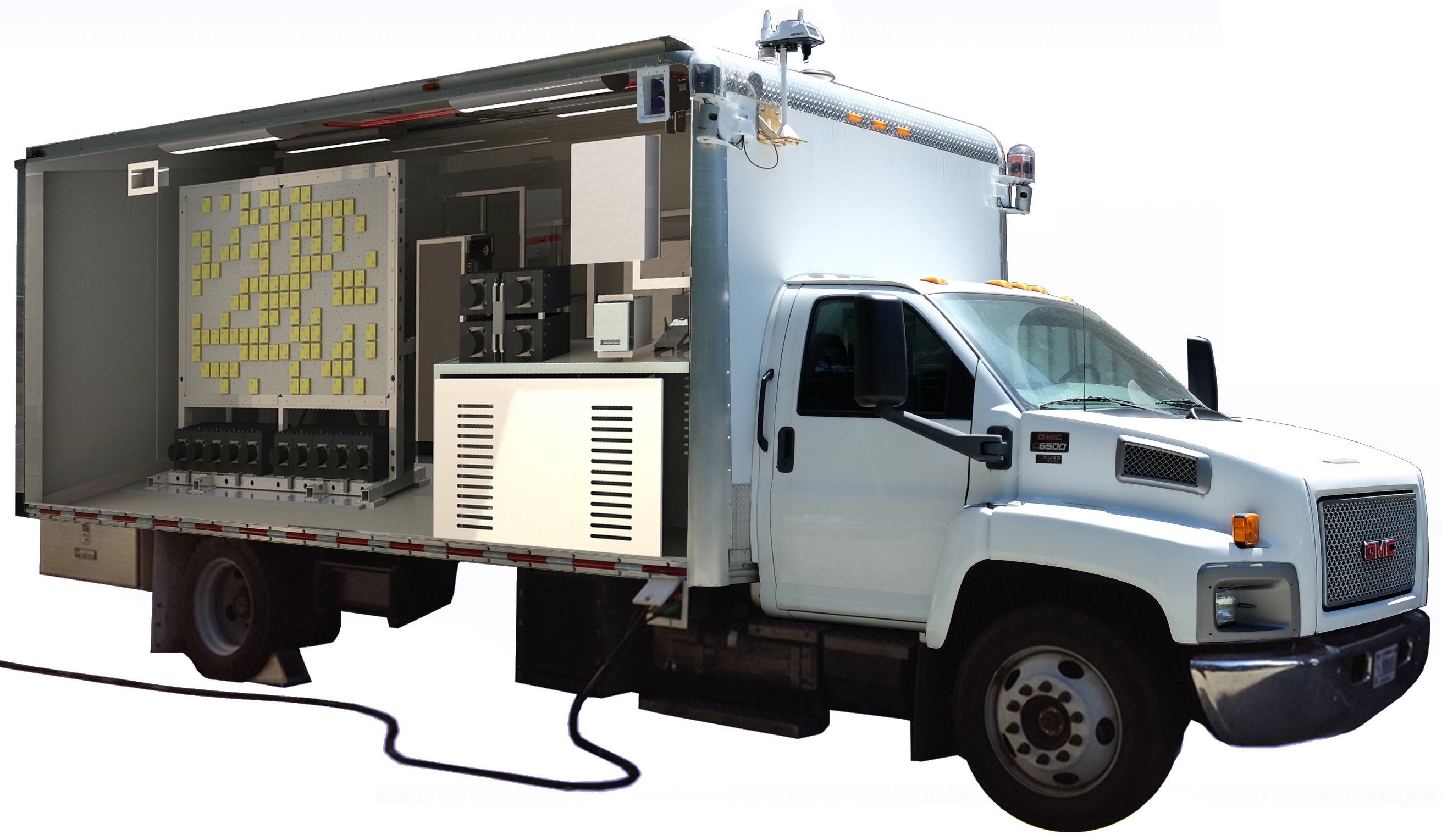}};
    \begin{scope}[x={(image.south east)}, y={(image.north west)}]
        \draw[red, thick, -latex] (0.18, 0.90) node [above, text width=3.8cm] {\footnotesize 1\,m$^2$ NaI(Tl) array behind lead coded mask} -- (0.19, 0.75);

        \node[red, text width=3.0cm] (ladybugs) at (0.85, 0.95) {\footnotesize Ladybug3 panoramic cameras};
        \draw[red, thick, -latex] (ladybugs) -- ++(-0.13, -0.11);
        \draw[red, thick, -latex] (ladybugs) -- ++(-0.35, -0.02);

        \draw[red, very thick, latex-latex] (0.193, 0.633) -- (0.489, 0.651) node [midway, below, yshift=-2, fill=white, opacity=0.8, text opacity=1, rounded corners=2pt, inner sep=2pt] {\footnotesize 445~cm};
        \draw[red, very thick, latex-latex] (0.49, 0.90) -- (0.489, 0.651) node [midway, right, xshift=1, fill=white, opacity=0.8, text opacity=1, rounded corners=2pt, inner sep=2pt] {\footnotesize 83~cm};
    \end{scope}
\end{tikzpicture}
\caption{A cutaway view of the RadMAP vehicle (from~\cite{bandstra_radmap:_2016}).
The locations of cameras and the NaI(Tl) array and the relevant dimensions between the two sensor systems have been indicated.\label{fig:radmap}}
\end{figure}

\begin{figure}[t!]
\centering
\includegraphics[width=0.99\columnwidth]{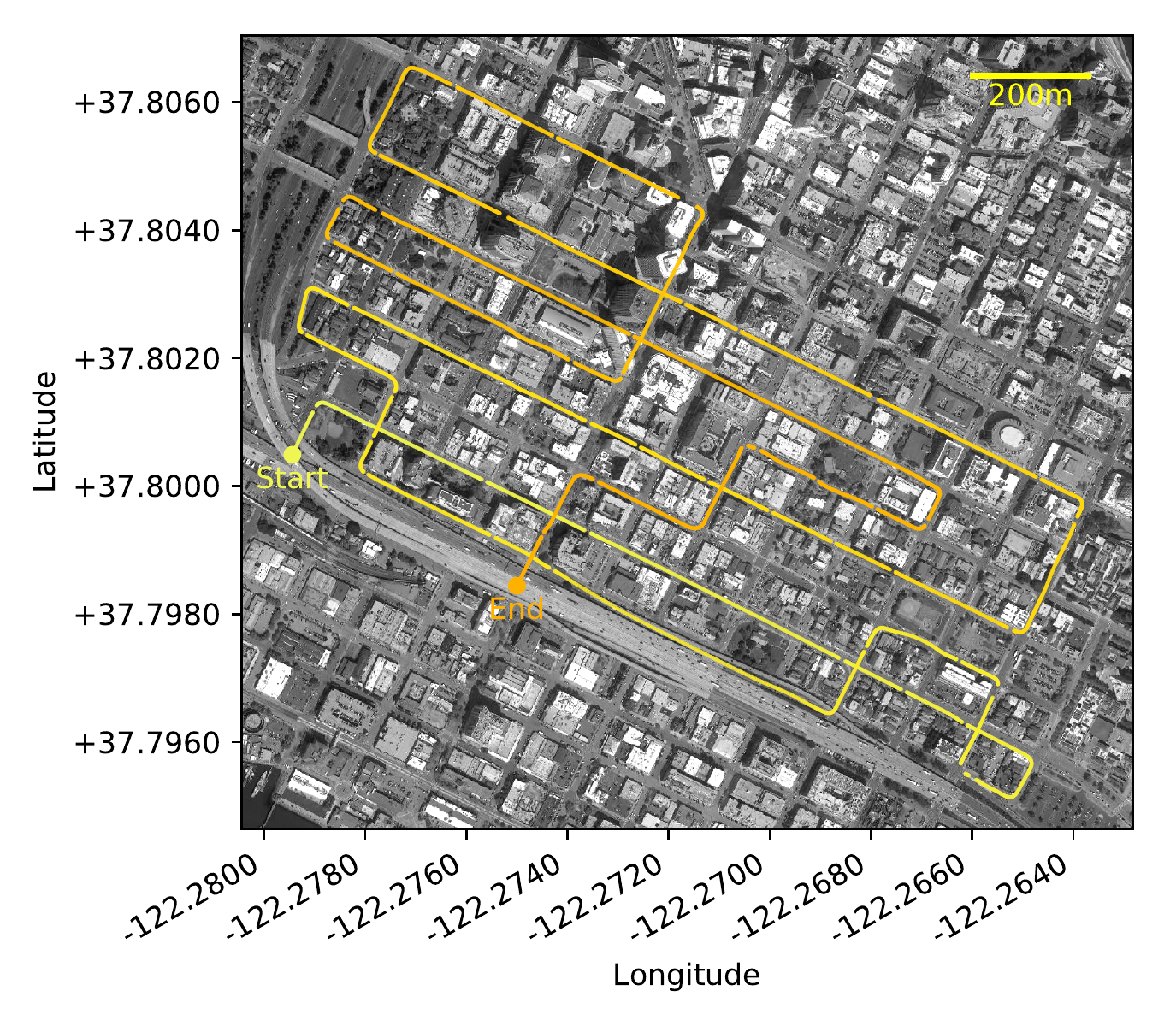}
\caption{The path taken by RadMAP during the evaluation dataset. (Map imagery: Google.)\label{fig:map}}
\end{figure}

The evaluation dataset used in this analysis consisted of data from multiple sensors on board the RadMAP vehicle~\cite{bandstra_radmap:_2016}.
The RadMAP data offer a unique opportunity to explore correlations between gamma-ray backgrounds and panoramic imagery because of its large (1\,m\(^2\)) NaI(Tl) detector array and two panoramic video cameras, which are shown in~\Fref{fig:radmap}.
One long, continuous set of data was chosen for this analysis, during which RadMAP traversed much of downtown Oakland, California (\Fref{fig:map}).
The data were taken on 18~August 2016 from 11:12:42 to 11:52:42~PDT (UTC-7), a total of \(40\)~minutes.
As in~\cite{bandstra_attribution_2020}, images from the two Ladybug3 panoramic cameras were down-sampled from the maximum rate of \(15\)~Hz to \(3\)~Hz so that the corresponding NaI(Tl) spectra had no fewer than approximately one count per bin on average, resulting in 7,190~images from each camera being considered.
The panoramic images from each camera were fused into a single panoramic image covering nearly the entire scene around the vehicle.
The rest of the preparation of the contextual data followed the same procedure described in~\cite{bandstra_attribution_2020} except for an improvement in the method used to align radiation data with imagery, which is described below.

Because the panoramic cameras were located approximately \(4.45\)\,m forward on the vehicle from the NaI(Tl) array, spectral and image data that are coincident in time do not necessarily (and often do not) represent the same location in space, and even a few meters can lead to a significant difference in background~\cite{aucott_routine_2013}.
The typical speed of the vehicle varied between \(0\) and \(10\)\,m/s, implying that there is an inherent time delay of at least 0.45\,s between the imagery and spectra that must be compensated for if they are to represent measurements of the same location.
Previously, this compensation was performed using data from RadMAP's Inertial Navigation System (INS), specifically the Global Positioning System (GPS) coordinates and heading,
whereas in this work, the spatial alignment was achieved by using RadMAP's INS and two LiDAR units to obtain a Simultaneous Localization and Mapping (SLAM) solution using Google Cartographer~\cite{google_cartog_2016}.
A SLAM solution consists of two simultaneously derived objects, the pose of the system (i.e., the 3-D position and orientation) at any timestamp during the measurement and a 3-D model of the environment, typically represented by a point cloud.
The SLAM pose, rotated and translated to fit the highest quality GPS points with a least squares optimization, is shown on the map in \Fref{fig:map}.

The spatial alignment was achieved by converting the timestamp of each panoramic image into a corresponding range of timestamps for the list-mode radiation data generated by the NaI(Tl) array.
First, the pose solution was used to calculate the distance traveled by the vehicle as a function of time, denoted \(s(t)\).
Next, each image was assumed to ``cover'' the time range \((t^{\mathrm{img}}_{0}, t^{\mathrm{img}}_{1}) = t^{\mathrm{img}} \pm \frac{\delta t}{2}\) where \(t^{\mathrm{img}}\) is the timestamp of the image and \(\delta t\) is the inverse of the frame rate (\(1/3\)\,s).
The corresponding NaI(Tl) array timestamps were calculated by solving \(s(t^{\mathrm{det}}_{0}) - 4.45~\mathrm{m} = s(t^{\mathrm{img}}_{0})\) and the analogous relationship for \(t^{\mathrm{img}}_{1}\).
Note that the detector time bins will not necessarily be of equal dwell time.

In order to ensure that the NaI(Tl) measurement was taken before the scene had changed, image-spectra pairs were only accepted if the delay between the image time and the NaI(Tl) time was less than \(2\)\,s (\(t^{\mathrm{det}}_{0} - t^{\mathrm{img}}_{0} \le 2\,\mathrm{s}\) and \(t^{\mathrm{det}}_{1} - t^{\mathrm{img}}_{1} \le 2\,\mathrm{s}\)), and if the duration of the NaI(Tl) measurement was less than \(0.5\)\,s (\(t^{\mathrm{det}}_{1} - t^{\mathrm{det}}_{0} \le 0.5~\mathrm{s}\)).
This cut reduced the number of images and spectra from 7,190 to 4,198, primarily due to times when RadMAP was stopped in traffic, often near intersections.


\section{Methodology}\label{sec:methods}
The main analysis of this work will be to relate gamma-ray spectral features with features derived from the panoramic images.
By ``feature'' we mean a scalar value that represents a more complex data structure through some encoding scheme to be determined.
``Spectral features'' (denoted \(y\)) will be one or more scalar values with units of counts per second that are derived from a single gamma-ray spectrum, and ``image features'' (denoted \(R\)) will be one or more scalar values with units of area derived from each image.

Spectral features and gamma-ray flux from the environment are assumed to be related through a simple model.
The model is based on three assumptions: (1) all spectral features and gamma-ray emissions are non-negative; (2) the intensity of a measured spectral feature is a linear combination of the intensity of emission from all visible material in the environment surrounding RadMAP; and (3) material that belongs to the same visual category has identical gamma-ray emission.
Assumption (1) must be true because of the physical properties of gamma-ray emission, while assumption (2) is only an approximation to reality because of the effects of scattering between visible objects, shielding of emission from objects that are not visible, and downscatter in the air.
The third assumption is the weakest, given the wide ranges of KUT concentrations in soil and building materials~\cite{trevisi_natural_2012}; however, it is possible that within a single urban area the variations in KUT might not be that large.
Since the model requires that each spectral feature be fully explained by a linear combination of non-negative fluxes, as a consequence, the model has an intercept of zero.
Here we will explain the model in anticipation of the analysis later on in~\Fref{sec:analysis}.

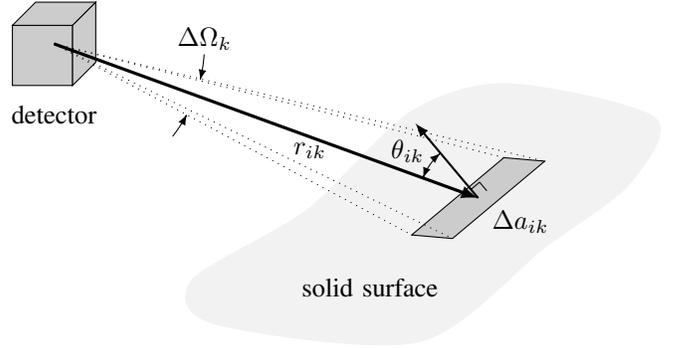
\begin{figure}[t!]
\begin{center}
\begin{tikzpicture}[scale=1]

    \pgfmathsetmacro{\detectorhalfx}{0.4}
    \pgfmathsetmacro{\detectorhalfy}{0.4}
    \pgfmathsetmacro{\detectorhalfz}{0.4}

    \pgfmathsetmacro{\patchhalfx}{0.0}
    \pgfmathsetmacro{\patchhalfy}{0.8}
    \pgfmathsetmacro{\patchhalfz}{0.5}

    \pgfmathsetmacro{\patchxoffset}{6.0}
    \pgfmathsetmacro{\rayangle}{-20}
    \pgfmathsetmacro{\patchrotation}{-30}

    \pgfmathsetmacro{\patchnormal}{1.3}
    \pgfmathsetmacro{\omegarad}{2}

    \draw[black, fill=black!20] (-\detectorhalfx,-\detectorhalfy,\detectorhalfz) -- ++({2*\detectorhalfx},0,0) -- ++(0,{2*\detectorhalfy},0) -- ++({-2*\detectorhalfx},0,0) -- cycle;
    \draw[black, fill=black!20] (-\detectorhalfx,\detectorhalfy,\detectorhalfz) -- ++({2*\detectorhalfx},0,0) -- ++(0,0,{-2*\detectorhalfz}) -- ++({-2*\detectorhalfx},0,0) -- cycle;
    \draw[black, fill=black!20] (\detectorhalfx,-\detectorhalfy,\detectorhalfz) -- ++(0,{2*\detectorhalfy},0) -- ++(0,0,{-2*\detectorhalfz}) -- ++(0,{-2*\detectorhalfy},0) -- cycle;

    \begin{scope}[rotate around={{\rayangle}:(0,0,0)}]
    \begin{scope}[shift={({\patchxoffset}, 0, 0)}]
    \begin{scope}[rotate around={{\patchrotation}:(0,0,0)}]
    \draw [draw=none, fill=black!5] plot [smooth cycle] coordinates {
        (0, -2.8, -2.5)
        (0, -2.8, 2.5)
            (0, -2.0, 4.2) (0, -0.5, 3.5) (0, 1.5, 3.8)
        (0, 2.2, 2.4)
        (0, 1.8, -1.8)
            (0, 0.5, -3.0) (0, -0.8, -2.8)};
    \coordinate (a) at (0,{-\patchhalfy},{-\patchhalfz});
    \coordinate (b) at (0,{\patchhalfy},{-\patchhalfz});
    \coordinate (c) at (0,{\patchhalfy},{\patchhalfz});
    \coordinate (d) at (0,{-\patchhalfy},{\patchhalfz});
    \draw[black, fill=black!20] (a) -- (b) -- (c) -- (d) -- cycle;
    \draw[black, thick, -{latex}] (0, 0, 0) -- ({-\patchnormal}, 0, 0);
    \draw[black] (-0.15, 0, 0) -- (-0.15, 0.15, 0) -- (0, 0.15, 0);
    \draw[black, {latex}-{latex}, domain=180:{180-\patchrotation}] plot ({0.6 * \patchnormal * cos(\x)}, {0.6 * \patchnormal * sin(\x)});
    \end{scope}
    \end{scope}
    \end{scope}

    \draw[black, very thick, -{latex}, rotate around={{\rayangle}:(0,0,0)}] (0,0) -- ({\patchxoffset}, 0);
    \draw[black, dotted, rotate around={{\rayangle}:(0,0,0)}] (0,0,0) -- (a);
    \draw[black, dotted, rotate around={{\rayangle}:(0,0,0)}] (0,0,0) -- (b);
    \draw[black, dotted, rotate around={{\rayangle}:(0,0,0)}] (0,0,0) -- (c);
    \draw[black, dotted, rotate around={{\rayangle}:(0,0,0)}] (0,0,0) -- (d);
    \draw[black, -{latex}, domain={\rayangle+16}:{\rayangle+6}] plot ({\omegarad * cos(\x)}, {\omegarad * sin(\x)});
    \draw[black, -{latex}, domain={\rayangle-18}:{\rayangle-8}] plot ({\omegarad * cos(\x)}, {\omegarad * sin(\x)});

    \node[below, align=center, text width=1.5cm] at (0, -0.7) {detector};
    \node[above, align=center] at (2, -0.2) {$\Delta \Omega_{k}$};
    \node[below, align=center] at (3.4, -1.2) {$r_{ik}$};
    \node[above, align=center] at (4.7, -1.7) {$\theta_{ik}$};
    \node[below, align=center, text width=4cm] at (6.2, -2.1) {$\Delta a_{ik}$};
    \node[below, align=center, text width=4cm] at (4.2, -3) {solid surface};

\end{tikzpicture}
\end{center}
\caption{Schematic of the geometric definitions used in equations~\ref{eq:linear_model1} and~\ref{eq:linear_model2} relating measurement~\(i\) to the emission from the surface element subtended by pixel \(k\).\label{fig:schematic}}
\end{figure}

We start out with the detector being surrounded by a series of solid surfaces and the sky, which we will simply model as a fictional solid surface at a distant radius.
The panoramic images divide the world around the detector into the small segments subtended by each image pixel, and each image pixel \(k\) subtends some solid angle element \(\Delta \Omega_k\).
During measurement \(i\), the pixel subtends some surface area element \(\Delta a_{ik}\) at distance \(r_{ik}\) and angle \(\theta_{ik}\) from its normal.
\Fref{fig:schematic} shows the geometry we are considering and \fref{tab:indices} summarizes the indices that will be used throughout.

\begin{table}[t!]
\caption{Description of the indices used in this manuscript.\label{tab:indices}}
\centering
\begin{tabularx}{\columnwidth}{cX}
    Index symbol & Dimension \\
    \midrule
    \(i\) & measurement \\
    \(j\) & emission type / feature type \\
    \(k\) & image pixel \\
    \(\ell \) & image label type \\
    \midrule
\end{tabularx}
\end{table}

We then assume gamma-ray emission of some type \(j\) is emitted from surface element \(\Delta a_{ik}\) with photon current \(\phi_{ijk}\) (photons per second per area).
An emission type may be comprised of a characteristic distribution of photon energies, so any energy-dependent quantity will need to be indexed by \(j\).
We will also assume that surfaces are solid objects that are multiple photon scatter lengths thick (so the differential photon current in units of photons per second per area per solid angle is proportional to \(\cos \theta_{ik}\)).
Letting \({\cal A}_{jk}\) be the effective area (geometric area times efficiency) of the NaI array for emission type \(j\) in the direction of image pixel \(k\), assuming the surfaces are in the far field (i.e., \(r_{ik}^2 \gg \Delta a_{ik}\) and \(r_{ik}^2 \gg\) the area of the array), and neglecting attenuation and scattering from the air, then the measured count rate of emission type \(j\) is obtained by summing over all of the surface elements:
\begin{align}
    y_{ij} &\approx \sum_{k \in \mathrm{pixels}} {\cal A}_{jk} \frac{\Delta a_{ik} \cos \theta_{ik}}{\pi r_{ik}^2} \phi_{ijk}  \label{eq:linear_model1} \\
    &\approx \sum_{k \in \mathrm{pixels}} {\cal A}_{jk} \frac{\Delta \Omega_{k}}{\pi} \phi_{ijk}. \label{eq:linear_model2}
\end{align}
Notably, the quantity \(\Delta a_{ik} \cos \theta_{ik} / r_{ik}^2\) loses any dependence on the geometry and orientation of the surface element to become simply the solid angle element \(\Delta \Omega_{k}\) subtended by each image pixel \(k\).
(Note that equation~(1) in our previous work~\cite{bandstra_attribution_2020} contains an erroneous factor of \(2 \pi \) in the denominator that we have corrected to \( \pi \) here.
This factor comes from the normalization for the differential photon current over the half unit sphere, which is \(\iint \cos \theta \, d\Omega = \int_{0}^{2\pi} \int_{0}^{\pi/2} \cos \theta \sin \theta \, d\theta \, d\phi = \pi \).)

We will identify the quantities \(y_{ij}\) as gamma-ray spectral features with units of counts per second.
These features are calculated independently of the panoramic imagery from the \(i\)th spectrum \(\mathbf{x}_i\), a vector of counts in each energy bin, and its time duration \(\Delta t_i\).
For example, spectral feature extraction could be done using a linear model, e.g., \(\mathbf{y}_i = \mathbf{U}^{\top} \mathbf{x}_i / \Delta t_i\) for spectrum \(\mathbf{x}_i\) and some matrix \(\mathbf{U}\).

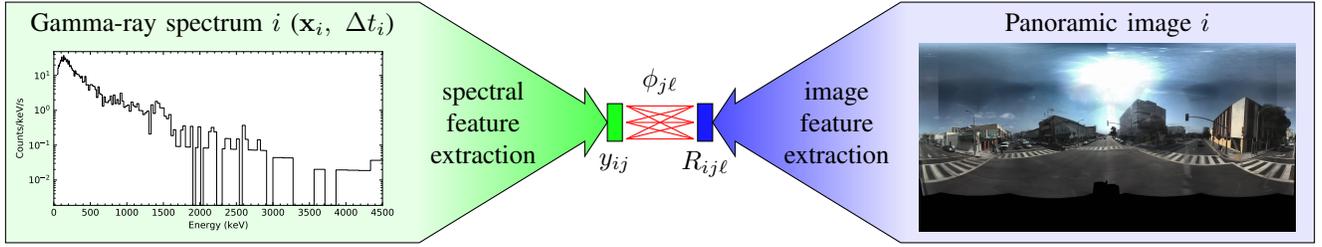
\begin{figure*}[t!]
\begin{center}
\begin{tikzpicture}[scale=1]

    \pgfmathsetmacro{\centerwidth}{1.0}
    \pgfmathsetmacro{\datawidth}{5.5}
    \pgfmathsetmacro{\dataheight}{3.2}
    \pgfmathsetmacro{\extractwidth}{2.5}
    \pgfmathsetmacro{\extractheight}{0.5}
    \pgfmathsetmacro{\smallwidth}{0.2}
    \pgfmathsetmacro{\extractwidth}{2.5}
    \pgfmathsetmacro{\arrowhead}{0.3}
    \pgfmathsetmacro{\arrowwidth}{0.2}
    \pgfmathsetmacro{\datalow}{{-\dataheight / 2}}
    \pgfmathsetmacro{\datahigh}{{\dataheight / 2}}
    \pgfmathsetmacro{\extractlow}{{-(\extractheight / 2)}}
    \pgfmathsetmacro{\extracthigh}{{(\extractheight / 2)}}
    \pgfmathsetmacro{\specextractrightedge}{{(-\centerwidth / 2) - \smallwidth}}
    \pgfmathsetmacro{\specextractleftedge}{{\specextractrightedge - \extractwidth}}
    \pgfmathsetmacro{\imgextractleftedge}{{(\centerwidth / 2) + \smallwidth}}
    \pgfmathsetmacro{\imgextractrightedge}{{\imgextractleftedge + \extractwidth}}
    \pgfmathsetmacro{\specdataleftedge}{{\specextractleftedge - \datawidth}}
    \pgfmathsetmacro{\imgdataleftedge}{{\imgextractrightedge}}

    \pgfdeclarehorizontalshading{specgradient}{100bp}{color(0bp)=(green!10); color(34bp)=(green!10); color(50bp)=(green!70)}
    \draw[shading=specgradient] (\specdataleftedge, \datalow) -- (\specdataleftedge, \datahigh) -- (\specextractleftedge, \datahigh) -- ({\specextractrightedge - \arrowhead}, \extracthigh) -- ({\specextractrightedge - \arrowhead}, {\extracthigh + \arrowwidth}) -- (\specextractrightedge, 0.0) -- ({\specextractrightedge - \arrowhead}, {\extractlow - \arrowwidth}) -- ({\specextractrightedge - \arrowhead}, \extractlow) -- (\specextractleftedge, \datalow) -- cycle;
    \draw[black, fill=green!90] ({\specextractrightedge}, \extractlow) -- ({\specextractrightedge}, \extracthigh) -- ({\specextractrightedge + \smallwidth}, \extracthigh) -- ({\specextractrightedge + \smallwidth}, \extractlow) -- cycle;
    \draw[red, .-.] (-0.45, 0.22) -- (0.45, 0.22);
    \draw[red, .-.] (-0.45, 0.22) -- (0.45, 0.0);
    \draw[red, .-.] (-0.45, 0.22) -- (0.45, -0.22);
    \draw[red, .-.] (-0.45, 0.0) -- (0.45, 0.22);
    \draw[red, .-.] (-0.45, 0.0) -- (0.45, 0.0);
    \draw[red, .-.] (-0.45, 0.0) -- (0.45, -0.22);
    \draw[red, .-.] (-0.45, -0.22) -- (0.45, 0.22);
    \draw[red, .-.] (-0.45, -0.22) -- (0.45, 0.0);
    \draw[red, .-.] (-0.45, -0.22) -- (0.45, -0.22);
    \draw[black, fill=blue!90] ({\imgextractleftedge - \smallwidth}, \extractlow) -- ({\imgextractleftedge - \smallwidth}, \extracthigh) -- ({\imgextractleftedge}, \extracthigh) -- ({\imgextractleftedge}, \extractlow) -- cycle;
    \pgfdeclarehorizontalshading{imggradient}{100bp}{color(0bp)=(blue!70); color(16bp)=(blue!10); color(50bp)=(blue!10)}
    \draw[shading=imggradient] (\imgdataleftedge, \datahigh) -- ({\imgdataleftedge + \datawidth}, \datahigh) -- ({\imgdataleftedge + \datawidth}, \datalow) -- (\imgdataleftedge, \datalow) -- ({\imgextractleftedge + \arrowhead}, \extractlow) -- ({\imgextractleftedge + \arrowhead}, {\extractlow - \arrowwidth}) -- (\imgextractleftedge, 0.0) -- ({\imgextractleftedge + \arrowhead}, {\extracthigh + \arrowwidth}) -- ({\imgextractleftedge + \arrowhead}, \extracthigh) -- cycle;
    \node[align=center, text width=1.7cm] at ({(\specextractleftedge + \specextractrightedge) / 2 - 0.4}, 0.0) {spectral feature extraction};
    \node[align=center, text width=1.7cm] at ({(\imgextractleftedge + \imgextractrightedge) / 2 + 0.4}, 0.0) {image feature extraction};

    \node[align=center] at ({\specextractrightedge + (\smallwidth / 2)}, {\extractlow - 0.3}) {$y_{ij}$};
    \node[align=center] at ({\imgextractleftedge - (\smallwidth / 2)}, {\extractlow - 0.3}) {$R_{ij\ell}$};
    \node[align=center] at (0.0, {\extracthigh + 0.3}) {$\phi_{j\ell}$};

    \node[align=center] at ({\specdataleftedge + (\datawidth / 2)}, {\datahigh - 0.3}) {Gamma-ray spectrum $i$ ($\mathbf{x}_i,~\Delta t_i$)};
    \node[align=center, inner sep=0pt] at ({\specdataleftedge + (\datawidth / 2) - 0.1}, -0.25)
        {\includegraphics[width=5.3cm]{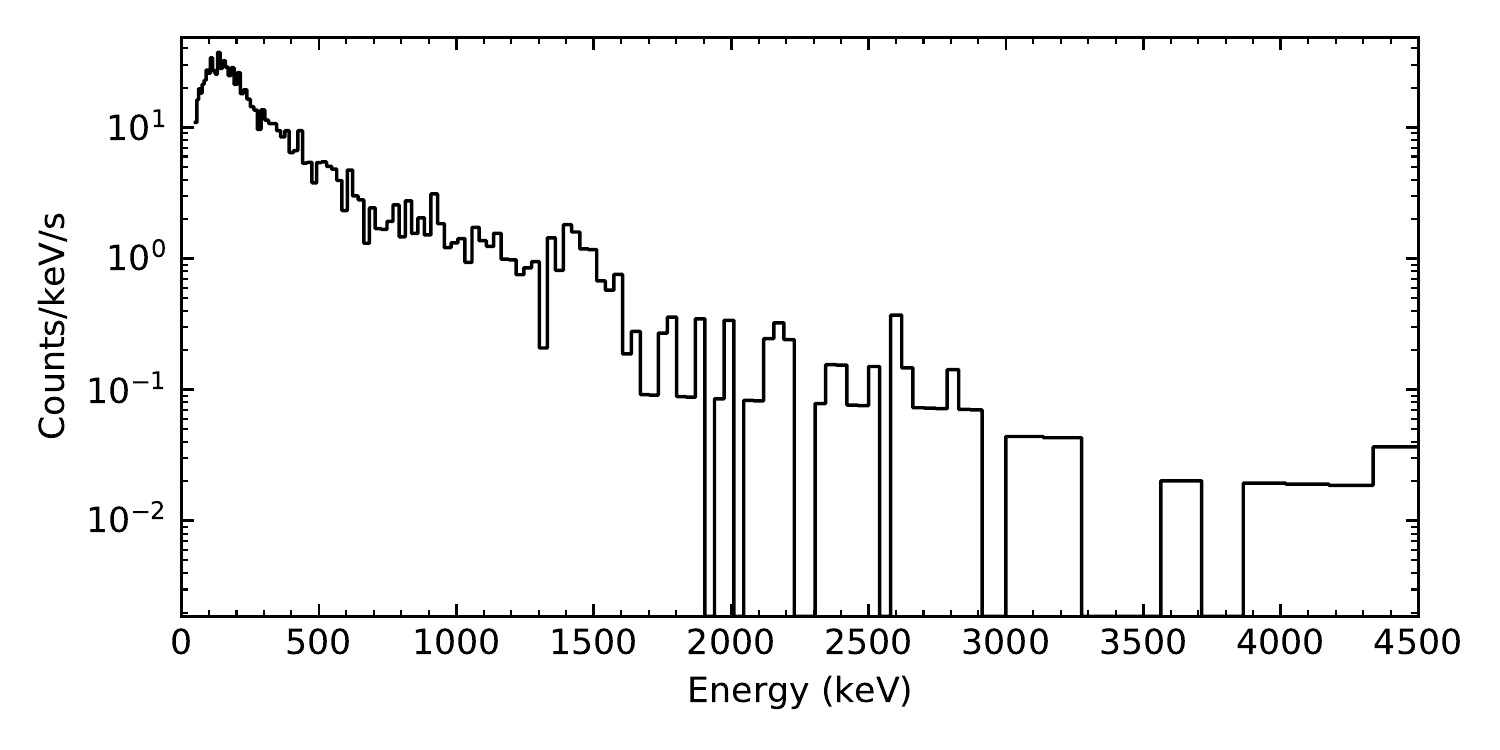}};

    \node[align=center] at ({\imgdataleftedge + (\datawidth / 2)}, {\datahigh - 0.3}) {Panoramic image $i$};
    \node[align=center, inner sep=0pt] at ({\imgdataleftedge + (\datawidth / 2)}, -0.2)
        {\includegraphics[width=5cm]{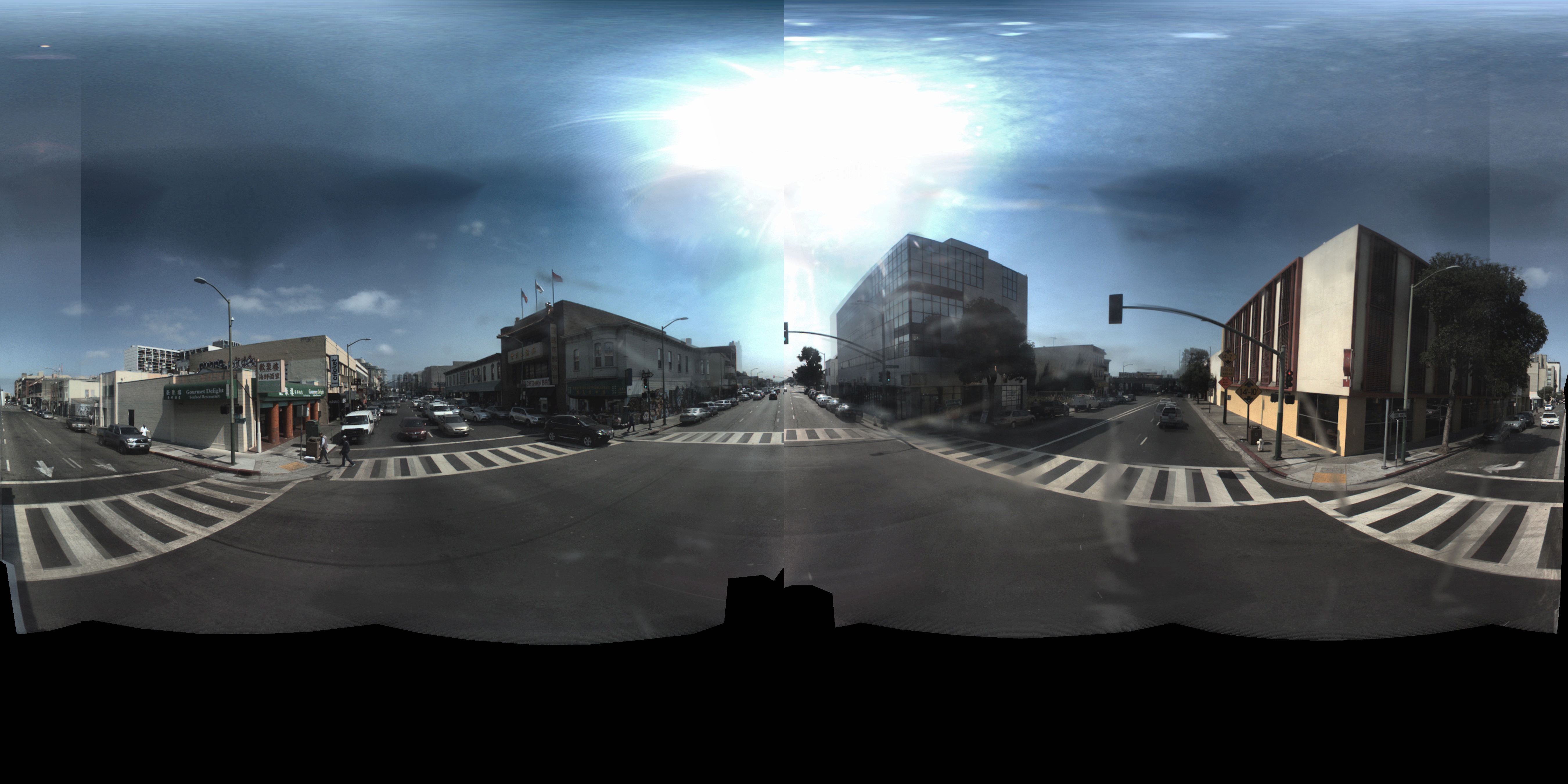}};

\end{tikzpicture}
\end{center}
\caption{A schematic of the process of extracting features from the spectra and images and then comparing the features through fitting a linear model with coefficients \(\phi_{j\ell}\).
The \(i\)th spectrum \(\mathbf{x}_i\) is a vector of counts in each energy bin, and \(\Delta t_i\) is its time duration.
Spectral features \(y_{ij}\) have units of counts per second, and image features \(R_{ij\ell}\) have units of area, leaving the photon currents \(\phi_{j\ell}\) with units of counts per second per area.\label{fig:featureextraction}}
\end{figure*}

Letting the semantic label for each pixel of every image be denoted \({\cal L}_{ik}\), and if we assume that the photon currents \(\phi_{ijk}\) are identical for identically labeled pixels across all the measurements (so that \(\phi_{ijk}\) only needs to be indexed by emission type \(j\) and label \(\ell\)), then this model further reduces to
\begin{align}
    y_{ij} &= \sum_{\ell \in \mathrm{labels}} \left[\sum_{k \in \mathrm{pixels}} {\cal A}_{jk} \frac{\Delta \Omega_{k}}{\pi} \delta_{{\cal L}_{ik}, \ell}\right] \phi_{j \ell} \\
    &\equiv \sum_{\ell \in \mathrm{labels}} R_{ij \ell} \phi_{j \ell}.\label{eq:linear_model}
\end{align}

Once a scheme for extracting spectral features \(y_{ij}\) and a scheme for labeling the panoramic images have been chosen, the image feature tensor \(R_{ij \ell}\), which describes the spectral response to each kind of image label, can be calculated and the linear model solved for the photon currents for each label \(\phi_{j \ell}\).
\Fref{fig:featureextraction} presents a conceptual model of this methodology.
The following sections will explain the calculation of \(y_{ij}\) (\Fref{sec:spectra}) and \(R_{ij\ell}\) (\Fref{sec:panos}), and the fits of the linear model will be examined in~\Fref{sec:analysis}.


\section{Gamma-ray spectral features from NMF}\label{sec:spectra}
There are many choices one could make to extract features from measured gamma-ray background spectra.
Simple examples of such features are the gross counts or counts in certain regions of each spectrum, while more complicated examples could use Principal Component Analysis (PCA)~\cite{runkle_analysis_2006} or convolutions with wavelets~\cite{stinnett_uncertainty_2017}.
To preserve the physical basis of the linear model, we are interested in non-negative features only.
Previous work has shown Non-negative Matrix Factorization (NMF)~\cite{paatero_positive_1994, lee_learning_1999} to be a useful full-spectrum representation for gamma-ray background spectra~\cite{bilton_non-negative_2019}, yielding physically relevant spectral features for vehicle- and aircraft-based systems~\cite{bandstra_attribution_2020, bandstra_modeling_2020}, so we will continue with that approach here.

Three different versions of NMF models will be discussed in more detail in the following subsections, but first are a few general details on how we prepared our dataset for NMF\@.
The NMF models generated in this section were trained using \(3\)-Hz NaI(Tl) spectra from RadMAP measured in the same area of downtown Oakland as the evaluation dataset, on 18 and 22~August 2016, and part of the training data includes the route in the evaluation dataset.
(Note that although there is temporal overlap with the evaluation dataset, the training dataset is larger and binned evenly in time.)
Three nuisance sources were scrubbed from the training set by coarsely binning the data into seven bins (boundaries at 50, 80, 100, 150, 300, 700, 1200, and 3060\,keV) and applying the Spectral Comparison Ratio Anomaly Detection (SCRAD) method~\cite{pfund_examination_2007, jarman_comparison_2008} with an exponential weighting parameter of~\(0.01\).
All three nuisances found were encountered on 18~August but were outside of the time range of the evaluation dataset.
The final training dataset consisted of 18,569~spectra totaling 6,190~seconds.


\subsection{NMF decomposition with Poisson loss (NR)}
The list-mode gamma-ray event data from the NaI(Tl) array on RadMAP were histogrammed in time and energy.
The energy bins were such that the widths were proportional to the square root of energy but broken into two spacing groups: 120\,bins from 50--3000\,keV and 10\,bins from 3000--4500\,keV\@.
The spectra are arranged as the columns of an \(m \times n\) matrix \(\mathbf{X}\), with \(m = 130\) the number of spectral bins and \(n = \)  18,569 the number of time intervals.

\begin{figure*}[t!]
\centering
\begin{tikzpicture}[scale=0.8]
    \pgfmathsetmacro{\a}{0.3}
    \pgfmathsetmacro{\xleft}{0.0}
    \pgfmathsetmacro{\xright}{\xleft + 6.0}
    \pgfmathsetmacro{\xtop}{3.0}
    \pgfmathsetmacro{\xbottom}{\xtop - 3.0}

    \pgfmathsetmacro{\wleft}{\xright + 2.5}
    \pgfmathsetmacro{\wright}{\wleft + 1.2}
    \pgfmathsetmacro{\wtop}{\xtop}
    \pgfmathsetmacro{\wbottom}{\xbottom}

    \pgfmathsetmacro{\hleft}{\wright + 1.5}
    \pgfmathsetmacro{\hright}{\hleft + \xright - \xleft}
    \pgfmathsetmacro{\htop}{0.5 * (\wtop + \wbottom) + 0.5 * (\wright - \wleft)}
    \pgfmathsetmacro{\hbottom}{\htop - (\wright - \wleft)}

    \pgfmathsetmacro{\xcenterx}{0.5 * (\xleft + \xright)}
    \pgfmathsetmacro{\xcentery}{0.5 * (\xbottom + \xtop)}
    \pgfmathsetmacro{\wcenterx}{0.5 * (\wleft + \wright)}
    \pgfmathsetmacro{\wcentery}{0.5 * (\wbottom + \wtop)}
    \pgfmathsetmacro{\hcenterx}{0.5 * (\hleft + \hright)}
    \pgfmathsetmacro{\hcentery}{0.5 * (\hbottom + \htop)}

    \draw [very thick] (\xleft + \a, \xtop) -- (\xleft, \xtop) -- (\xleft, \xbottom) -- (\xleft + \a,\xbottom);
    \draw [very thick] (\xright - \a, \xtop) -- (\xright, \xtop) -- (\xright, \xbottom) -- (\xright - \a,\xbottom);

    \draw [very thick] (\wleft + \a, \wtop) -- (\wleft, \wtop) -- (\wleft, \wbottom) -- (\wleft + \a,\wbottom);
    \draw [very thick] (\wright - \a, \wtop) -- (\wright, \wtop) -- (\wright, \wbottom) -- (\wright - \a,\wbottom);

    \draw [very thick] (\hleft + \a, \htop) -- (\hleft, \htop) -- (\hleft, \hbottom) -- (\hleft + \a,\hbottom);
    \draw [very thick] (\hright - \a, \htop) -- (\hright, \htop) -- (\hright, \hbottom) -- (\hright - \a,\hbottom);

    \pgfmathsetmacro{\r}{0.20}  
    \pgfmathsetmacro{\e}{0.05}  
    \pgfmathsetmacro{\b}{0.08}  

    \pgfmathsetmacro{\nxx}{int(floor((\xright - \xleft - 2 * \b + 2 * \e) / (\r + 2 * \e)))}
    \pgfmathsetmacro{\nxy}{int(floor((\xtop - \xbottom - 2 * \b + 2 * \e) / (\r + 2 * \e)))}
    \pgfmathsetmacro{\nwx}{int(floor((\wright - \wleft - 2 * \b + 2 * \e) / (\r + 2 * \e)))}
    \pgfmathsetmacro{\nwy}{int(floor((\wtop - \wbottom - 2 * \b + 2 * \e) / (\r + 2 * \e)))}
    \pgfmathsetmacro{\nhx}{int(floor((\hright - \hleft - 2 * \b + 2 * \e) / (\r + 2 * \e)))}
    \pgfmathsetmacro{\nhy}{int(floor((\htop - \hbottom - 2 * \b + 2 * \e) / (\r + 2 * \e)))}

    \pgfmathsetmacro{\oxx}{0.5 * (\xright - \xleft - 2 * \b + 2 * \e - \nxx * (\r + 2 * \e))}
    \pgfmathsetmacro{\oxy}{0.5 * (\xtop - \xbottom - 2 * \b + 2 * \e - \nxy * (\r + 2 * \e))}
    \pgfmathsetmacro{\owx}{0.5 * (\wright - \wleft - 2 * \b + 2 * \e - \nwx * (\r + 2 * \e))}
    \pgfmathsetmacro{\owy}{0.5 * (\wtop - \wbottom - 2 * \b + 2 * \e - \nwy * (\r + 2 * \e))}
    \pgfmathsetmacro{\ohx}{0.5 * (\hright - \hleft - 2 * \b + 2 * \e - \nhx * (\r + 2 * \e))}
    \pgfmathsetmacro{\ohy}{0.5 * (\htop - \hbottom - 2 * \b + 2 * \e - \nhy * (\r + 2 * \e))}

    \foreach \x in {1,...,\nxx}
        \foreach \y in {1,...,\nxy}
            \fill [black!10] ({\xleft + \oxx + \b + (\x - 1) * \r + 2 * (\x - 1) * \e}, {\xbottom + \oxy + \b + (\y - 1) * \r + 2 * (\y - 1) * \e}) rectangle ++(\r, \r);

    \foreach \x in {1,...,\nwx}
        \foreach \y in {1,...,\nwy}
            \fill [black!10] ({\wleft + \owx + \b + (\x - 1) * \r + 2 * (\x - 1) * \e}, {\wbottom + \owy + \b + (\y - 1) * \r + 2 * (\y - 1) * \e}) rectangle ++(\r, \r);

    \foreach \x in {1,...,\nhx}
        \foreach \y in {1,...,\nhy}
            \fill [black!10] ({\hleft + \ohx + \b + (\x - 1) * \r + 2 * (\x - 1) * \e}, {\hbottom + \ohy + \b + (\y - 1) * \r + 2 * (\y - 1) * \e}) rectangle ++(\r, \r);

    \node at (\xcenterx, \xcentery) {\LARGE $\mathbf{X}$};
    \node at ({0.5 * (\xright + \wleft)}, \xcentery) {\huge $\approx$};
    \node at (\wcenterx, \wcentery) {\LARGE $\mathbf{W}$};
    \node at ({0.5 * (\wright + \hleft)}, \xcentery) {\huge $\cdot$};
    \node at (\hcenterx, \hcentery) {\LARGE $\mathbf{H}$};

    \pgfmathsetmacro{\c}{0.15}

    \draw [decorate, decoration={brace, mirror, amplitude=10pt}]
    (\xleft + \c, \xbottom - \c) -- (\xright - \c, \xbottom - \c) node [black, midway, yshift=-0.6cm, font=\footnotesize\linespread{0.9}\selectfont] {columns are spectra ($\mathbf{x}_i$)};

    \draw [decorate, decoration={brace, mirror, amplitude=10pt}]
    (\wleft + \c, \wbottom - \c) -- (\wright - \c, \wbottom - \c) node [black, midway, yshift=-0.6cm, align=center, text width=4.0cm, font=\footnotesize\linespread{0.9}\selectfont] {columns are components ($\mathbf{w}_j$)};

    \draw [decorate, decoration={brace, mirror, amplitude=10pt}]
    (\hleft + \c, \hbottom - \c) -- (\hright - \c, \hbottom - \c) node [black, midway, yshift=-0.6cm, text width=4.0cm, font=\footnotesize\linespread{0.9}\selectfont] {columns are \(\mathbf{h}_i\) and rows are \(\mathbf{h}_j^{\top}\)};

    \draw [-{Latex}] (\xleft - \a, \xtop - \oxy - \b) -- (\xleft - \a, \xbottom + \b) node [midway, left, xshift=-0.3cm, text centered, anchor=center, rotate=90, font=\footnotesize\linespread{0.9}\selectfont] {spectral bins (length $m$)};

    \draw [-{Latex}] (\xleft + \oxx + \b, \xtop + \a) -- (\xright - \b, \xtop + \a) node [midway, above, yshift=0.1cm, text centered, font=\footnotesize\linespread{0.9}\selectfont] {measurement (index $i$, length $n$)};

    \draw [-{Latex}] (\wleft + \owx + \b, \wtop + \a) -- (\wright - \b, \wtop + \a) node [midway, above, yshift=0.1cm, text centered, text width=4.0cm, font=\footnotesize\linespread{0.9}\selectfont] {component\\ (index $j$, length $d$)};

    \draw [-{Latex}] (\hleft + \ohx + \b, \htop + \a) -- (\hright - \b, \htop + \a) node [midway, above, yshift=0.1cm, text centered, font=\footnotesize\linespread{0.9}\selectfont] {measurement (index $i$, length $n$)};
\end{tikzpicture}
\caption{Diagram of the NMF decomposition (\fref{eq:nmf}) describing the matrix dimensions and the interpretation of the columns of each of the matrices. \label{fig:nmf}}
\end{figure*}
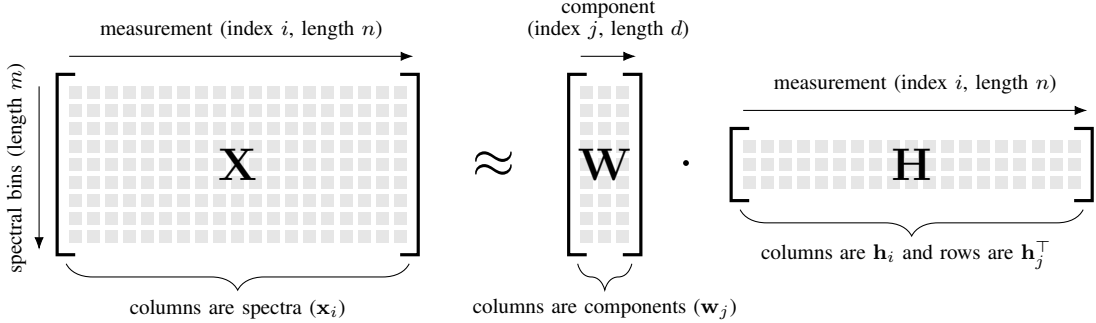

A \(d\)-component NMF decomposition is a linear decomposition of the spectra of the form
\begin{equation}\label{eq:nmf}
    \mathbf{X} \approx \mathbf{W} \mathbf{H},
\end{equation}
where \(\mathbf{W}\) is an \(m \times  d\) matrix whose columns are the spectral components and \(\mathbf{H}\) is a \(d \times n\) matrix whose rows are the component weights.
\Fref{fig:nmf} shows a diagram of the matrices in~\fref{eq:nmf}.
\Fref{eq:nmf} is solved by minimizing the negative log likelihood of \( \mathbf{X} \) given \( \mathbf{\hat{X}} \equiv \mathbf{W} \mathbf{H} \) assuming Poisson statistics:
\begin{equation}\label{eq:nll}
    -\log L(\mathbf{X} | \mathbf{W}, \mathbf{H}) = \sum\left(\mathbf{\hat{X}} - \mathbf{X} \odot \log(\mathbf{\hat{X}}) + \log \mathbf{X}!\right),
\end{equation}
where \( \odot \) denotes element-wise multiplication, the natural logarithm and factorial are applied element-wise, and the sum is over all matrix elements.
To find an NMF solution, the multiplicative update rules from refs.~\cite{lee_learning_1999, lee_algorithms_2001} can be used:
\begin{align}
    \mathbf{W} &\leftarrow \mathbf{W} \odot \left(\frac{\left(\frac{\mathbf{X}}{\mathbf{W} \mathbf{H}}\right) \cdot \mathbf{H}^T}{\mathbf{1}_{m, n} \cdot \mathbf{H}^T}\right) \label{eq:update_W} \\
    \mathbf{H} &\leftarrow \mathbf{H} \odot \left(\frac{\mathbf{W}^T \cdot \left(\frac{\mathbf{X}}{\mathbf{W} \mathbf{H}}\right)}{\mathbf{W}^T \cdot \mathbf{1}_{m, n}}\right) \label{eq:update_H}
\end{align}
where \(\mathbf{1}_{m,n}\) is an \(m \times n\) matrix of ones.
To preserve the normalization of the columns of \(\mathbf{W}\), the following renormalization is done at each step:
\begin{align}
    \mathbf{D} &\equiv \mathrm{diag}(\mathbf{W}^{\top} \cdot \mathbf{1}_{m}) \label{eq:norm_1} \\
    \mathbf{W} &\leftarrow \mathbf{W} \mathbf{D}^{-1} \label{eq:norm_3}
\end{align}
where \(\mathbf{1}_{m}\) is a length-\(m\) column vector of ones and \(\mathrm{diag}\) creates a diagonal matrix from a column vector.

A series of NMF models were trained using the multiplicative update rules without any further regularization.
Since the multiplicative update rules are only guaranteed to find a local, not necessarily global, optimum~\cite{lee_algorithms_2001}, the initialization of the model can influence the final result, and in the high-dimensionality cases considered here and in related work, the initialization appears to always affect the result.
These models were initialized by setting the components to the mean spectral shape, with small random numbers between \(0\) and \(10^{-6}\) added to break the degeneracy between them.
In addition, since NMF models have no preferred order of the components, for ease of comparison between models, the components were sorted in order of increasing variance of their weights, i.e., by the row-wise variance of \(\mathbf{H}\).
This same NMF approach was taken in~\cite{bandstra_modeling_2020}.

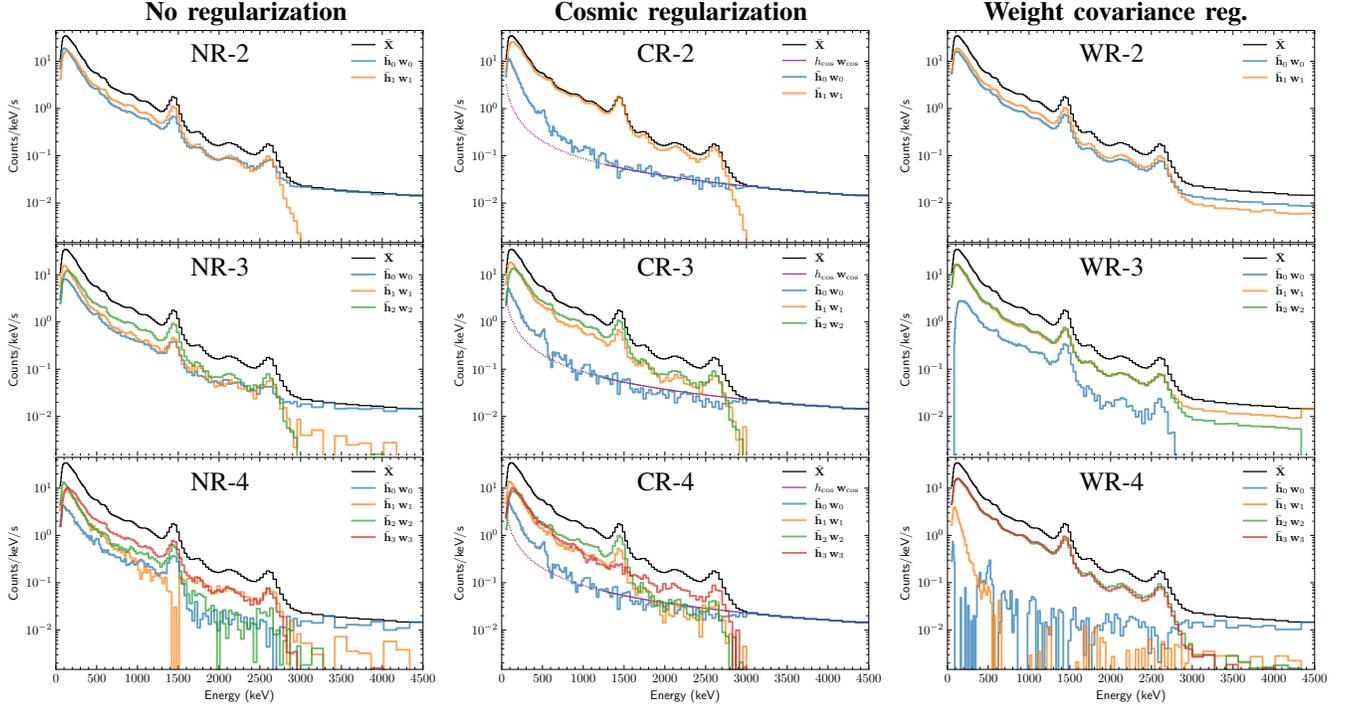
\begin{figure*}[t!]
  \centering
    \begin{tabularx}{0.9\textwidth}{@{}YYY@{}}
        \textbf{No regularization} & \textbf{\ \ \ Cosmic regularization} & \textbf{\ \ \ \ \ \ Weight covariance reg.}
    \end{tabularx}
    \begin{tikzpicture}
        \draw (0, 0) node[inner sep=0] {\includegraphics[clip, trim=0.2cm 1.3cm 0.3cm 0.35cm, width=0.32\textwidth]{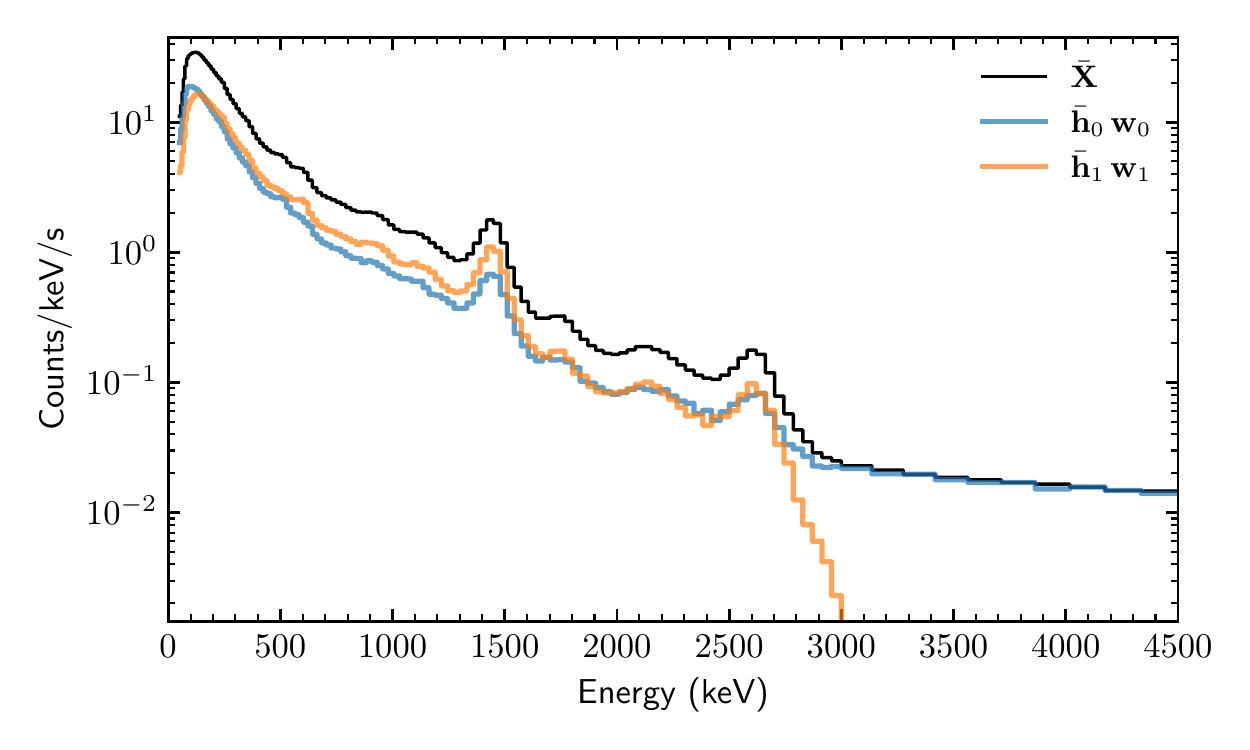}};
        \draw (0, 1.1) node {NR-2};
    \end{tikzpicture}
    \begin{tikzpicture}
        \draw (0, 0) node[inner sep=0] {\includegraphics[clip, trim=0.2cm 1.3cm 0.3cm 0.35cm, width=0.32\textwidth]{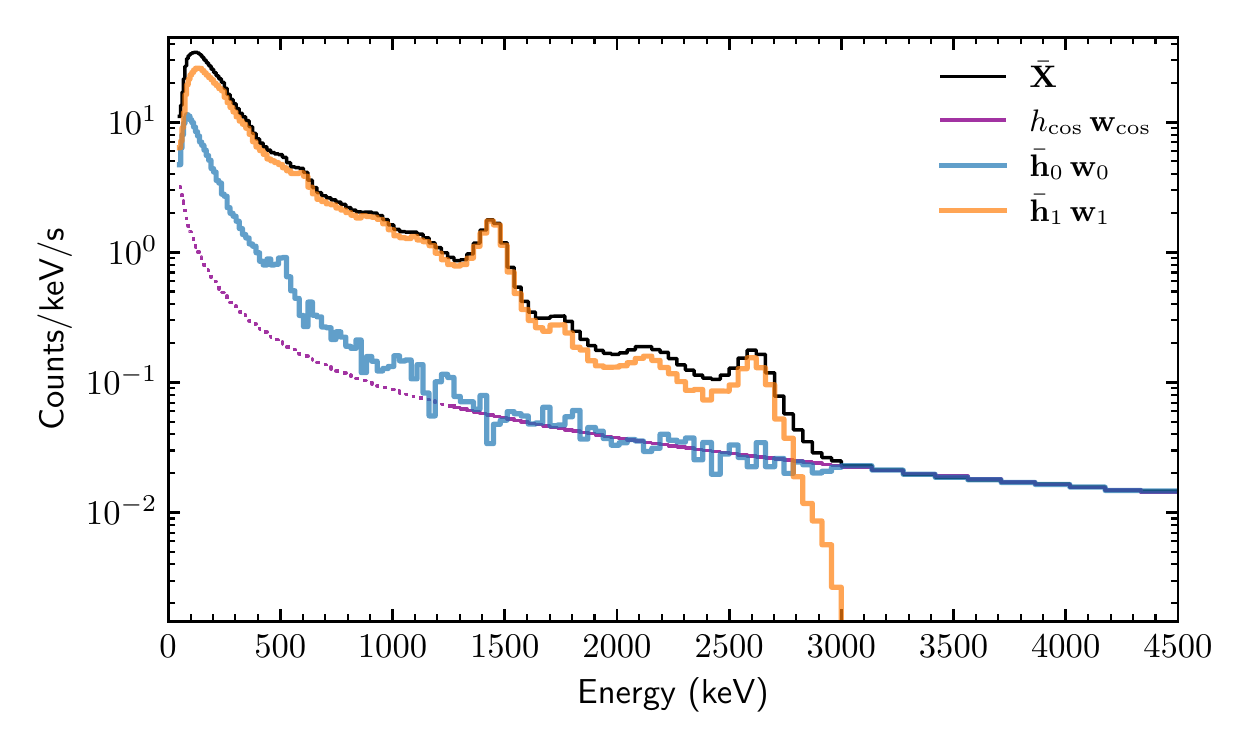}};
        \draw (0, 1.1) node {CR-2};
    \end{tikzpicture}
    \begin{tikzpicture}
        \draw (0, 0) node[inner sep=0] {\includegraphics[clip, trim=0.2cm 1.3cm 0.3cm 0.35cm, width=0.32\textwidth]{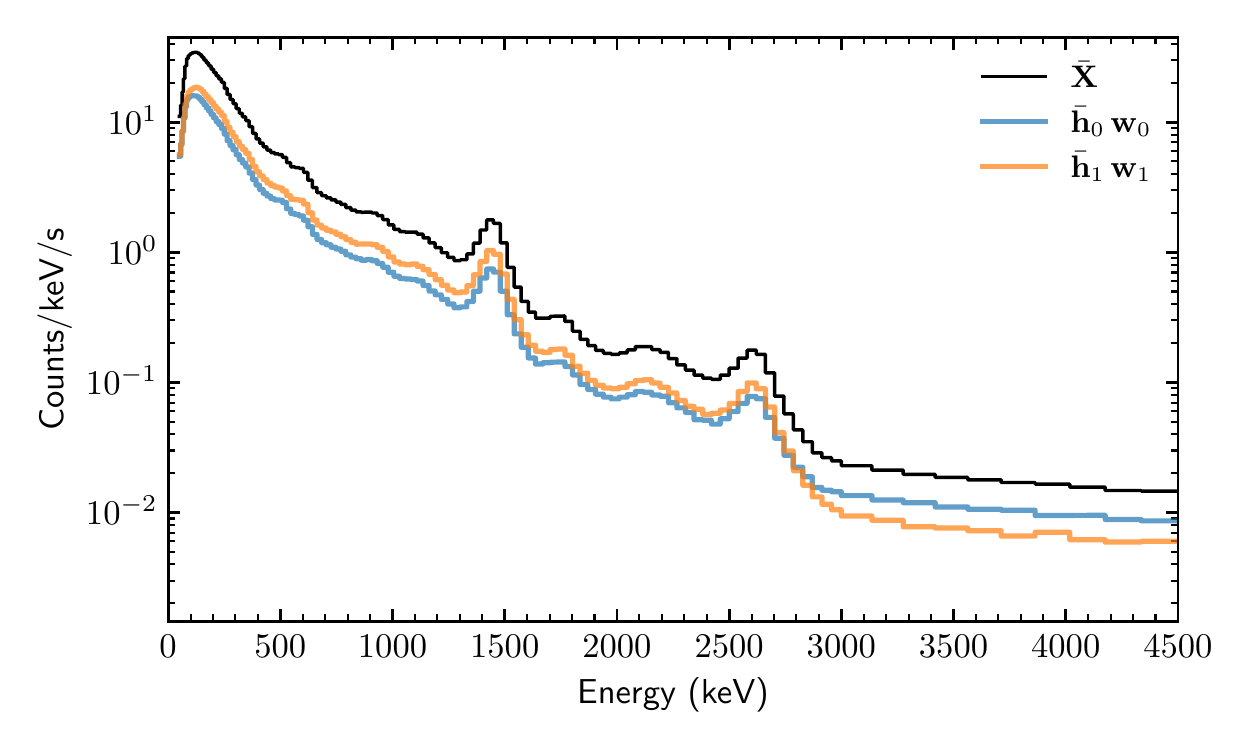}};
        \draw (0, 1.1) node {WR-2};
    \end{tikzpicture}\\
    \begin{tikzpicture}
        \draw (0, 0) node[inner sep=0] {\includegraphics[clip, trim=0.2cm 1.3cm 0.3cm 0.35cm, width=0.32\textwidth]{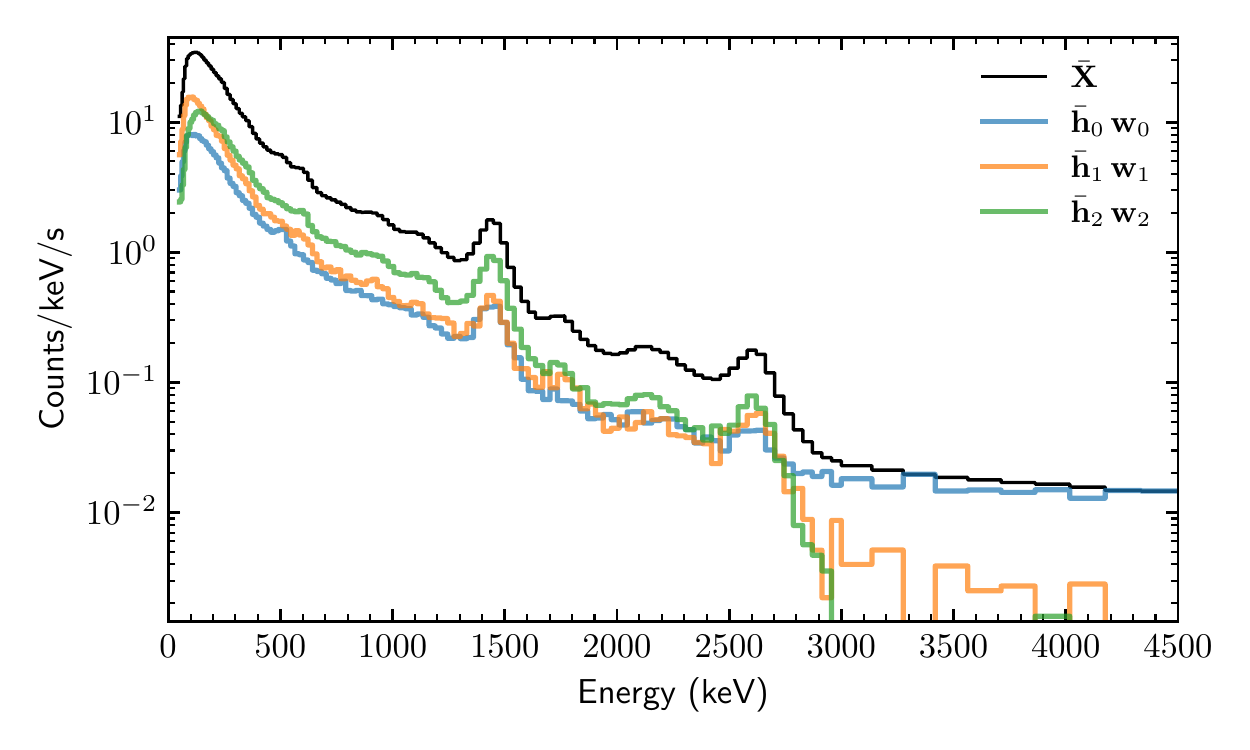}};
        \draw (0, 1.1) node {NR-3};
    \end{tikzpicture}
    \begin{tikzpicture}
        \draw (0, 0) node[inner sep=0] {\includegraphics[clip, trim=0.2cm 1.3cm 0.3cm 0.35cm, width=0.32\textwidth]{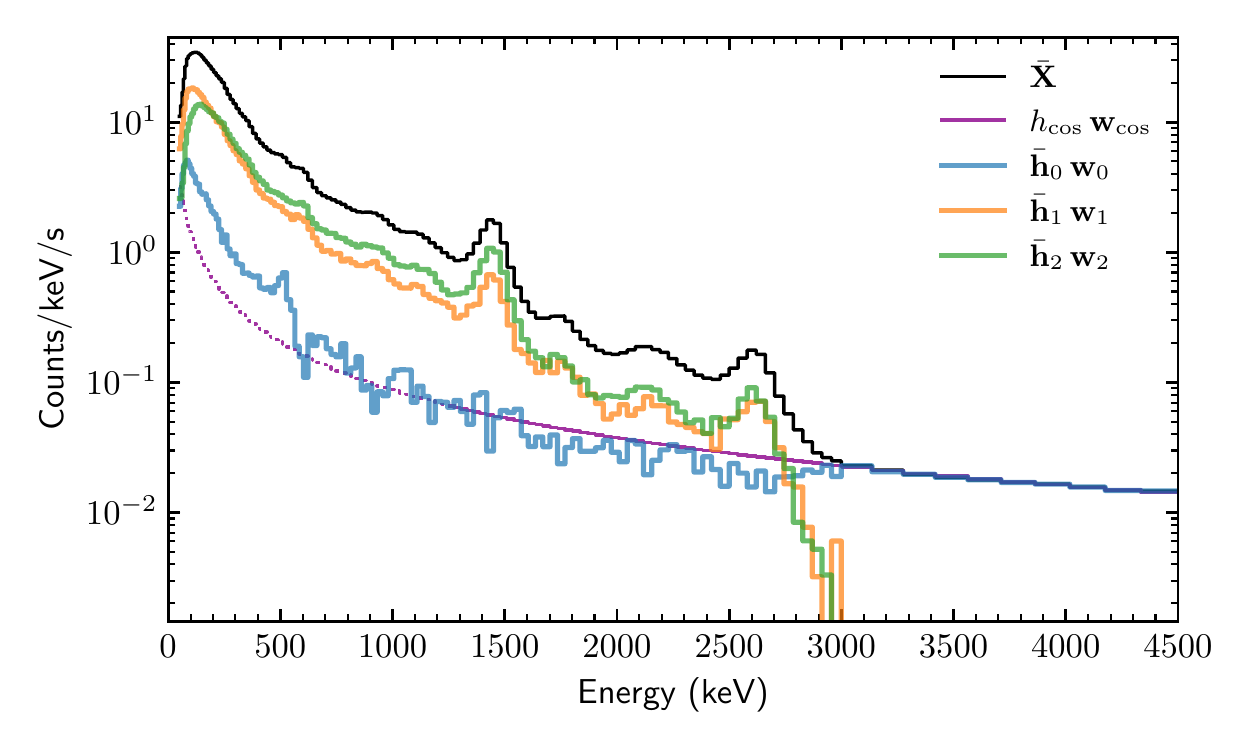}};
        \draw (0, 1.1) node {CR-3};
    \end{tikzpicture}
    \begin{tikzpicture}
        \draw (0, 0) node[inner sep=0] {\includegraphics[clip, trim=0.2cm 1.3cm 0.3cm 0.35cm, width=0.32\textwidth]{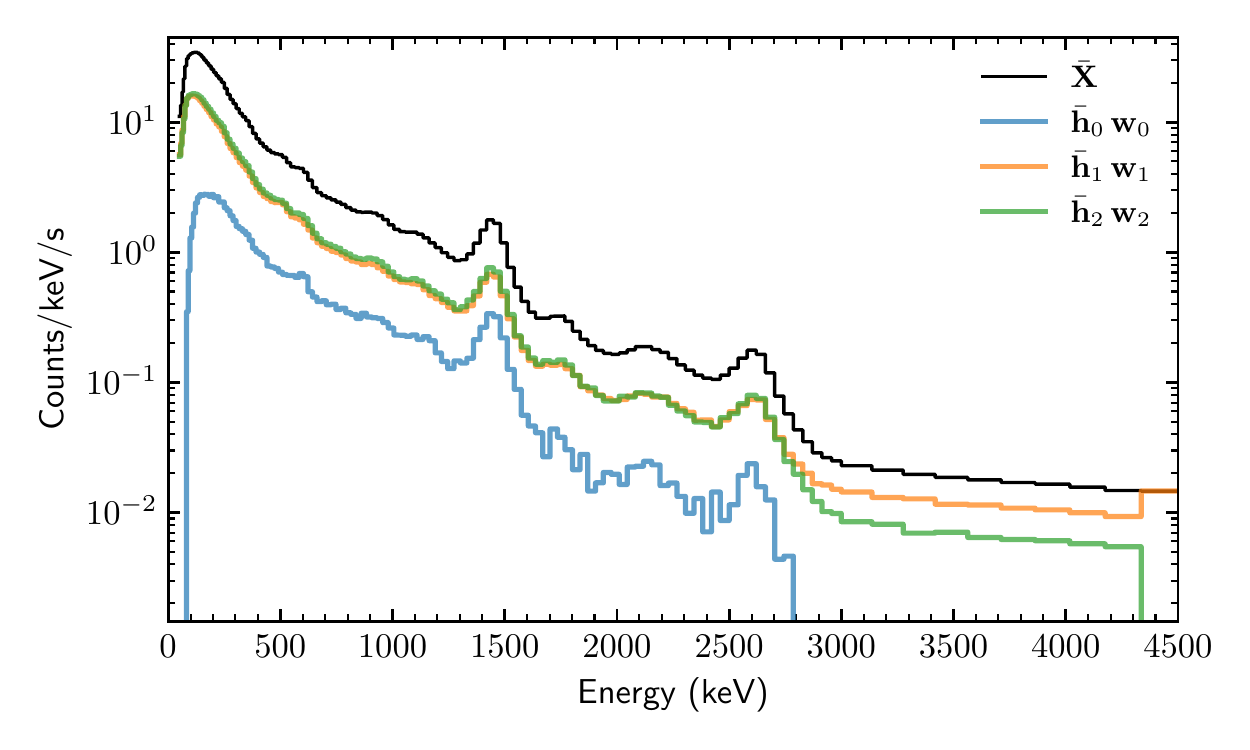}};
        \draw (0, 1.1) node {WR-3};
    \end{tikzpicture}\\
    \begin{tikzpicture}
        \draw (0, 0) node[inner sep=0] {\includegraphics[clip, trim=0.2cm 0 0.3cm 0.35cm, width=0.32\textwidth]{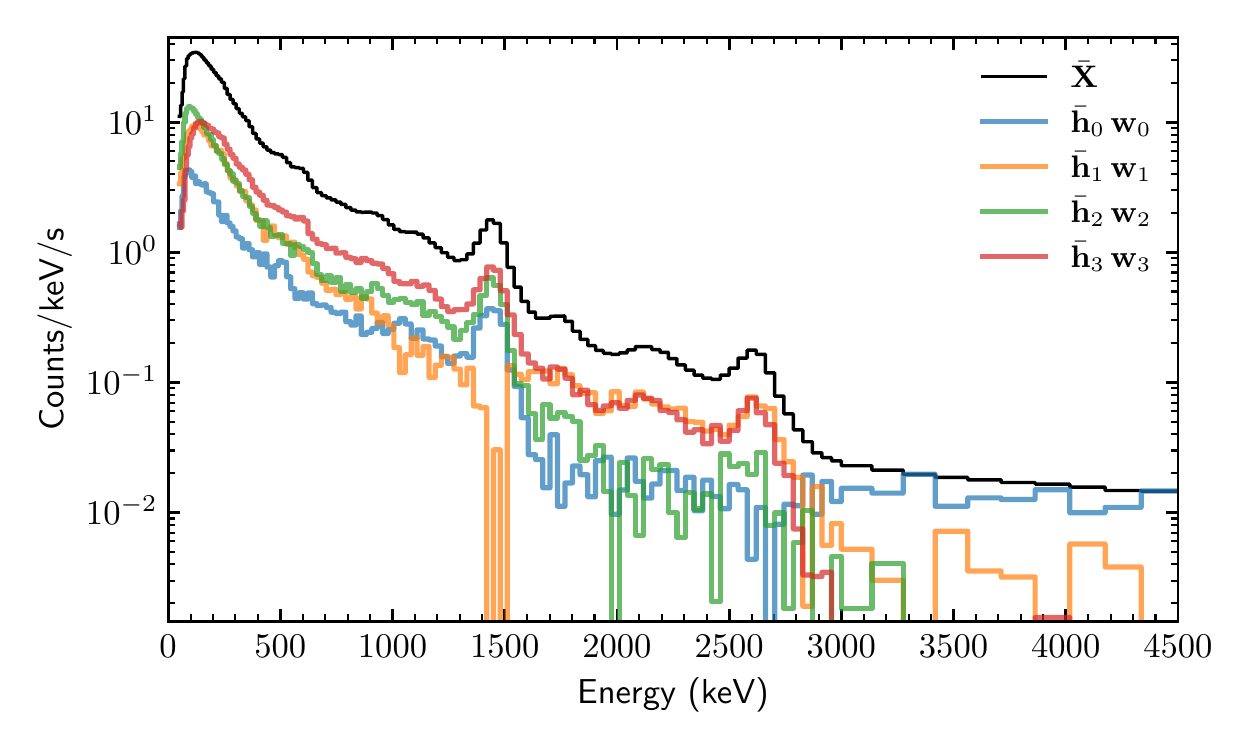}};
        \draw (0, 1.4) node {NR-4};
    \end{tikzpicture}
    \begin{tikzpicture}
        \draw (0, 0) node[inner sep=0] {\includegraphics[clip, trim=0.2cm 0 0.3cm 0.35cm, width=0.32\textwidth]{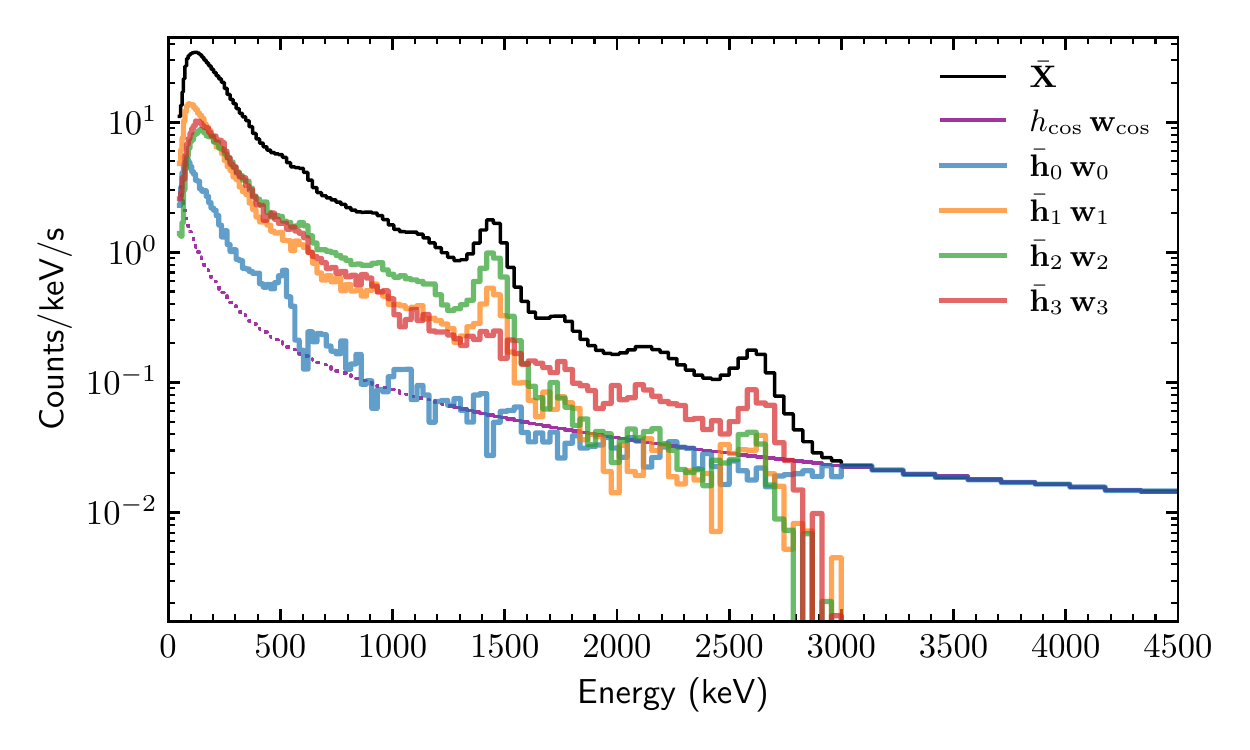}};
        \draw (0, 1.4) node {CR-4};
    \end{tikzpicture}
    \begin{tikzpicture}
        \draw (0, 0) node[inner sep=0] {\includegraphics[clip, trim=0.2cm 0 0.3cm 0.35cm, width=0.32\textwidth]{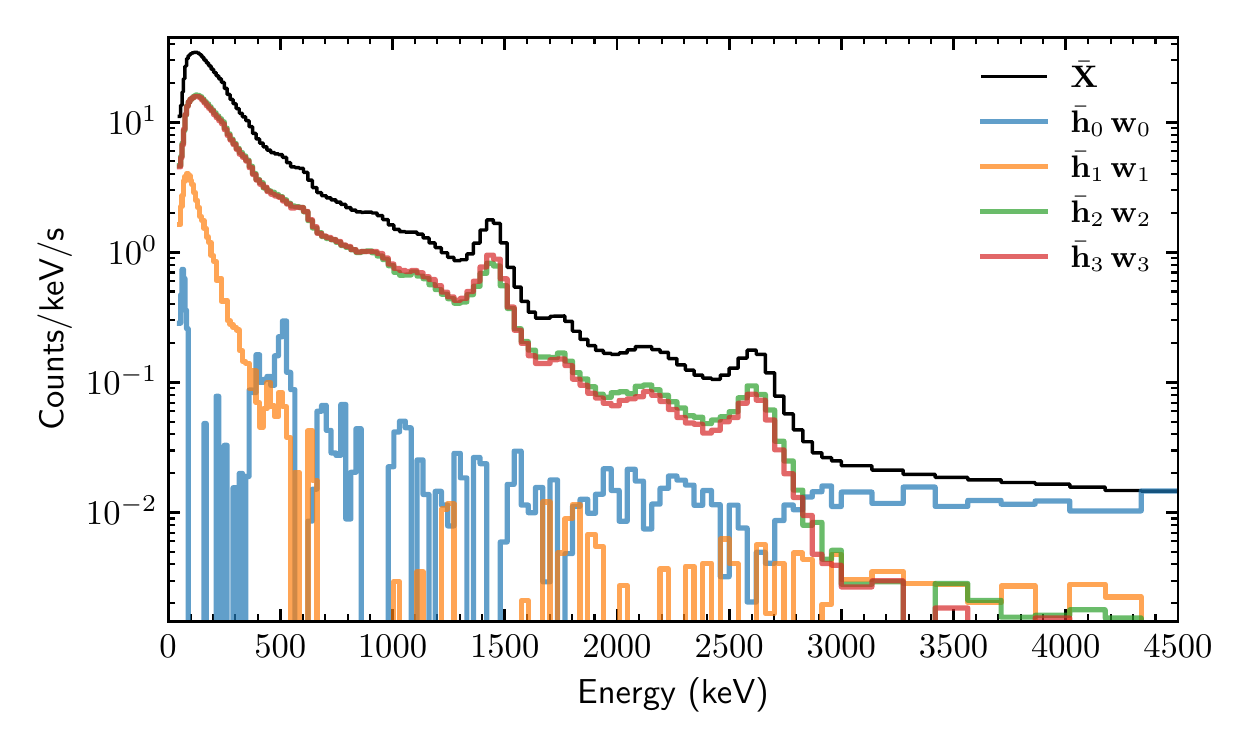}};
        \draw (0, 1.4) node {WR-4};
    \end{tikzpicture}
\caption{The spectral components found by the various NMF approaches explored for 2, 3, and 4 components.
In each plot, the average measured spectrum in the training dataset (black) is compared to the NMF components, each scaled by their mean weight.
In addition, for the CR-\(d\) models, the power law used in the regularization function is shown.\label{fig:nmf_spec_all}}
\end{figure*}

The models that result from this procedure we will denote NR-\(d\) (for \textit{no regularization}, with \(d\) components).
The first column of \Fref{fig:nmf_spec_all} shows the results of these unregularized models for \(d=\)2 to 4~components when allowed to converge until the difference in \(-(\log L) / n\) between subsequent iterations is less than the arbitrary level of \(10^{-9}\).
Of note is that for all three models, exactly one component contains the cosmic continuum above 3\,MeV, and each of these components also uniquely displays a slight 511\,keV line, which is expected from cosmic emission.
For all models, this component is always the component with the lowest variance of its weights (i.e., component~0).
Also worth noting is that all of the components have different shapes for the low energy continuum ``roll off'' around 100\,keV.
For the 2- and 3-component models, all of the components display all of the prominent lines from the KUT background sources, whereas for NR-4 they do not --- e.g., component~0 lacks prominent \isot{U}{238} series lines, component~2 lacks the prominent \isot{Th}{232} series line at 2614\,keV, and component~1 lacks the \isot{K}{40} line at 1460\,keV.
Additionally, particularly for NR-4, different ratios of the KUT background sources are clearly seen, with higher \isot{K}{40} in component~3 and higher \isot{U}{238} and \isot{Th}{232} series in component~1.
These observations imply that NMF is able to capture spectral features that arise from physics, but the various spectral features are not necessarily correlated across the different decompositions.
For example, the high energy cosmic continuum is present in component~0 for all the NR models, but the component that has the highest mean count rate at 100~keV is different for each model (components~0, 1, and 2, respectively).
Evidently the dynamic range of KUT backgrounds encountered is not enough to more cleanly separate the KUT spectra from each other.

\subsection{Cosmic component regularization (CR)}
On its own, NMF does not contain any physics; it is a mathematical method for finding structure in the training data.
However, physical priors and constraints, if expected to have relevance to the model, can be encoded in the training process in the form of regularization functions.
Here we will present one such physical prior --- namely, that one of the components should represent cosmic ray-induced gamma-ray emission, and the remainder only terrestrial emission.

Attempting to isolate a cosmic component may be worthwhile for the following reasons.
First, we know that a cosmic component is present in the data because any detector exposed to the atmosphere will measure some background due to cosmic rays.
Second, the shape of the cosmic emission can be partially estimated from the region above 3\,MeV because there are negligible contributions in that spectral region from terrestrial emission.
Third, repeatedly training with randomly initialized NMF models usually results in a model where the spectrum above 3\,MeV is nearly entirely contained in only one NMF component (e.g., the NR models in~\Fref{fig:nmf_spec_all} all display this behavior).

Regularization to isolate the KUT components from each other was also attempted, but these results were less conclusive than for the cosmic component.
Profiles for each of the KUT emission types were derived from simulations and used in a similar manner to what will be described for the cosmic regularization, however these attempts have led to non-physical shapes at low energies, e.g., components where the counts in the spectrum below 300~keV decrease with decreasing energy more rapidly than expected.
Another issue with KUT regularization is that the training dataset itself appears to comprise an area with a limited range in the ratios of K, U, and T, so there is not enough data to support cleanly separating those background sources from each other.

Here we will initialize an NMF component meant to represent cosmic emission and describe the regularization and final results from training such models.
The models that result from this treatment will be denoted CR-\(d\).


\paragraph{Initialization}
The cosmic component has been observed to consist of a continuum described by a power law with an index of \(\approx 1.3\) and 511\,keV emission from pair production~\cite{sandness_accurate_2009}.
The power-law index can be estimated from the spectrum above 3\,MeV because energy deposition events above the \isot{Tl}{208} line at 2614\,keV are almost entirely due to cosmic rays and cosmic-induced gamma rays.
A weighted least squares fit of the data in the ten spectral bins above 3\,MeV was performed using a power-law model, resulting in a power-law index of \(1.22\).
Although we do not expect the entire cosmic spectrum to follow this power-law shape because changing detector efficiencies will cause a ``roll off'' at the lowest energies and we have not yet included the expected 511\,keV emission, we nevertheless extrapolated this power-law fit to all spectral bins and normalized it to unity to generate the provisional cosmic component \(\mathbf{w}_{\mathrm{cos}}\).
The first column of \(\mathbf{W}\) was initialized to \(\mathbf{w}_{\mathrm{cos}}\).

\begin{figure}[t!]
  \begin{center}
    \includegraphics[width=0.95\columnwidth]{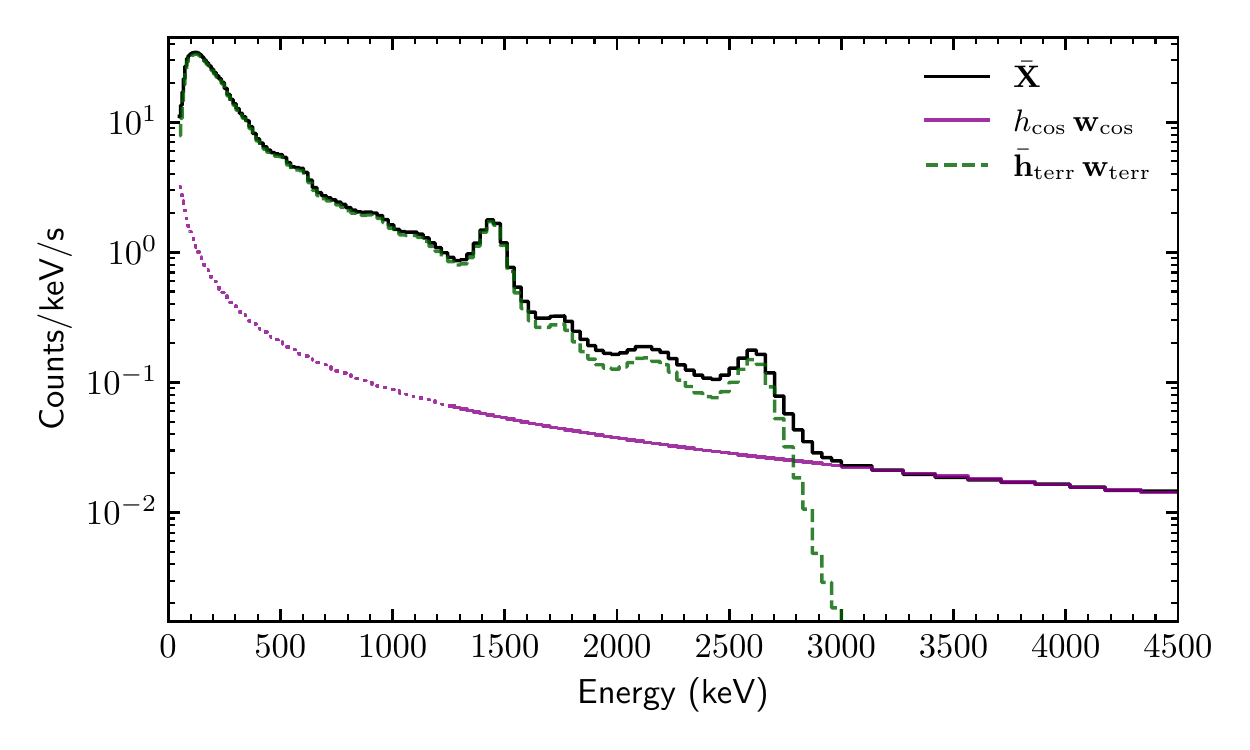}
  \end{center}
\caption{The average background spectrum in the training data shown with the provisional cosmic and terrestrial components used to initialize the NMF model.
The solid line denotes the portion of the cosmic power law used for the regularization function \(f_1\).\label{fig:cosmic_fit}}
\end{figure}

For the sake of initialization, an ansatz was made that the count rate of the cosmic component be approximately constant for all measurements.
The constant weight \(h_{\mathrm{cos}}\) was calculated as the average number of total counts above 3\,MeV in \(\mathbf{X}\) divided by the sum of the portion of \(\mathbf{w}_{\mathrm{cos}}\) above 3\,MeV\@.
The entire first row of \(\mathbf{H}\) was initialized to \(h_{\mathrm{cos}}\).
The provisional cosmic spectrum \(\mathbf{w}_{\mathrm{cos}}\) times the average cosmic counts \(h_{\mathrm{cos}}\) is compared to the average spectrum in \Fref{fig:cosmic_fit}.

After the first column of \(\mathbf{W}\) was initialized to \(\mathbf{w}_{\mathrm{cos}}\), the following procedure was followed to initialize the remaining \(d-1\) components.
First, the non-cosmic spectrum (assumed to be largely terrestrial) was estimated:
\begin{align}
    \mathbf{X}_{\mathrm{terr}} &= \mathbf{X} - \left(\mathbf{w}_{\mathrm{cos}} \otimes \mathbf{1}_{n} \right) h_{\mathrm{cos}},
\end{align}
where \( \otimes \) denotes the outer product.
Any negative values encountered were clipped to zero.
The terrestrial rates \(\mathbf{h}_{\mathrm{terr}}\) were calculated as the sum over the columns of \(\mathbf{X}_{\mathrm{terr}}\), while the terrestrial component \(\mathbf{w}_{\mathrm{terr}}\) was calculated by summing \(\mathbf{X}_{\mathrm{terr}}\) over its rows and dividing by the sum of \(\mathbf{h}_{\mathrm{terr}}\).
\Fref{fig:cosmic_fit} shows the resulting terrestrial component \(\mathbf{w}_{\mathrm{terr}}\) times the average terrestrial counts \(\mathbf{\bar{h}}_{\mathrm{terr}}\) compared to the average spectrum.

For all \(d \ge 2\), the remaining \(d-1\) columns of \(\mathbf{W}\) were initialized with \(\mathbf{w}_{\mathrm{terr}}\), and the remaining \(d-1\) rows of \(\mathbf{H}\) were initialized with \(\mathbf{h}_{\mathrm{terr}} / (d - 1)\).
Small random numbers between \(0\) and \(10^{-6}\) were added to these rows and columns to avoid degeneracy between these components.

\paragraph{Cosmic regularization functions}
With the provisional shapes of the cosmic and terrestrial components now estimated, we would like regularization functions that preserve the power-law shape of the cosmic component during training and keep the majority of events above 3\,MeV in that component.
Two regularization functions were added to the loss function:
\begin{align}
    \Lambda(\mathbf{W}, \mathbf{H}) &= -\log L(\mathbf{X} | \mathbf{W}, \mathbf{H}) + \alpha_1 n f_1(\mathbf{W}) + \alpha_2 n f_2(\mathbf{W})
\end{align}
where \( \alpha_1 \) and \( \alpha_2 \) are dimensionless regularization parameters, \(n\) is used to account for the number of training spectra, and \(f_1\) and \(f_2\) are the additional penalty functions.

The \(f_1\) regularization term is used to compare the shape of the first column of \(\mathbf{W}\) with \(\mathbf{w}_{\mathrm{cos}}\).
We chose to use the symmetric Kullback-Leibler (KL) divergence~\cite{kullback_information_1951} since it is a natural metric for comparing the shape of two normalized, non-negative distributions.
For example, this divergence has been used in the analysis of hyperspectral data to match NMF components with laboratory-measured spectra~\cite{pauca_nonnegative_2006}.
The symmetric KL~divergence between normalized vectors \(\mathbf{p}\) and \(\mathbf{q}\) takes the form:
\begin{equation}
    D_S(\mathbf{p}, \mathbf{q}) = \sum_i (p_i - q_i) \log\left(\frac{p_i}{q_i}\right)
\end{equation}
Since we desire that the cosmic component maintain its power-law shape at high energies, but we are uncertain what the shape will be at low energies (in fact, the NR models indicate we should anticipate a 511\,keV peak feature), we only apply the penalty to the components of the first column of \(\mathbf{W}\) that are above 1250\,keV.
The exact form of \(f_1\) and the positive and negative terms in its gradient (\(\boldsymbol\nabla_1^{+}\) and \(-\boldsymbol\nabla_1^{-}\)) are given in Appendix~\ref{sec:append}.

The \(f_2\) regularization term is used to suppress the contributions above 3\,MeV by any components except for the first column of \(\mathbf{W}\).
This regularization is needed as pressure to keep the cosmic contribution entirely in the first column of \(\mathbf{W}\), since \(f_1\) on its own does not guarantee that desired property.
For \(f_2\) we chose the sum of \(\mathbf{W}\) over the nominally terrestrial rows and spectral bins above 3\,MeV.
The exact form of \(f_2\) and its gradient (\(\boldsymbol\nabla_2^{+}\)) are given in Appendix~\ref{sec:append}.

In the presence of regularization terms, the multiplicative update rules for NMF can be modified to include the gradients of the regularization terms (e.g.,~\cite{becker_adaptive_2015}).
The multiplicative update rule for \(\mathbf{W}\) (\fref{eq:update_W}) is changed to
\begin{align}
    \mathbf{W} &\leftarrow \mathbf{W} \odot \left(\frac{ \left(\frac{\mathbf{X}}{\mathbf{W} \mathbf{H}}\right) \cdot \mathbf{H}^T + \alpha_1 n \boldsymbol\nabla_1^{-}}{\mathbf{1}_{m,n} \cdot \mathbf{H}^T + \alpha_1 n \boldsymbol\nabla_1^{+} + \alpha_2 n \boldsymbol\nabla_2^{+}}\right), \label{eq:update_W_modified}
\end{align}
where \(\boldsymbol\nabla^{+}\) are the positive and \(\boldsymbol\nabla^{-}\) the negative parts of the gradients (see appendix).
This change means that the NMF solution continues to approximately follow the negative gradient of \( \Lambda \) while maintaining non-negativity, however, values of \( \alpha_1 \) and \( \alpha_2 \) that are too large can lead to an overprioritization of the regularization functions relative to the Poisson likelihood.
To keep the regularizations from having an undue influence early on during the training, we delayed the application of the regularization terms until after 500~iterations of the multiplicative update rules so that gradients of the Poisson loss were first able to stabilize.
We chose \( \alpha \) values by starting with small values and increasing them until there was a small but noticeable effect on the shapes of the components relative to their unregularized counterparts after 10,000 iterations.
We found that \(\alpha_1 = 10^{-2}\) and \(\alpha_2 = 10^{+1}\) were suitable to meet this goal.


\paragraph{Cosmic component training results}
The final results of cosmic component initialization and regularization are shown in the middle column of~\Fref{fig:nmf_spec_all} as CR-\(d\) for \(d=\)~2 to 4.
These models show that component~0 is consistent in shape even as the number of NMF components is increased.
Even though the regularization is only applied above 1250\,keV, the shape of component~0 below that energy is remarkably similar across the models, and all components show the expected 511\,keV emission line from atmospheric positrons, which was not an engineered feature.

Also of note is that, similar to the NR models, as the number of components is increased, different KUT ratios are captured by the components, although now the KUT variations are all relegated to the non-cosmic components.
For example, in CR-4, component~3 has a high amount of \isot{U}{238} and \isot{Th}{232} series lines relative to the other components.
Some of the non-cosmic components even resemble the NR components, such as CR-3 components~1 and 2, which look similar to NR-3 components~1 and 2, respectively.


\subsection{Maximizing the covariance of the NMF weights (WR)}
Finally, a third class of NMF models, which we shall refer to as weight covariance regularization (WR-\(d\)), was generated.
Rather than using physics to inform the spectral shape (\(\mathbf{W}\)), this regularization applies a desired mathematical property to the temporal evolution of the weights (\(\mathbf{H}\)).
For example, if we assume that the features arise from spatially distinct sources of radiation in the environment (e.g., soil versus buildings), we might expect them to evolve somewhat independently of one another over time.

To put this property in mathematical language, we calculated the row-wise mean of the weights \(\mathbf{H}\), which is \(\frac{1}{n} \mathbf{H} \mathbf{1}_{n,n}\), and then the row-wise covariance of \(\mathbf{H}\):
\begin{align}
    \mathrm{var}[\mathbf{H}] &= \frac{1}{n} \left(\mathbf{H} - \frac{1}{n} \mathbf{H} \mathbf{1}_{n,n} \right) \left(\mathbf{H} - \frac{1}{n} \mathbf{H} \mathbf{1}_{n,n} \right)^{\top} \\
    &= \frac{1}{n} \mathbf{H} \mathbf{C} \mathbf{C}^{\top} \mathbf{H}^{\top} \\
    &= \frac{1}{n} \mathbf{H} \mathbf{C} \mathbf{H}^{\top},
\end{align}
where \(\mathbf{C} = \mathbf{I} - \frac{1}{n} \mathbf{1}_{n,n}\) and we have used the fact that \(\mathbf{C}\) is both symmetric and idempotent, meaning \(\mathbf{C} \mathbf{C}^{\top} = \mathbf{C}^2 = \mathbf{C}\).

We want \(\mathrm{var}[\mathbf{H}]\) to be as diagonal as possible, since that would mean the rows of \(\mathbf{H}\) vary independently.
There are multiple ways to apply regularizations to achieve this goal.
Following a similar constraint in local NMF (LNMF)~\cite{li_learning_2001}, the trace of the covariance matrix was maximized.
This choice allows the regularization function and its gradient (calculated symbolically using~\cite{matrixcalculus}) to take the forms:
\begin{align}
    f_3(\mathbf{H}) &= -\frac{1}{n} \mathrm{tr} ( \mathbf{H} \mathbf{C} \mathbf{H}^{\top} ) \\
    \frac{\partial f_3}{\partial \mathbf{H}} &= - \frac{2}{n} \mathbf{H} \mathbf{C},
\end{align}
and the positive and negative parts of the gradient become \(\boldsymbol\nabla_3^{+} = \frac{2}{n^2} \mathbf{H} \mathbf{1}_{n,n}\) and \(\boldsymbol\nabla_3^{-} = \frac{2}{n} \mathbf{H}\).
As before, the NMF models were trained using a modified multiplicative update rule (the analogous version of \fref{eq:update_W_modified} for \(\mathbf{H}\), including weighting \(f_3\) with a factor of \(n\)).
This regularization was applied to the same training data with a coefficient \(\alpha_3 = 10^{-7}\), arrived at using the same heuristic as previously.
Once again, the regularization was not applied until after 500~iterations of the multiplicative update rules.

The results of applying the weight covariance regularization to the training data for 2--4 components are shown in~\Fref{fig:nmf_spec_all}.
Compared with the NR and CR models, the cosmic continuum above 3\,MeV is shared more between multiple components, except for WR-4 where component~0 contains most of it.
The two components of WR-2 look more similar to each other than those of the other two-component models (these components will be examined further in~\Fref{sec:examination}).
Component~0 of WR-3 seems to contain mostly KUT emission, while the other two components of WR-3 are similar mixtures of KUT but differ in the low energy shape.
Finally, WR-4 seems to break the spectra into cosmic emission (component~0), a sharp low energy continuum that resembles skyshine (i.e., emission downscattered in the air from distant sources, which has its largest contributions below 200\,keV~\cite{sandness_accurate_2009}) in component~1, and two similar KUT components (2 and 3) that seem to differ mostly in the relative amount of \isot{K}{40}.


\subsection{Extracting spectral features using the NMF models}\label{sec:weight_unc}
Finally, all the NMF models \(\mathbf{W}\) (\(\mathbf{W}_{\mathrm{NR-2}}\), etc.) obtained in the previous sections using the training data were applied to the evaluation dataset from 18~August 2016 to generate new NMF decompositions (e.g., \(\mathbf{H}_{\mathrm{NR-2}}\)).
These decompositions were performed by holding each model's component matrix \(\mathbf{W}\) fixed and minimizing~\fref{eq:nll} to obtain the weight matrix \(\mathbf{H}\), i.e., by performing repeated applications of the multiplicative update rule for \(\mathbf{H}\) (\fref{eq:update_H}).
The various regularization functions were only used during the respective training processes and not during this step.
The final result for a \(d\)-component NMF model is that the \(m \times n\) matrix of spectra (\(m = 130\) and \(n = \) 4,198) is reduced to the \(d \times n\) matrix of weights \(\mathbf{H}\), or in other words that the spectral data are reduced to \(d\)~feature vectors of length~\(n\).

In addition, the NMF weights have measurement uncertainties due to Poisson statistics that will be needed to properly weight the different models later when being fit to the image features.
To estimate the uncertainty of each element of \(\mathbf{H}\), we use the Fisher information in the following manner.
For each measurement \(\mathbf{x}_i\) (the \(i\)th column of \(\mathbf{X}\)), we will call the \(d\) corresponding best-fit weights \(\mathbf{h}_i\), i.e., the \(i\)th column of \(\mathbf{H}\).
Since \(\mathbf{h}_i\) was determined using maximum likelihood, which is to say:
\begin{align}
    \mathbf{h}_i &= \argmax_{\mathbf{h}}\ \log L(\mathbf{x}_i | \mathbf{W} \mathbf{h}),
\end{align}
then the covariance of \(\mathbf{h}_i\) can be approximated using the inverse of the observed Fisher information matrix:
\begin{align}
    \mathbf{F}_i &\equiv -\left.\frac{\partial^2 \log L}{\partial\mathbf{h} \partial\mathbf{h}}\right|_{\mathbf{h}_i} \\
    &= \mathbf{W}^{\top} \mathrm{diag}\left(\frac{\mathbf{x}_i}{(\mathbf{W} \mathbf{h}_i)^2}\right) \mathbf{W}.
\end{align}
So the covariance of \(\mathbf{h}_i\) is approximately
\begin{align}
    \mathrm{var}[\mathbf{h}_i] &\approx \mathbf{F}_i^{-1},
\end{align}
and for simplicity the standard deviation of each element of \(\mathbf{h}\) was approximated as the square root of the diagonal elements:
\begin{align}
    \boldsymbol\sigma_{\mathbf{h}_i} &\approx \sqrt{\mathrm{diag}\left( \mathbf{F}_i^{-1} \right)}.
\end{align}
Then call \(\boldsymbol\Sigma_{\mathbf{H}}\) the matrix with the same dimension as \(\mathbf{H}\) where column~\(i\) is \(\boldsymbol\sigma_{\mathbf{h}_i}\).
These uncertainties are used in~\Fref{sec:analysis}.


\section{Features from panoramic imagery}\label{sec:panos}
In~\cite{bandstra_attribution_2020}, the panoramic images were labeled according to known visual classes at the FtIG MOUT facility, such as \textit{asphalt}, \textit{concrete}, \textit{gravel}, \textit{red building}, \textit{brown building}, etc.
This class selection was motivated by the small variety of visual classes at the facility as well as the extensive ground truth measurements of those materials, which revealed relative uniformity within several of the visual classes~\cite{swinney_methodology_2018}.

\begin{figure}[t!]
  \centering
    \setlength{\fboxsep}{0pt}
    \fbox{\includegraphics[width=0.85\columnwidth]{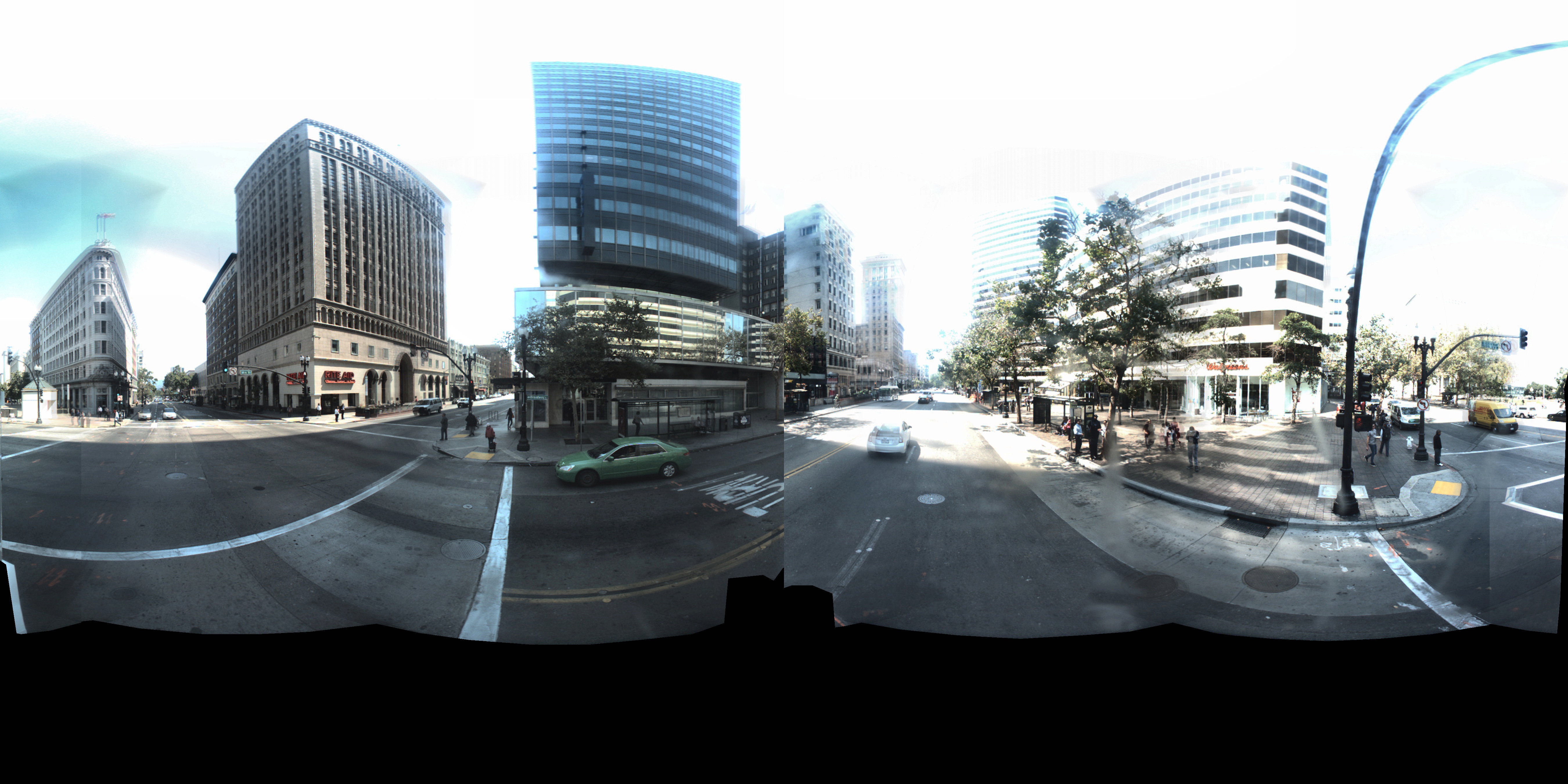}}\\
    \vspace{1pt}
    \fbox{\includegraphics[width=0.85\columnwidth]{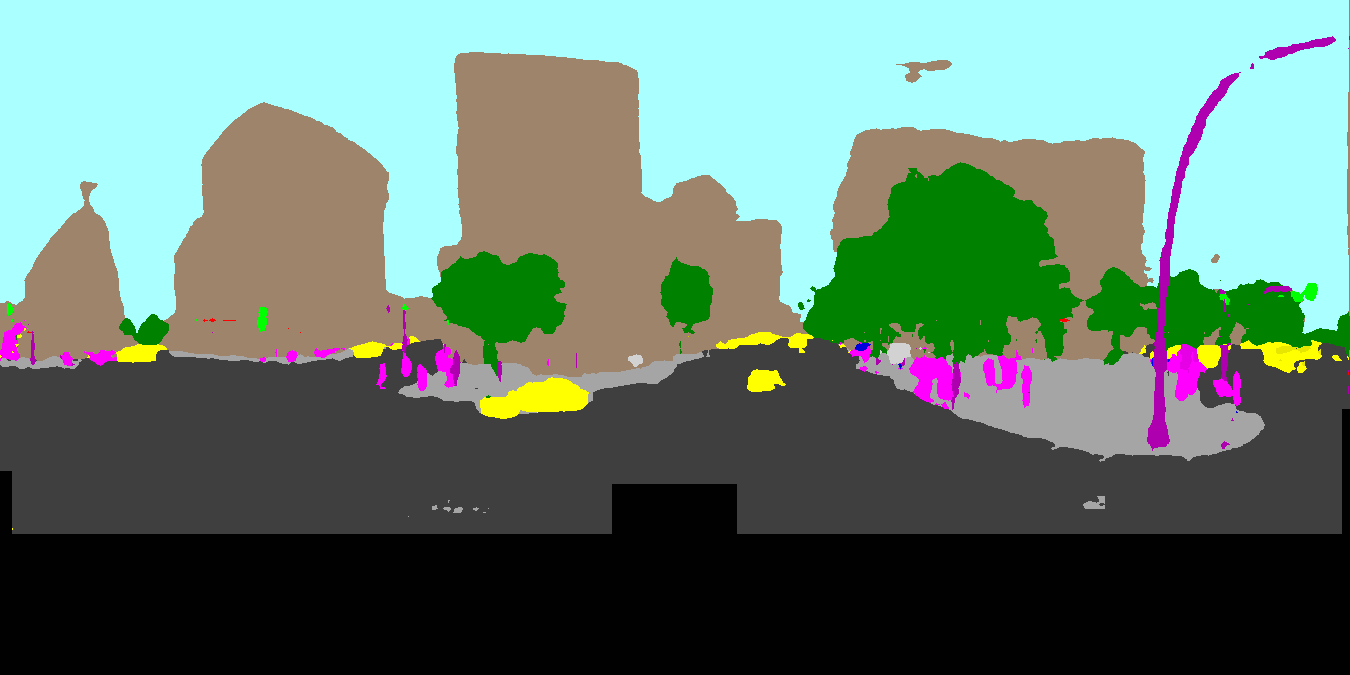}}\\
    \vspace{1pt}
    \fbox{\includegraphics[width=0.85\columnwidth]{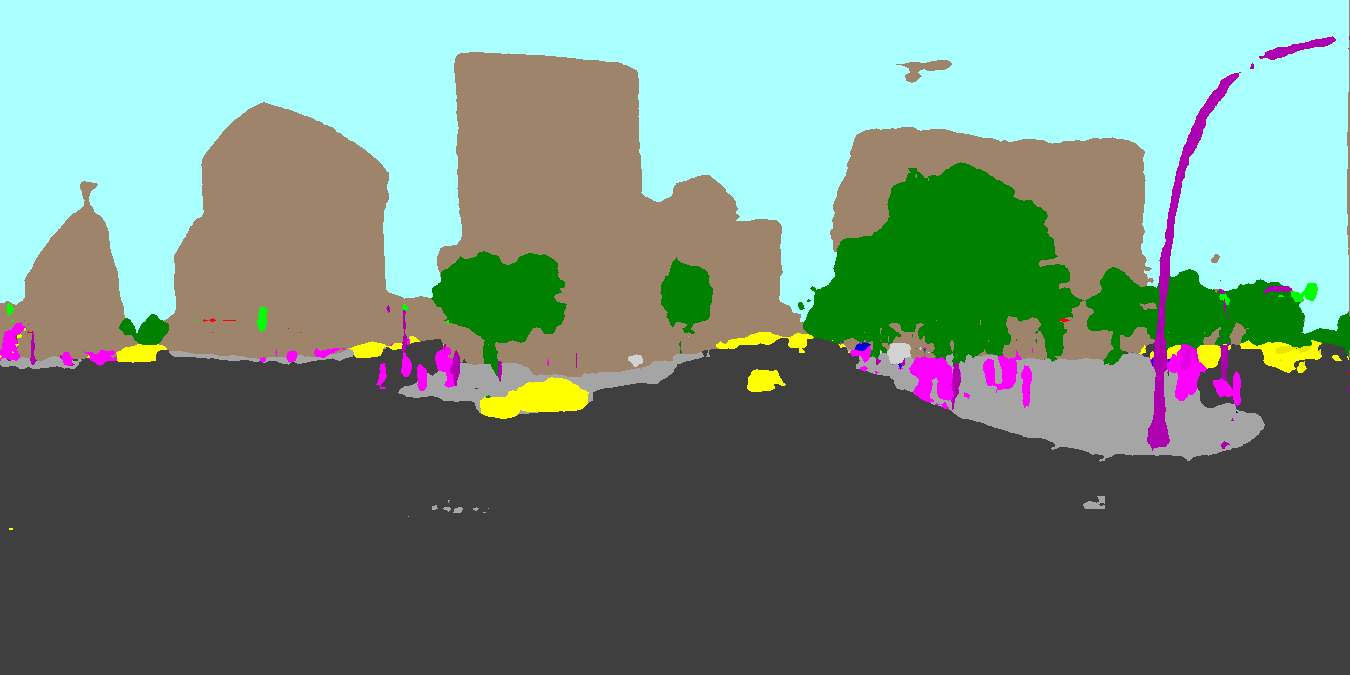}}
\caption{The creation of semantically labeled panoramic images.
The original stitched panoramic image (top) is labeled using the DeepLabv3+ model (middle).
The classes that can be seen in this image are \textit{sky} (light blue), \textit{building} (brown), \textit{road} (dark gray), \textit{sidewalk} (light gray), \textit{vegetation} (green), \textit{car} (yellow), \textit{person} (magenta), \textit{pole} (purple), and \textit{traffic sign} (light green).
The foreground region is masked out in the middle image.
The bottom image is the result of using the vehicle pose to transform nearby labeled images to estimate the missing foreground region and thus obtain a label for all \( 4\pi \) steradians around the vehicle.\label{fig:pano_pipeline}}
\end{figure}

For this dataset, since no ground truth exists but also because there are many possible visual classes, we decided to use other tools to label the imagery.
To do so, we took advantage of recent advances in machine learning for the semantic segmentation of urban scenes by using a DeepLabv3+ model~\cite{chen_encoder-decoder_2018} trained on the Cityscapes dataset~\cite{cordts_cityscapes_2016}.
The model applies 19~separate labels: flat horizontal regions (\textit{road}, \textit{sidewalk}), humans (\textit{person}, \textit{rider}), vehicles (\textit{car}, \textit{truck}, \textit{bus}, \textit{train}, \textit{motorcycle}, \textit{bicycle}), vertical structures (\textit{building}, \textit{wall}, \textit{fence}), small structures (\textit{pole}, \textit{traffic light}, \textit{traffic sign}), nature (\textit{vegetation}, \textit{terrain}), and \textit{sky}~\cite{cordts_cityscapes_2016}.
The results of the segmentation of one RadMAP panoramic image from the Oakland dataset is shown in~\Fref{fig:pano_pipeline}.
Although the model was trained on standard projection images, the results show that the model can still correctly identify many features of the panoramic images, so no retraining of the network was performed (although ideally such work should be performed in the future).
The accuracy of the model observed elsewhere was 82.1\% when measured as pixel intersection-over-union and averaged over all the classes in the evaluation~\cite{chen_encoder-decoder_2018}.
The misidentifications present in~\Fref{fig:pano_pipeline} are fairly typical of the quality of labeling observed throughout the present work (e.g., sometimes clouds are mislabeled as \textit{building} or \textit{vegetation}, which may be due to the panoramic projection).
As in~\cite{bandstra_attribution_2020}, labeled images close in time and the relative motion of RadMAP were used to fill in the masked-out foreground region at the bottoms of all the images.
Because some classes were rare, the number of classes was reduced from 19 to eight as summarized in~\fref{tab:classes}.

\begin{table}[t!]
\caption{The reduced set of eight classes made by combining DeepLabv3+ image classes.\label{tab:classes}}
\centering
\begin{tabularx}{\columnwidth}{XX}
    Reduced classes & DeepLabv3+ classes \\
    \midrule
    road & road \\
    sidewalk & sidewalk \\
    building & building, wall \\
    vegetation & vegetation \\
    terrain & terrain \\
    sky & sky \\
    vehicles and people & car, truck, bus, train, motorcycle, bicycle, person, rider \\
    other & fence, pole, traffic light, traffic sign \\
    \midrule
\end{tabularx}
\end{table}

With labeled images providing \({\cal L}_{ik}\), the image feature tensor \(R_{ij\ell}\) can be calculated using~\fref{eq:linear_model}.
As in~\cite{bandstra_attribution_2020}, when calculating these features, the solid angle element of each pixel \(\Delta \Omega_{k}\) was increased in the bottom half of the image to account for the vertical displacement of the cameras, which were appreciably higher than the array.
In addition, the effective area \({\cal A}_{jk}\) was derived from Monte Carlo simulations as described in~\cite{bandstra_attribution_2020}, where the shapes of the NMF components were used to form a weighted sum of the effective areas calculated at a variety of discrete energies.


\section{Correlations between image and spectral features}\label{sec:analysis}
This section will explain how weighted non-negative least squares was used to fit the linear model in~\fref{eq:linear_model} to the evaluation dataset, which figures of merit were used to analyze the results, and what those results might indicate about the connections between the spectra and contemporaneous imagery.


\subsection{Fitting NMF weights with the linear model}\label{sec:linear_model_nmf}
The linear model in~\fref{eq:linear_model} was fit in the following manner.
For each NMF model, the linear model was fit to each of the rows of \(\mathbf{H}\) independently of one another.
Specifically, for a given NMF model \(\mathbf{W} \mathbf{H}\) (fit to the evaluation dataset according to~\fref{sec:weight_unc}), the following steps were performed for each feature \(j\):

\begin{enumerate}
    \item For feature \(j\), choose the \(j\)th column of \(\mathbf{W}\) (\(\mathbf{w}_j\)) and the \(j\)th row of \(\mathbf{H}\) (\(\mathbf{h}_j\)).
    \item Estimate the effective area \({\cal A}_{jk}\) for spectral shape \(\mathbf{w}_j\) and the image feature tensor \(R_{ij\ell}\) as described earlier. Since \(j\) is fixed, call the resulting 2-D matrix \(\mathbf{R}_j\).
    \item With \(j\) fixed, the 2-D spectral feature tensor \(y_{ij}\) in~\fref{eq:linear_model} is a 1-D feature vector \(\mathbf{y}_j\). Identify the count rates \(\mathbf{h}_j / \boldsymbol\Delta\mathbf{t} = \mathbf{y}_j\) as this feature vector.
    \item Using~\Fref{sec:weight_unc}, estimate the uncertainty of  \(\mathbf{y}_j\) and call it \(\boldsymbol\sigma_j = \boldsymbol\sigma_{\mathbf{h}_j} / \boldsymbol\Delta\mathbf{t}\), where \(\boldsymbol\sigma_{\mathbf{h}_j}\) is the \(j\)th column of \(\boldsymbol\Sigma_{\mathbf{H}}\).
    \item Fit the linear model \(\mathbf{y}_j = \mathbf{R}_j \boldsymbol\phi_j\) using non-negative least squares weighted by \(\boldsymbol\sigma_j\), finally obtaining the photon currents \(\boldsymbol\phi_j\) and the fit \(\hat{\mathbf{y}}_j\).
\end{enumerate}

The weighted non-negative least squares fit was performed using Lasso (least absolute shrinkage and selection operator) regression, which is implemented in \texttt{scikit-learn}~\cite{scikit-learn}.
Lasso performs non-negative linear regression while trying to minimize the number of nonzero coefficients in the solution, i.e., enforcing sparsity on the solution.
The sparsity parameter was empirically set to \(0.1\), which had the effect of forcing the fits to use approximately four out of the eight total image features.


\subsection{Results of linear model fits}\label{sec:linear_results}
In order to assess the quality and physical relevance of each fit, multiple statistics were considered.

First, in order to understand the prominence of individual image features within each fit, the fraction of the fit from each image feature was calculated.
For measurement~\(i\), NMF feature~\(j\), and image feature~\(\ell\), this fraction is
\begin{align}
    f_{j\ell} &= \frac{\sum_i R_{ij\ell} \phi_{j\ell}}{\sum_i \hat{y}_{ij}}.
\end{align}

In addition to the fraction made up by each image feature, three goodness-of-fit metrics were considered to assess how well the fits represent the data.
The reduced chi-squared statistic (\(\chi^2_{\nu}\)) was chosen as a standard statistical measure of goodness of fit, however a model with large weight uncertainties (\(\boldsymbol\sigma_{\mathbf{h}}\)) can lead to an acceptable \(\chi^2_{\nu}\) value but an uninformative model.
To also provide a sense of how well the model follows trends in the data, the Pearson's correlation coefficient was calculated between the image features and a smoothed version of the data.
Finally, the usefulness of the model in predicting the count rates of each feature using the imagery was calculated using the root mean squared deviation (RMSD).

The reduced chi-squared statistic is
\begin{align}
    \left(\chi^2_{\nu}\right)_j &= \frac{1}{n} \sum_i \frac{(y_{ij} - \hat{y}_{ij})^2}{\sigma_{ij}^2}
\end{align}
and has an expected value of unity for a model that perfectly describes the data.
Technically, the degrees of freedom are not \(n\) but could be as few as \(n - 8\) depending on the sparsity of the linear model, but since \(n\) is so large this difference was ignored.

The second metric is Pearson's correlation coefficient \(r\), to measure how closely the model count rate correlates with the count rate from the NMF weights:
\begin{align}
    r_j &= \frac{\sum_i (\tilde{y}_{ij} - \mu[\tilde{\mathbf{y}}_{j}]) (\hat{y}_{ij} - \mu[\hat{\mathbf{y}}_{j}])}{\sqrt{ \sum_i (\tilde{y}_{ij} - \mu[\tilde{\mathbf{y}}_{j}])^2 } \sqrt{ \sum_i (\hat{y}_{ij} - \mu[\hat{\mathbf{y}}_{j}])^2 }},
\end{align}
where \(\tilde{\mathbf{y}}_j\) is obtained by smoothing \(\mathbf{y}_j\) using a boxcar kernel of width~5 and \(\mu[\cdot]\) is the sample mean over all \(n\) measurements.
This filtering, which smooths the data over a timescale of at least \(1.67\)\,s, was done to reduce the influence of statistical fluctuations on~\(r\).
The values~\(r\) can take range from \(-1\) (perfectly anti-correlated) to \(+1\) (perfectly correlated).

The third metric is the root mean squared deviation (RMSD), for a measure of each model's prediction error:
\begin{align}
    \mathrm{RMSD}_j &= \sqrt{\frac{1}{n} \sum_i (\tilde{y}_{ij} - \hat{y}_{ij})^2}.
\end{align}
where \(\tilde{\mathbf{y}}_j\) is once again the smoothed \(\mathbf{y}_j\).

\begin{figure*}[t!]
\centering
\includegraphics[width=0.97\textwidth]{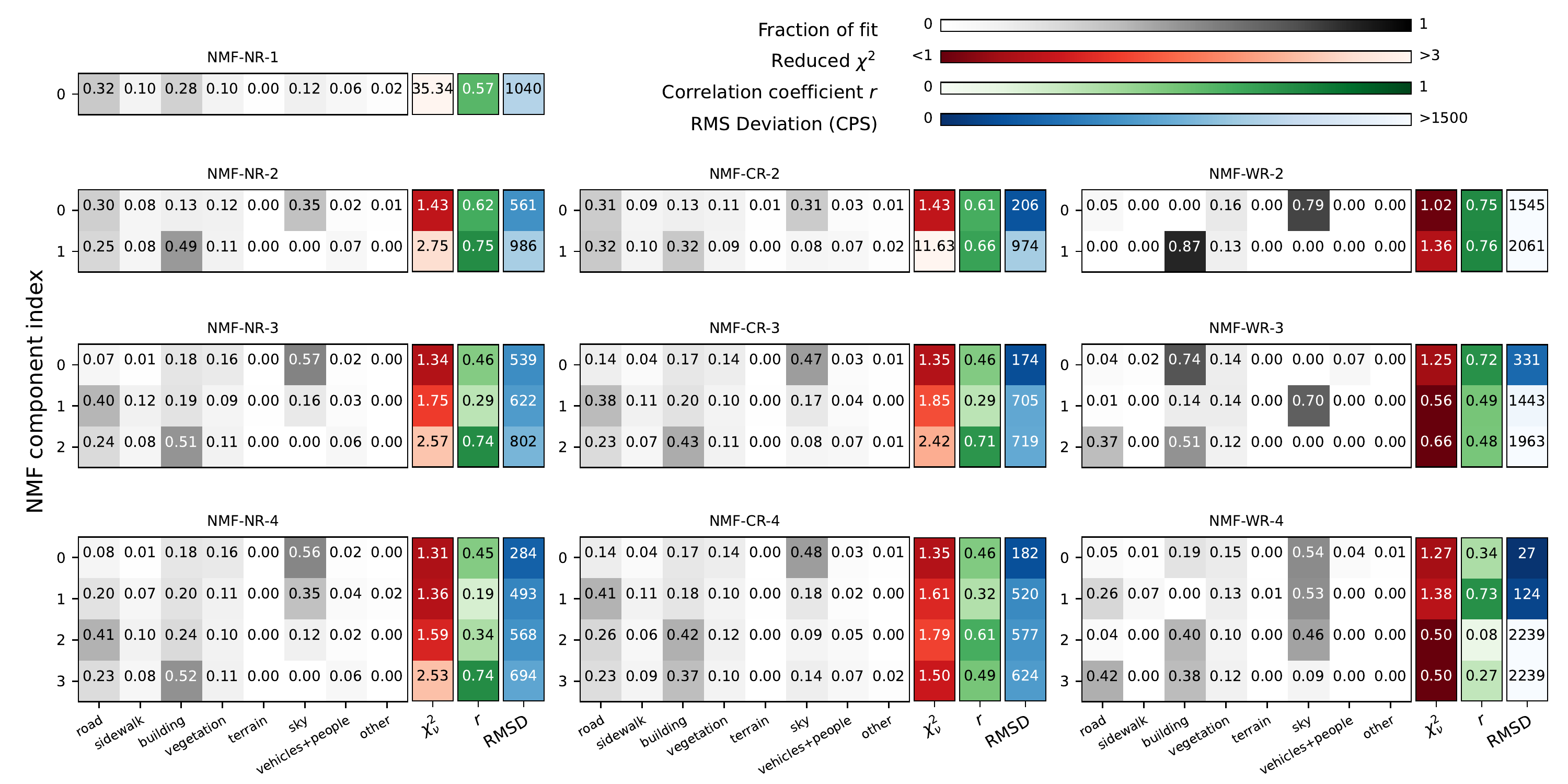}
\caption{Results of Lasso fits of the evaluation RadMAP dataset for all of the NMF models considered.
Shown in grayscale are the fractions of each fit made up by the image class, along with the goodness of fit as measured by the reduced \(\chi^2\) statistic (red), correlation coefficient (green), and root-mean-squared deviation (blue).
Darker shades indicate better performance or stronger correlation, whichever is appropriate.
The goodness-of-fit metrics represent how well the temporal evolution of the spectral features matches the linear model.\label{fig:fits_all}}
\end{figure*}

The results of fitting all of the NMF models is shown in~\Fref{fig:fits_all}, which displays the fraction of the fits made up by each of the image features as well as the goodness-of-fit metrics.
It is immediately apparent that the vast majority of the fits are not objectively good (i.e., the null hypothesis can be rejected with high confidence), since all but five of the \(\chi^2_{\nu}\) values are too large given the number of degrees of freedom, which, at approximately 4,198, would require \(\chi^2_{\nu} > 1.05\) to reject the null hypothesis at the \(10^{-2}\)~level).
In addition, the RMSD values, which range from 27 to over 2,200~counts per second, indicate that the fits frequently under-predict and/or over-predict the spectral features.
(For comparison, the mean gross count rate is 10,056 counts~per~second.)
Even the five models with acceptable~\(\chi^2_{\nu}\) values have some of the largest RMSD values.
This comparison of RMSD and \(\chi^2_{\nu}\) values reveals that the NMF-WR models, probably as a side effect of the regularization itself, are creating inflated values of \(\boldsymbol\sigma_{\mathbf{h}}\) and thus appear to be good fits in a \(\chi^2_{\nu}\) sense but in fact have a high prediction error (RMSD value).
From these metrics we can conclude that the linear model approach of relating spectral and image features is not statistically correct, and a more complex model or different features would be needed for a greater goodness of fit and lower prediction error.

On the other hand, the correlation coefficient \(r\) reveals that all models have positive correlations between the image features and the NMF-derived spectral features, ranging from weak (0.08) to somewhat strong (0.76).
Of particular interest are those models indicative of strong correlations, and especially those with both the largest \(r\) values and the largest fractions of the fits made up of a single image feature \( f_{j\ell}\).
Though the overall fit may be statistically or predictively poor, if a large fraction of the fit is made up of a single image feature, that is an indication that the spectral data still correlate with that image feature in a meaningful sense.
For example, each model except \(d = 1\) has at least one component with a fit comprised of a significant fraction (\(>\)30\%) from \textit{sky}.
The component with the largest fraction from \textit{sky} is always the component that contains the largest amount of emission over 3\,MeV, as can be seen in~\Fref{fig:nmf_spec_all}, and it also tends to be the lowest-variance component (0).
This finding comports with the prediction that atmospheric emission should have some connection to the \textit{sky} feature, and that of the natural background sources, cosmic emission has the most stable flux and composition.

Further examining the fractions from \textit{sky}, we can see that cosmic regularization, which was designed to produce an NMF component~0 that has a shape consistent with cosmic emission, does not enhance this fraction in the  CR-\(d\) models when compared to the other models --- component~0 in the equivalent NR-\(d\) and WR-\(d\) models counterintuitively have a similar or greater fraction from \textit{sky}.
The process of regularization, however, does force the component's spectral shape to be consistent across different values of \(d\).

Besides \textit{sky}, other features that have significant fractions are \textit{road} and \textit{building}.
Of particular note is that \textit{building} has some fraction values near and over 50\% (e.g., NR-3 component~2 and WR-2 component~1).
The components with large fractions of these image features tend to have low fractions from \textit{sky}, suggesting that these correlations are due to a difference between atmospheric and terrestrial emission.

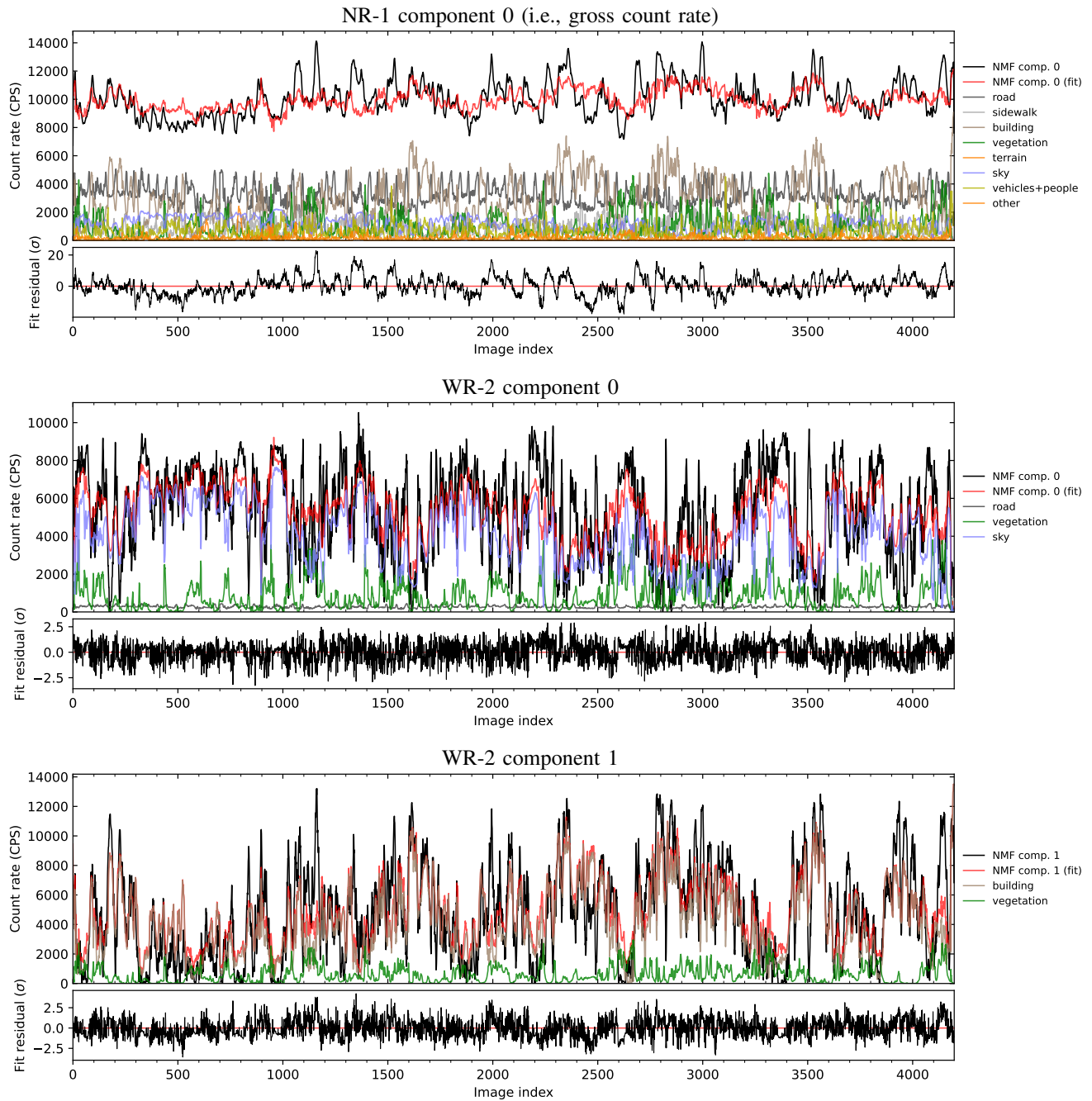
\begin{figure*}[t!]
\centering
\begin{tikzpicture}
    \draw (0, 0) node[inner sep=0] {\includegraphics[width=0.99\textwidth]{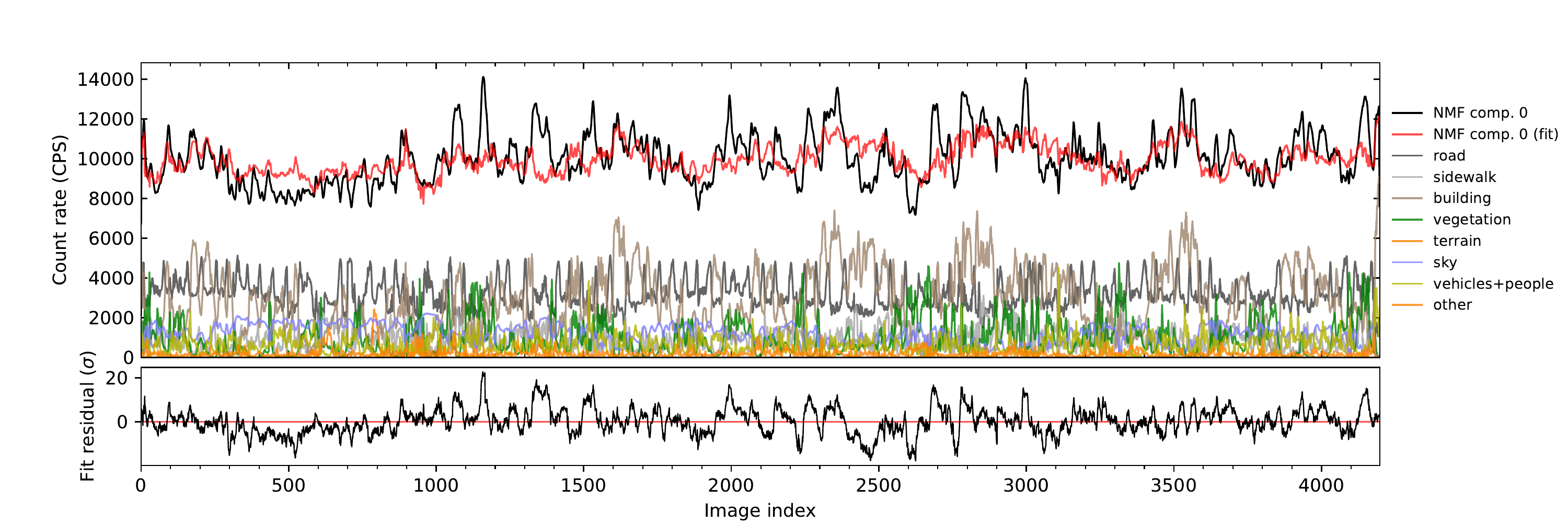}};
    \draw (0, 2.5) node {NR-1 component 0 (i.e., gross count rate)};
\end{tikzpicture}\\
\begin{tikzpicture}
    \draw (0, 0) node[inner sep=0] {\includegraphics[width=0.99\textwidth]{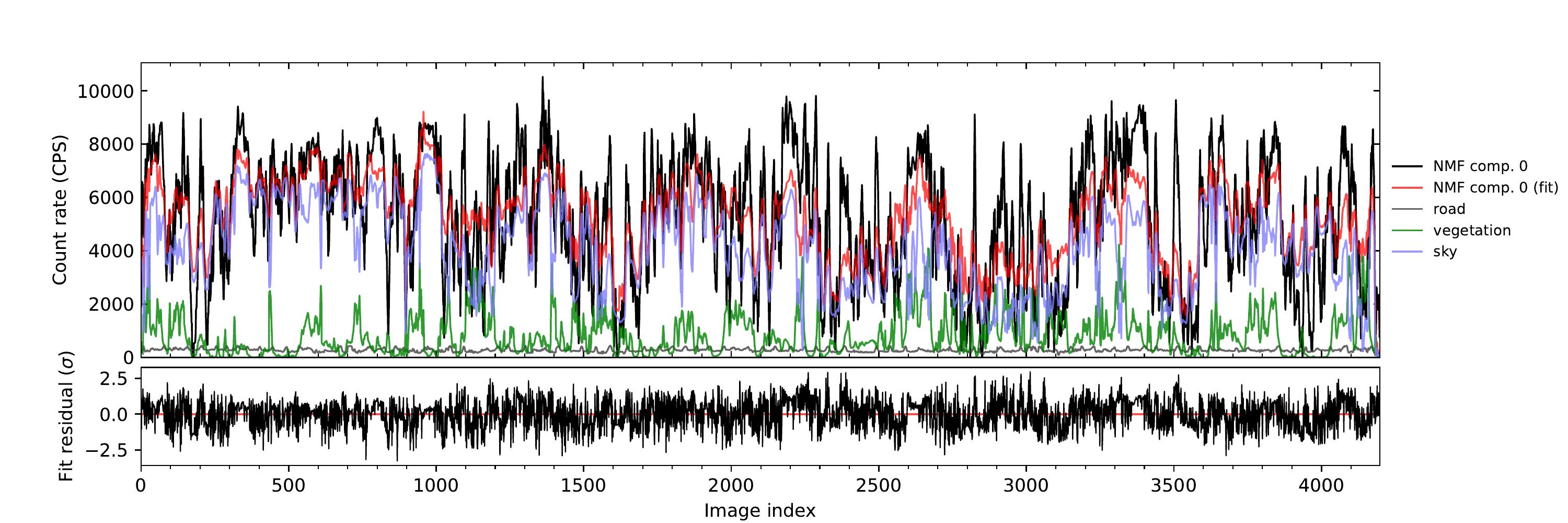}};
    \draw (0, 2.5) node {WR-2 component 0};
\end{tikzpicture}\\
\begin{tikzpicture}
    \draw (0, 0) node[inner sep=0] {\includegraphics[width=0.99\textwidth]{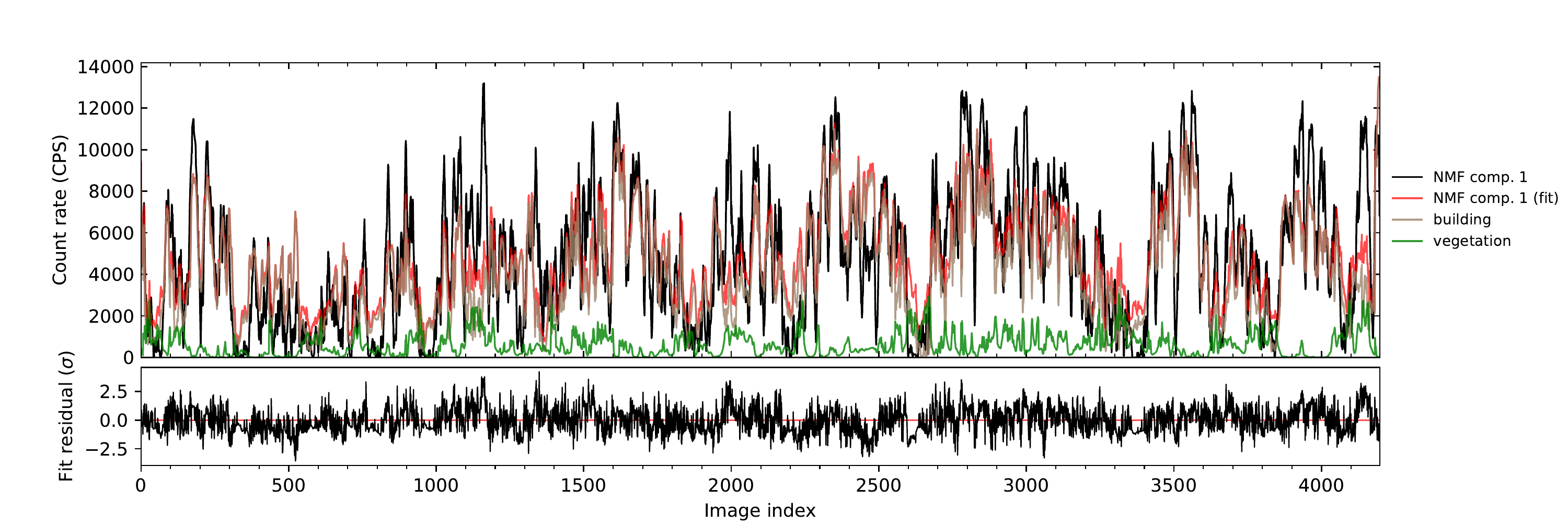}};
    \draw (0, 2.5) node {WR-2 component 1};
\end{tikzpicture}
\caption{Results of Lasso fits to NMF component weights from model WR-2 (middle and bottom).
For comparison, the Lasso fits to the single component of NR-1, which are the gross counts, is also shown (top).
The spectral features from NMF, shown in black, are smoothed with a boxcar filter of width~5 to improve clarity, but the residuals are not smoothed.
Image features with coefficients of zero (due to Lasso's sparsity regularization) are not shown. \label{fig:fits_nmf2}}
\end{figure*}

The models exhibiting the largest fractions for both \textit{sky} and \textit{building} are the 2- and 3-component NMF models with weight covariance regularization (WR-2 and WR-3).
We will examine WR-2 in further detail, since each of its two components has a large fraction associated with a single image feature.
For this model, both NMF components have large fractions explained by image features --- component~0 has a 69\% fraction from \textit{sky}, and component~1 has an 81\% fraction from \textit{building}.
These fits and their breakdown by image feature are shown in~\Fref{fig:fits_nmf2}, with a comparison to fit to the simplest model, NMF-NR-1, which is just the gross count rate.
Although there are significant departures between the data and the fits (which is the subject of~\Fref{sec:examination}), the general trends in both are strong.

\subsection{Examination of WR-2}\label{sec:examination}
The WR-2 model will be further examined here, both for its spectral component shapes and when the linear model fit when the linear model fit is poorest.

\begin{figure}[t!]
\centering
\includegraphics[width=0.95\columnwidth]{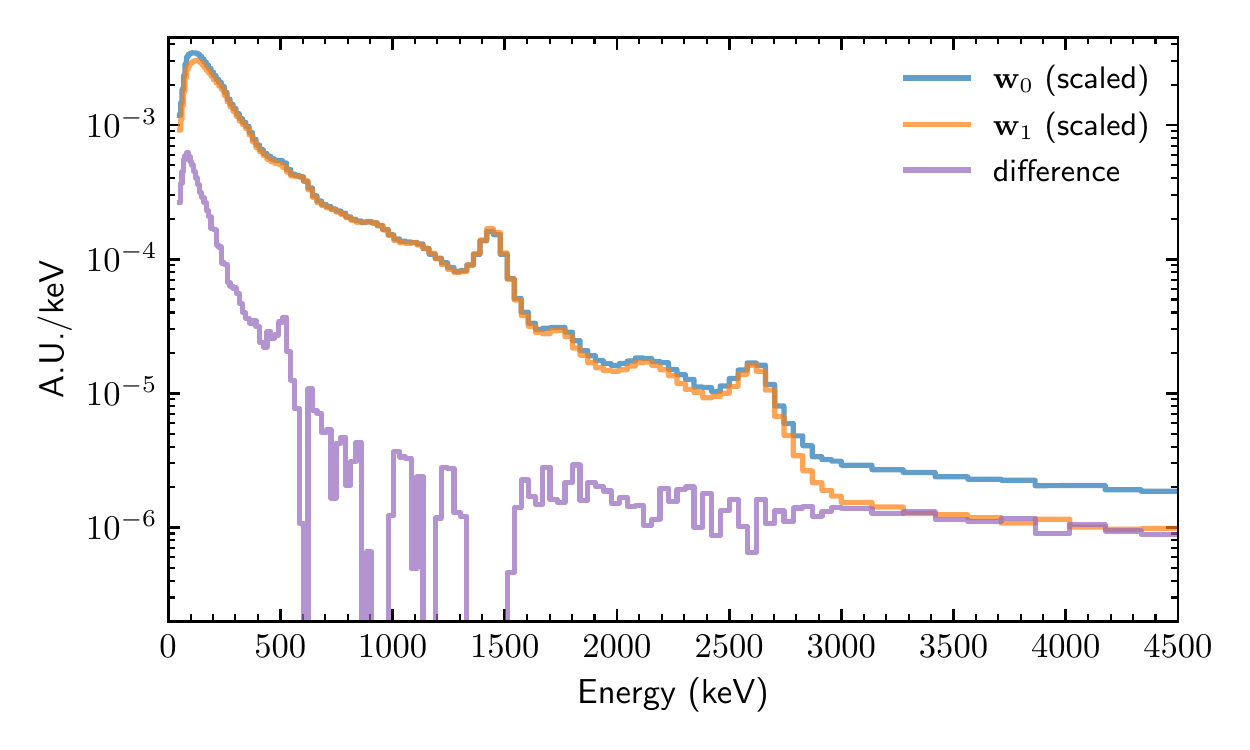}
\caption{The difference between the two components of NMF-WR-2, after scaling the components to be the same in the 1460\,keV region.
The resulting spectrum resembles the cosmic components of the NMF-CR models.\label{fig:nmf2_scaled}}
\end{figure}

\paragraph{Spectral component shapes}
The two components of this NMF model are shown in~\Fref{fig:nmf_spec_all}.
\Fref{fig:nmf2_scaled}, which shows the two components rescaled to match in the region around 1460\,keV, reveals that they have a similar shape between 500 and 2800\,keV\@.
The main contrasts between the components are that component~0 has relatively increased rates above 3\,MeV, below 200\,keV, and also near 511\,keV.
To draw these differences into sharper contrast, the two scaled components are subtracted from each other and shown in~\Fref{fig:nmf2_scaled}.
The shape of this difference spectrum is strikingly similar to the cosmic-regularized component (component~0) of the CR-\(d\) models in~\Fref{fig:nmf_spec_all}.
This finding suggests that the difference between the two WR-2 components is that component~0 contains an additional amount of cosmic and skyshine emission in addition to the terrestrial, skyshine, and cosmic emission shared by both components.
Therefore, component~0 may represent emission that has relatively more ``distant'' emission, and component~1 may represent emission from relatively more ``nearby'' emission, although both include a mixture of all emission types.
Though different kinds of regularizations have been applied to obtain many different NMF models, this cosmic and skyshine spectrum and its association with \textit{sky} consistently emerges from this analysis.

\paragraph{Discrepancies between fits and NMF weights}\label{sec:discrepancies}
Examining the weights of WR-2, it is clear from~\Fref{fig:fits_nmf2} that there are certain regions where the fit and the data significantly diverge.
The largest discrepancies were examined to see if there were consistent explanations for why the models may have failed in a manner consistent with their assumed (more distant versus more nearby) origins.
The twelve regions containing the largest RMSD values were identified and the index of the maximum was recorded.
In many, but not all, cases, a plausible explanation could be found for the differences.

\begin{figure}[t!]
  \centering
    \setlength{\fboxsep}{0pt}
    \fbox{\includegraphics[width=0.85\columnwidth]{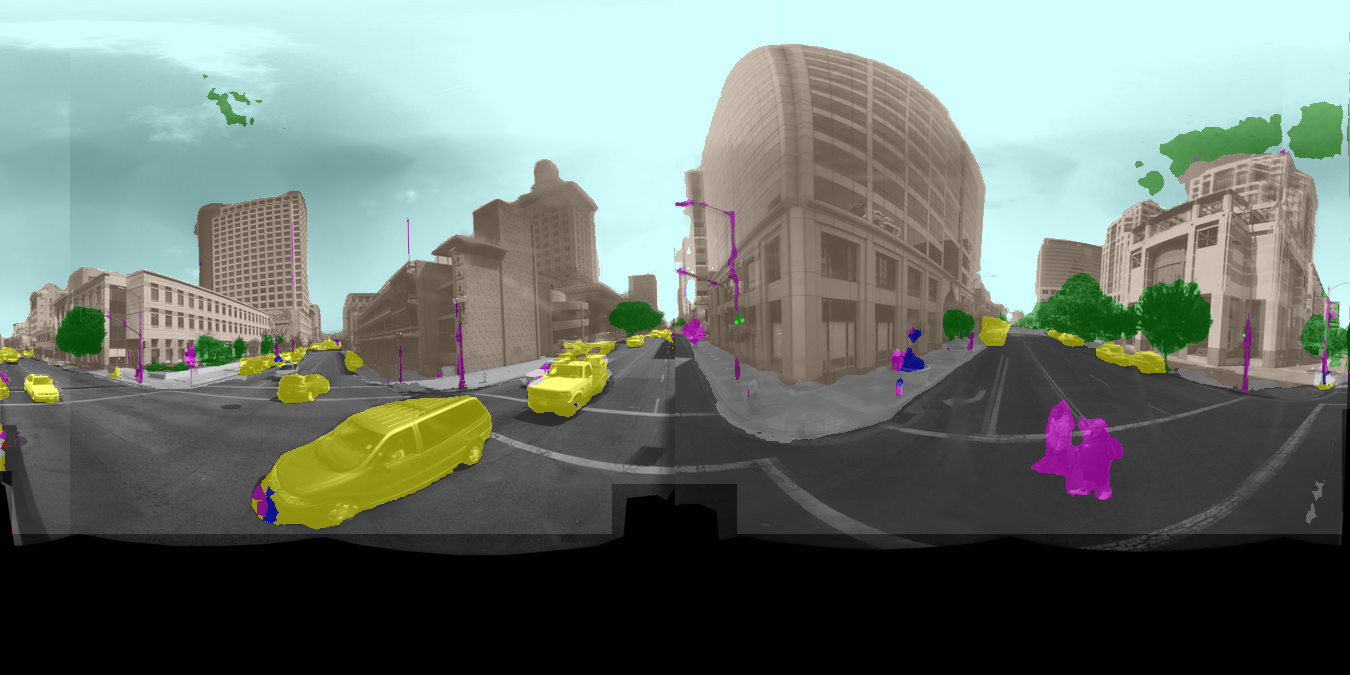}}\\
    \vspace{1pt}
    \fbox{\includegraphics[width=0.85\columnwidth]{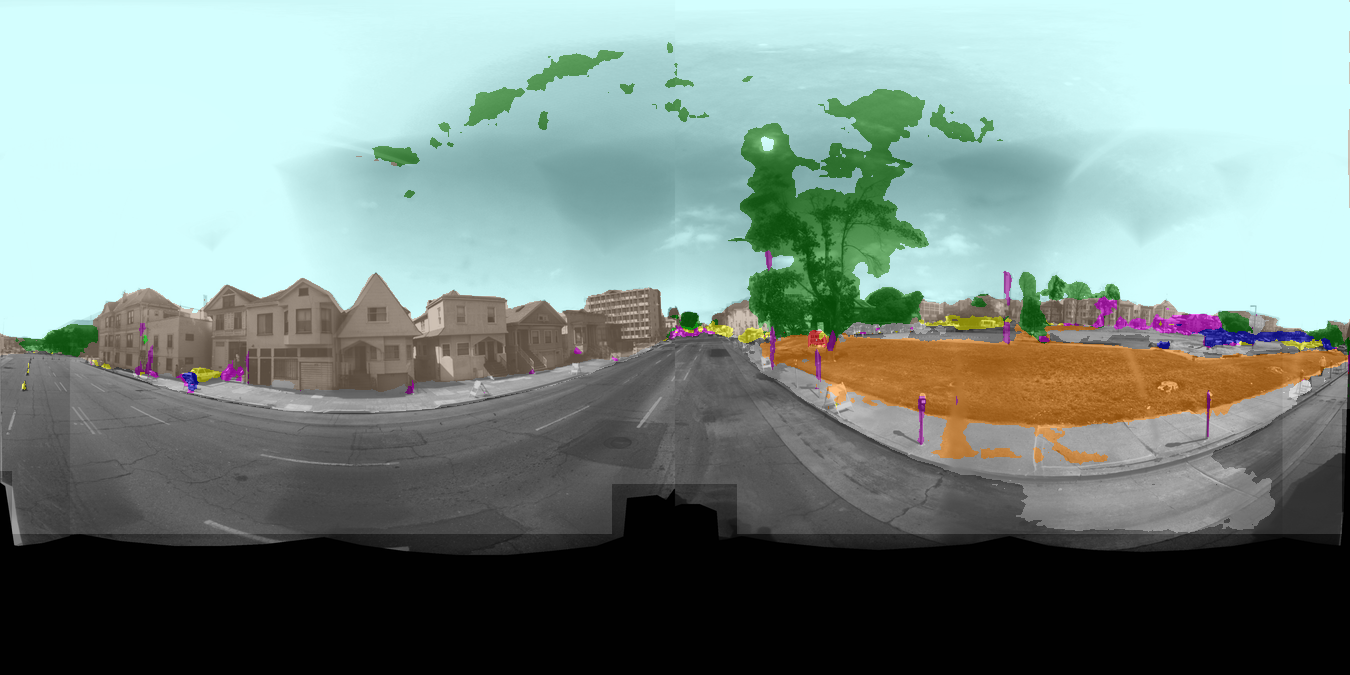}}\\
    \vspace{1pt}
    \fbox{\includegraphics[width=0.85\columnwidth]{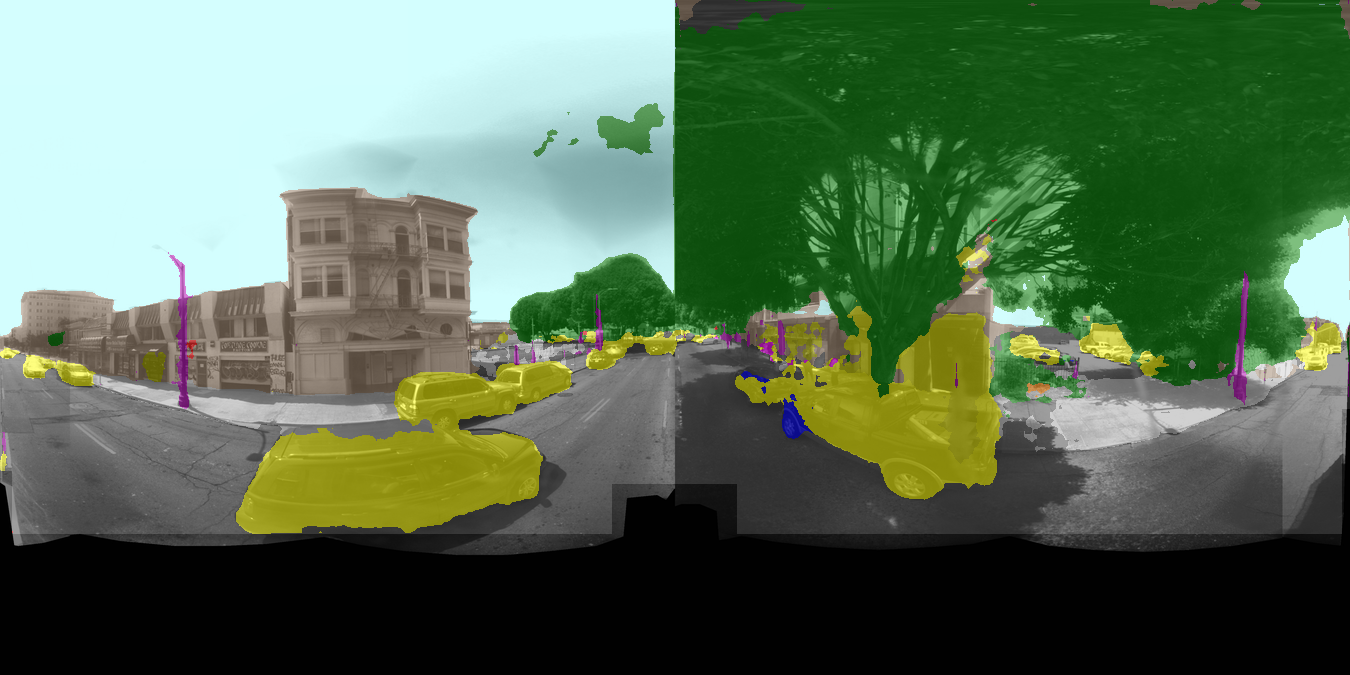}}
\caption{Panoramic images at the times when the linear model under-predicts the spectral features from component~0 of WR-2.
Intersections (top) and plazas or open fields (middle --- the orange color is the \textit{terrain} label) are common scenarios that occur when the model under-predicts.
One large discrepancy was seen when a large tree obscured the sky behind it (bottom).
DeepLabv3+ labeling is overlaid as a transparency.\label{fig:discrepancies0}}
\end{figure}

For component~0, the twelve largest discrepancies where the fit under-predicted the NMF weights were found and examined (these discrepancies are in raw count rate, not \(\sigma\) residual).
In nine of these cases, the panoramic images showed that the vehicle was either in the center of an intersection or adjacent to a large open lot or plaza (see top and middle images in~\Fref{fig:discrepancies0} for examples).
In these scenarios, buildings are farther away from the sides of the vehicle, which breaks up the typical ``urban canyon'' scenario found in the data set, and so more distant, downscattered terrestrial emission such as skyshine might be expected due to greater exposure to the distant roads and terrain.
At the same time, the portion of the image subtended by \textit{sky} may increase somewhat when in these locations, but the increase might not fully capture what is actually a three-dimensional phenomenon that should include, e.g., scattering in the air from distant terrestrial emission (see~\cite{salathe_determining_2021} for the development of a 3D emission model).
In one of the other cases, a large tree obscured the open sky behind it, thus leading to a much lower estimated contribution from \textit{sky} than if the tree were not there, presumably without commensurate attenuation by the tree (bottom images in~\Fref{fig:discrepancies0}).
In two other cases there were no clear explanations.

The twelve largest discrepancies where the fit over-predicted the component~0 weights were also examined.
In six of these cases, the vehicle was directly adjacent to a large building on its right side, which is the side of the NaI array with a larger effective area.
The other six cases offered no obvious common features, and in general it is not known what the main reasons for these over-predictions are.

\begin{figure}[t!]
  \centering
    \setlength{\fboxsep}{0pt}
    \fbox{\includegraphics[width=0.85\columnwidth]{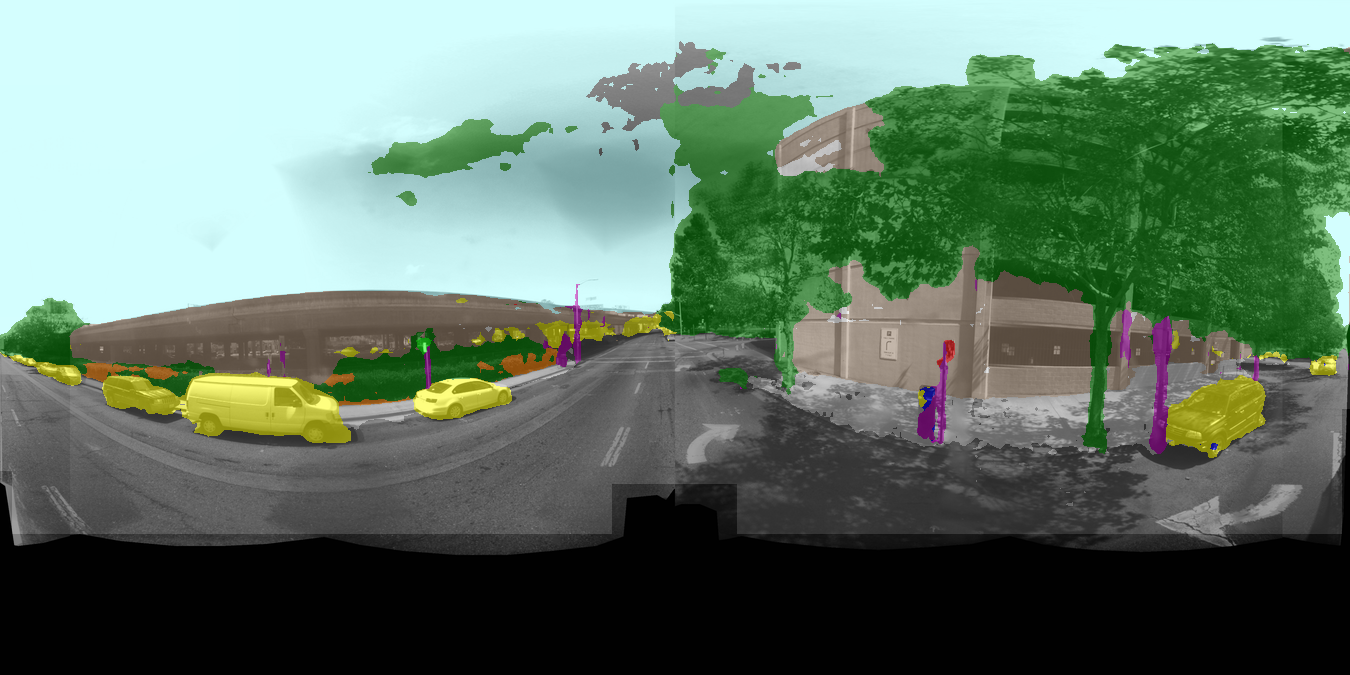}}\\
    \vspace{1pt}
    \fbox{\includegraphics[width=0.85\columnwidth]{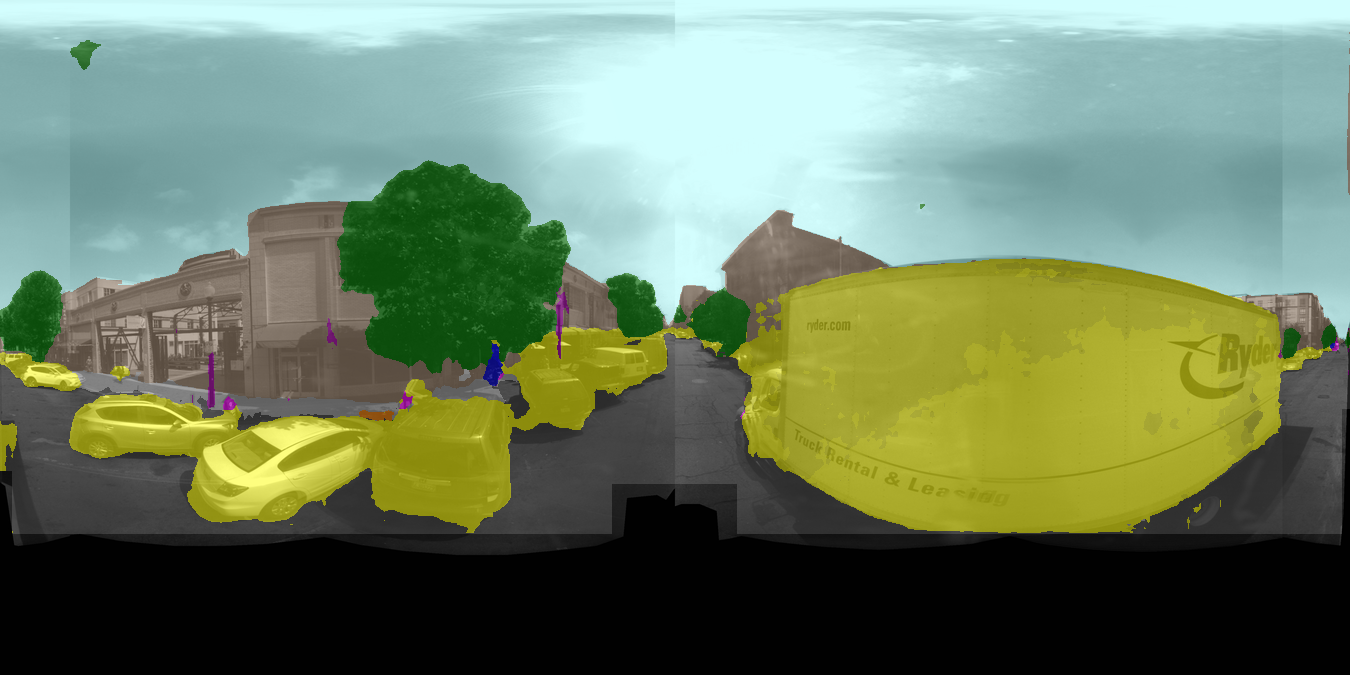}}\\
    \vspace{1pt}
    \fbox{\includegraphics[width=0.85\columnwidth]{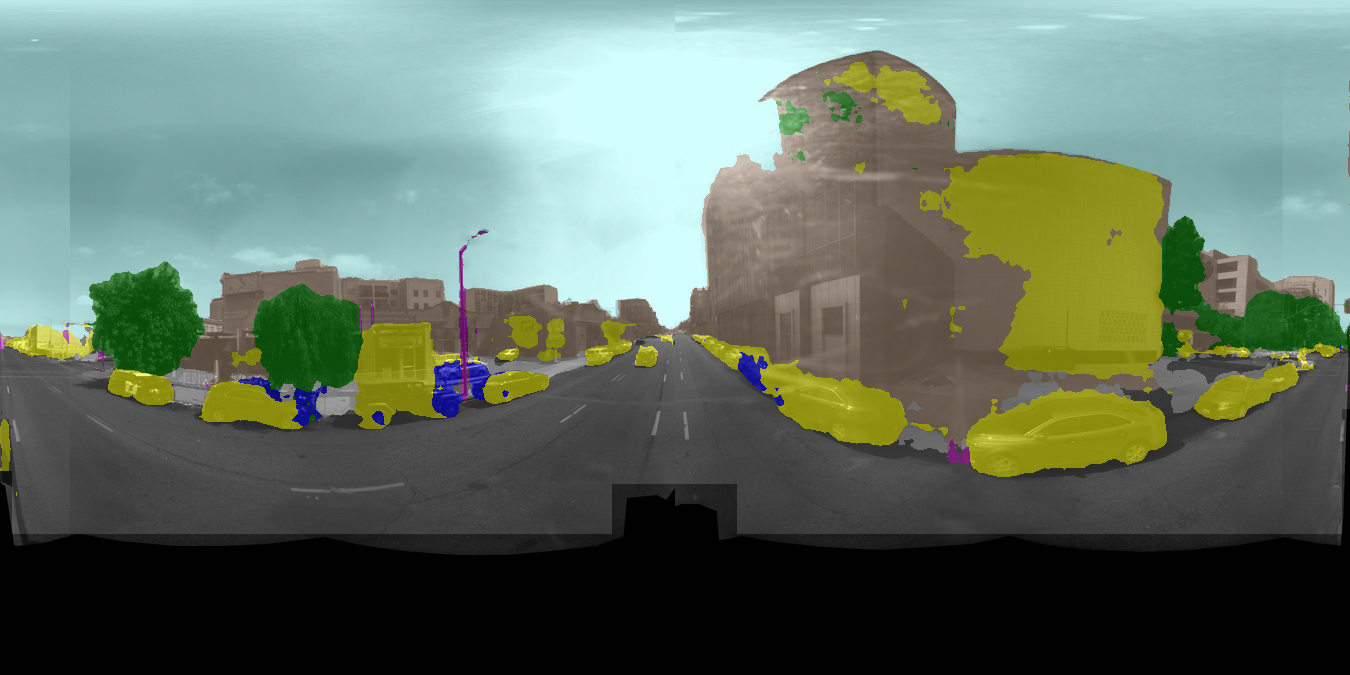}}
\caption{Panoramic images at the times when the linear model under-predicts the spectral features from component~1 of WR-2.
Trees (top) or large vehicles (middle) can visually obscure emission from buildings, as can errors in the labeling of buildings (bottom).
DeepLabv3+ labeling is overlaid as a transparency.\label{fig:discrepancies1}}
\end{figure}

For component~1, the same inspections were performed of the top twelve discrepancies caused by model underprediction.
There were five cases where a large foreground object obscured a building behind it (four were trees and one was a large truck).
Two such examples are shown in the top and middle images of~\Fref{fig:discrepancies1}.
It is likely that those particular objects do not fully attenuate the emission from the buildings that is passing through them, and since the model coefficients have low photon currents from \textit{vegetation} and \textit{vehicles and people}, there is a net deficit in the model prediction.
These cases thus reveal another weakness in using 2D imagery alone instead of a 3D model (see~\cite{salathe_determining_2021}) that could account for material present behind the foreground objects.
Two other cases involved encounters with two different facades of the same building, a possible indication that that particular building likely comprises material with higher concentrations of radioactive material than other buildings in the area.
The last two cases involved the DeepLabv3+ model's mislabeling of large parts of the buildings as \textit{vegetation} or \textit{vehicles and people}, thus decreasing the linear model's prediction (e.g., the bottom image of~\Fref{fig:discrepancies1}).

Of the dozen cases where the model over-predicted component~1, five involved passing by wooden buildings, which were less common in the downtown area than concrete and brick buildings and possibly indicates that the lower emissivity of buildings constructed from wood results in poor fits in this model.
Three cases involved the right side of the vehicle being very close to (non-wooden) building facades, which could be an indication that those buildings were lower in activity than the average building.
The other scenarios offered no clear commonalities.

\section{Discussion}\label{sec:discuss}
The work presented here shows that features derived from measurements of the gamma-ray background in an urban area can be correlated with some fidelity to features derived from panoramic visual imagery of the same.
In particular, the WR-2 model decomposition of the measured spectra resulted in one component that is strongly associated with the presence of buildings, and another that is strongly associated with the visibility to the sky.
Although all of the model fits had high prediction error and only a few were statistically acceptable, the analysis presented here nevertheless finds connections between spectral and image features similar to what has previously been seen with the same mobile system but in a smaller and more radiologically uniform urban environment~\cite{bandstra_attribution_2020}.

The shapes of the spectral components from NMF suggest there may be a reason for the associations, since component~0's spectrum shows what is likely to be more cosmic and skyshine emission than component~1, and visibility to the sky should be a rough proxy for exposure to cosmic and skyshine emission.
In addition, nearby KUT emission, captured at a higher proportion by component~1, should increase when there are larger surfaces of KUT emission around the vehicle, a rough proxy for which is the size of nearby buildings.
These same associations were noted in a separate RadMAP dataset at Fort Indiantown Gap, Pennsylvania~\cite{bandstra_attribution_2020}, and a similar separation of gamma-ray emission into nearby and distant emission has also been observed in airborne data~\cite{bandstra_modeling_2020}.
The separation of urban gamma-ray backgrounds into nearby (\textit{building}-dominated) and distant (\textit{sky}-dominated) emission may therefore be a general phenomenon.
The connection between cosmic emission and the \textit{sky} feature is also reminiscent of the connection between measured fast neutron backgrounds and the fraction of the sky visible and not shielded by buildings~\cite{iyengar_systematic_2015, davis_fast_2017, glick_deployment_2021}.
The results of this work suggest an analogous result for gamma-ray backgrounds.

Another result of this work is that in all of the models, little to no emission is fit to clutter: the classes \textit{vehicles and people} and \textit{other}.
The fact that they contribute so little to the fits comports with the notion their presence largely serves to attenuate background emission coming from the buildings or roads behind or beneath them, which is a known impact of such clutter on these systems~\cite{stewart_understanding_2018}.
Of course, the rare exception for clutter is when vehicle cargo is particularly radioactive or a pedestrian has undergone a nuclear medicine treatment, but such events have been screened for and are not in this dataset.

Obviously, models that tie gamma-ray backgrounds more strongly to surrounding imagery may exist than what is shown here.
This work takes the brute-force approach of generating various sets of gamma-ray features, using NMF models that have been tailored by regularization approaches that were inspired by physics or statistical considerations.
These gamma-ray features are then compared to image features that were derived using a single semantic segmentation approach with fixed visual categories.
This approach results in us only examining a small parameter space within a high dimensional dataset.
Correlations were found to exist between some of these feature sets, but there could easily be other correlations that exist outside of the small set of features examined here.
For example, different kinds of buildings (e.g., brick versus concrete versus wooden) might cluster together into different and predictable levels of KUT flux that the DeepLabv3+ model is currently blind to since it collapses all building types into one class.
But even for the same type of material, the KUT activities can vary by orders of magnitude~\cite{trevisi_natural_2012}, and therefore any model that seeks to make a unique KUT activity prediction based on imagery alone is going to have difficulty.
Also, the correlations might break down outside of this relatively small region of this particular city, if, for example, the flux from the road bed or nearby soil is much larger in another location, or if the magnitude and composition of surface fluxes among buildings becomes more variable.
Other radiological effects that might not be readily noticed in imagery are those due to weather, especially the increased background due to radon progeny during rainfall events, and while the data analyzed herein did not include any rainfall, we posit other contextual sensors could be incorporated into a model to address such complications.
To be as robust and accurate as possible, any model like this would have to be trained and deployed in the same local area, but some phenomena such as sky-related features would likely persist even if a model were broadly trained.

One area for improvement over this work would be to extend the model beyond the simple linear model used here.
A trivial extension would be to include an intercept in the linear model, which would allow the model to capture spectral emission that does not change in time and may not be attributable to imagery.
Some nonlinear models may also be useful.

Another area for improvement would be in the simultaneous engineering and correlation of spectral and image features.
We attempted a small move in this direction via regularization with NMF\@.
Training on the evaluation dataset, we used a regularization term that minimized the symmetric KL divergence between component~0's weights and the \textit{sky} image feature, and also the symmetric KL divergence between component~1's weights and the \textit{building} image feature.
The resulting 2-component model closely resembled NMF-WR-2, and had only slightly higher fractions \(f\) for the two image classes.

The simultaneous engineering and correlation of spectral and image features is perhaps a task better performed by deep convolutional neural networks (CNNs).
Indeed, CNNs have already been used for this task with some initial success~\cite{kaffine_background_2017}.
This work could provide guidance to the design of future CNNs as well as in aiding the interpretability of a trained CNN model.
For example, a model that takes its inspiration from the DeepLabv3+ model for its initial layers, and with one or more fully connected layers at the end to predict the spectrum, might be a promising direction to pursue (one fully connected layer is equivalent to a linear model like NMF).
Of course, one would want to explore such a model not only in Oakland, but in many other cities to understand the transferability of the knowledge, since different cities have different background distributions (e.g.,~\cite{mitchell_gamma-ray_2015}).

Given the complexity of urban scenes, a model that can generate a spectrum based only on a single image may be possible but may not be robust.
Instead, another potential avenue of research is to explore machine learning models that update the current background estimate using both recently measured spectra and the changing context around the system.
A reinforcement learning model, that means a model that is trained to make predictions based on the recent state and new data, could be appropriate here.
Such a model could be more flexible in generic urban scenes by learning universal trends in spectral variability that correspond to noticeable trends in imagery, such as learning that a vehicle pulling up next to the system will depress the background count rate by a certain amount.
Beyond these suggested machine learning approaches, there likely exist numerous other approaches that could leverage the observations made herein in effort to improve radiological anomaly detection algorithms' efficacy when confronted with the realistic spectral and temporal variability encountered when operating in urban environments.

A limitation of this current work is the small size of the dataset examined, which consisted of 40~minutes in the same neighborhood of a single city.
To draw wider conclusions on the types of correlations present in urban backgrounds, more data from more neighborhoods and cities is needed.
RadMAP data can provide some of this need, although its coverage is limited to certain portions of the San Francisco Bay Area.
Larger datasets will be crucial for the training of CNNs, which typically require enormous datasets due to their multitude of parameters.
It is unknown whether a single model (of any type) could be useful for any urban area, or whether its domain would be limited to a single city or even neighborhood.

The ultimate application of the types of models discussed here would be to improve the performance of a detection algorithm by leveraging visual imagery.
Whether this is achieved by creating an algorithm that directly predicts background, or whether algorithms simply leverage context to inform alarming behavior is unclear.
However it is arrived at, a model that is able to ingest contextual information such as panoramic imagery and improve algorithm performance beyond a state-of-the-art would potentially revolutionize the urban search problem.


\bibliographystyle{IEEEtran}
\bibliography{radmap_inversion2}

\begin{thebibliography}{10}
\providecommand{\url}[1]{#1}
\csname url@samestyle\endcsname
\providecommand{\newblock}{\relax}
\providecommand{\bibinfo}[2]{#2}
\providecommand{\BIBentrySTDinterwordspacing}{\spaceskip=0pt\relax}
\providecommand{\BIBentryALTinterwordstretchfactor}{4}
\providecommand{\BIBentryALTinterwordspacing}{\spaceskip=\fontdimen2\font plus
\BIBentryALTinterwordstretchfactor\fontdimen3\font minus
  \fontdimen4\font\relax}
\providecommand{\BIBforeignlanguage}[2]{{%
\expandafter\ifx\csname l@#1\endcsname\relax
\typeout{** WARNING: IEEEtran.bst: No hyphenation pattern has been}%
\typeout{** loaded for the language `#1'. Using the pattern for}%
\typeout{** the default language instead.}%
\else
\language=\csname l@#1\endcsname
\fi
#2}}
\providecommand{\BIBdecl}{\relax}
\BIBdecl

\bibitem{hjerpe_statistical_2001}
\BIBentryALTinterwordspacing
T.~Hjerpe, R.~R. Finck, and C.~Samuelsson,
  ``\BIBforeignlanguage{english}{Statistical {Data} {Evaluation} in {Mobile}
  {Gamma} {Spectrometry}: {An} {Optimization} of {On}-line {Search}
  {Strategies} in the {Scenario} of {Lost} {Point} {Sources}},''
  \emph{\BIBforeignlanguage{english}{Health Physics}}, vol.~80, no.~6, pp.
  563--570, Jun. 2001. [Online]. Available:
  \url{https://journals.lww.com/health-physics/Abstract/2001/06000/STATISTICAL_DATA_EVALUATION_IN_MOBILE_GAMMA.6.aspx}
\BIBentrySTDinterwordspacing

\bibitem{aage_search_2003}
\BIBentryALTinterwordspacing
H.~K. Aage and U.~Korsbech, ``Search for lost or orphan radioactive sources
  based on {NaI} gamma spectrometry,'' \emph{Applied Radiation and Isotopes},
  vol.~58, no.~1, pp. 103--113, Jan. 2003. [Online]. Available:
  \url{http://www.sciencedirect.com/science/article/pii/S0969804302002221}
\BIBentrySTDinterwordspacing

\bibitem{ziock_large_2004}
\BIBentryALTinterwordspacing
K.-P. Ziock, W.~Craig, L.~Fabris, R.~Lanza, S.~Gallagher, B.~K.~P. Horn, and
  N.~Madden, ``Large {Area} {Imaging} {Detector} for {Long}-{Range}, {Passive}
  {Detection} of {Fissile} {Material},'' \emph{IEEE Transactions on Nuclear
  Science}, vol.~51, no.~5, pp. 2238--2244, Oct. 2004. [Online]. Available:
  \url{https://ieeexplore.ieee.org/document/1344316}
\BIBentrySTDinterwordspacing

\bibitem{mitchell_mobile_2009}
\BIBentryALTinterwordspacing
L.~Mitchell, B.~F. Phlips, W.~Johnson, E.~A. Wulf, A.~Hutcheson, C.~Lister,
  K.~Bynum, B.~Leas, and G.~Guadagno, ``Mobile {Imaging} and {Spectroscopic}
  {Threat} {Identification} ({MISTI}): {System} {Overview},'' in \emph{2009
  {IEEE} {Nuclear} {Science} {Symposium} {Conference} {Record} ({NSS}/{MIC})},
  Oct. 2009, pp. 110--118. [Online]. Available:
  \url{https://ieeexplore.ieee.org/document/5401849}
\BIBentrySTDinterwordspacing

\bibitem{penny_dual-sided_2011}
\BIBentryALTinterwordspacing
R.~D. Penny, W.~E. Hood, R.~M. Polichar, F.~H. Cardone, L.~G. Chavez, S.~G.
  Grubbs, B.~P. Huntley, R.~A. Kuharski, R.~T. Shyffer, L.~Fabris, K.~P. Ziock,
  S.~E. Labov, and K.~Nelson, ``A dual-sided coded-aperture radiation detection
  system,'' \emph{Nuclear Instruments and Methods in Physics Research Section
  A: Accelerators, Spectrometers, Detectors and Associated Equipment}, vol.
  652, no.~1, pp. 578--581, Oct. 2011. [Online]. Available:
  \url{http://www.sciencedirect.com/science/article/pii/S0168900211002634}
\BIBentrySTDinterwordspacing

\bibitem{zelakiewicz_sorisstandoff_2011}
\BIBentryALTinterwordspacing
S.~Zelakiewicz, R.~Hoctor, A.~Ivan, W.~Ross, E.~Nieters, W.~Smith, D.~McDevitt,
  M.~Wittbrodt, and B.~Milbrath, ``{SORIS}—{A} standoff radiation imaging
  system,'' \emph{Nuclear Instruments and Methods in Physics Research Section
  A: Accelerators, Spectrometers, Detectors and Associated Equipment}, vol.
  652, no.~1, pp. 5--9, Oct. 2011. [Online]. Available:
  \url{http://www.sciencedirect.com/science/article/pii/S0168900211004244}
\BIBentrySTDinterwordspacing

\bibitem{curtis_simulation_2020}
\BIBentryALTinterwordspacing
J.~C. Curtis, R.~J. Cooper, T.~H. Joshi, B.~Cosofret, T.~Schmit, J.~Wright,
  J.~Rameau, D.~Konno, D.~Brown, F.~Otsuka, E.~Rappeport, M.~Marshall, and
  J.~Speicher, ``\BIBforeignlanguage{english}{Simulation and validation of the
  {Mobile} {Urban} {Radiation} {Search} ({MURS}) gamma-ray detector
  response},'' \emph{\BIBforeignlanguage{english}{Nuclear Instruments and
  Methods in Physics Research Section A: Accelerators, Spectrometers, Detectors
  and Associated Equipment}}, vol. 954, p. 161128, Feb. 2020. [Online].
  Available:
  \url{http://www.sciencedirect.com/science/article/pii/S0168900218310453}
\BIBentrySTDinterwordspacing

\bibitem{aucott_routine_2013}
\BIBentryALTinterwordspacing
T.~Aucott, M.~Bandstra, V.~Negut, D.~Chivers, R.~Cooper, and K.~Vetter,
  ``Routine {Surveys} for {Gamma}-{Ray} {Background} {Characterization},''
  \emph{IEEE Transactions on Nuclear Science}, vol.~60, no.~2, pp. 1147--1150,
  Apr. 2013. [Online]. Available:
  \url{https://ieeexplore.ieee.org/document/6496304}
\BIBentrySTDinterwordspacing

\bibitem{archer_systematic_2015}
\BIBentryALTinterwordspacing
D.~E. Archer, D.~E. Hornback, J.~O. Johnson, A.~D. Nicholson, B.~W. Patton,
  D.~E. Peplow, T.~M. Miller, and B.~Ayaz-Maierhafer,
  ``\BIBforeignlanguage{english}{Systematic {Assessment} of {Neutron} and
  {Gamma} {Backgrounds} {Relevant} to {Operational} {Modeling} and {Detection}
  {Technology} {Implementation}},'' Oak Ridge National Lab. (ORNL), Oak Ridge,
  TN (United States), Tech. Rep. ORNL/TM-2014/687, Jan. 2015. [Online].
  Available: \url{https://www.osti.gov/biblio/1185844}
\BIBentrySTDinterwordspacing

\bibitem{jarman_comparison_2008}
\BIBentryALTinterwordspacing
K.~D. Jarman, R.~C. Runkle, K.~K. Anderson, and D.~M. Pfund, ``A comparison of
  simple algorithms for gamma-ray spectrometers in radioactive source search
  applications,'' \emph{Applied Radiation and Isotopes}, vol.~66, no.~3, pp.
  362--371, Mar. 2008. [Online]. Available:
  \url{http://www.sciencedirect.com/science/article/pii/S0969804307002886}
\BIBentrySTDinterwordspacing

\bibitem{aucott_effects_2014}
\BIBentryALTinterwordspacing
T.~Aucott, M.~Bandstra, V.~Negut, J.~Curtis, D.~Chivers, and K.~Vetter,
  ``Effects of {Background} on {Gamma}-{Ray} {Detection} for {Mobile}
  {Spectroscopy} and {Imaging} {Systems},'' \emph{IEEE Transactions on Nuclear
  Science}, vol.~61, no.~2, pp. 985--991, Apr. 2014. [Online]. Available:
  \url{https://ieeexplore.ieee.org/document/6786359}
\BIBentrySTDinterwordspacing

\bibitem{runkle_photon_2009}
\BIBentryALTinterwordspacing
R.~C. Runkle, L.~E. Smith, and A.~J. Peurrung, ``The photon haystack and
  emerging radiation detection technology,'' \emph{Journal of Applied Physics},
  vol. 106, no.~4, p. 041101, Aug. 2009. [Online]. Available:
  \url{https://aip.scitation.org/doi/full/10.1063/1.3207769}
\BIBentrySTDinterwordspacing

\bibitem{sandness_accurate_2009}
\BIBentryALTinterwordspacing
G.~A. Sandness, J.~E. Schweppe, W.~K. Hensley, J.~D. Borgardt, and A.~L.
  Mitchell, ``Accurate {Modeling} of the {Terrestrial} {Gamma}-{Ray}
  {Background} for {Homeland} {Security} {Applications},'' in \emph{2009 {IEEE}
  {Nuclear} {Science} {Symposium} {Conference} {Record} ({NSS}/{MIC})}, Oct.
  2009, pp. 126--133. [Online]. Available:
  \url{https://ieeexplore.ieee.org/document/5401843}
\BIBentrySTDinterwordspacing

\bibitem{trevisi_natural_2012}
\BIBentryALTinterwordspacing
R.~Trevisi, S.~Risica, M.~D’Alessandro, D.~Paradiso, and C.~Nuccetelli,
  ``\BIBforeignlanguage{english}{Natural radioactivity in building materials in
  the {European} {Union}: a database and an estimate of radiological
  significance},'' \emph{\BIBforeignlanguage{english}{Journal of Environmental
  Radioactivity}}, vol. 105, pp. 11--20, Feb. 2012. [Online]. Available:
  \url{http://www.sciencedirect.com/science/article/pii/S0265931X11002402}
\BIBentrySTDinterwordspacing

\bibitem{cosofret_utilization_2014}
\BIBentryALTinterwordspacing
B.~R. Cosofret, K.~Shokhirev, P.~Mulhall, D.~Payne, and B.~Harris,
  ``Utilization of advanced clutter suppression algorithms for improved
  standoff detection and identification of radionuclide threats,'' in
  \emph{Proceedings of {SPIE}}, vol. 9073, 2014, pp. 907\,316--907\,316--13.
  [Online]. Available: \url{http://dx.doi.org/10.1117/12.2049831}
\BIBentrySTDinterwordspacing

\bibitem{pfund_improvements_2016}
\BIBentryALTinterwordspacing
D.~M. Pfund, K.~K. Anderson, R.~S. Detwiler, K.~D. Jarman, B.~S. McDonald,
  B.~D. Milbrath, M.~J. Myjak, N.~C. Paradis, S.~M. Robinson, and M.~L.
  Woodring, ``Improvements in the method of radiation anomaly detection by
  spectral comparison ratios,'' \emph{Applied Radiation and Isotopes}, vol.
  110, pp. 174--182, Apr. 2016. [Online]. Available:
  \url{http://www.sciencedirect.com/science/article/pii/S0969804315304024}
\BIBentrySTDinterwordspacing

\bibitem{tandon_detection_2016}
\BIBentryALTinterwordspacing
P.~Tandon, P.~Huggins, R.~Maclachlan, A.~Dubrawski, K.~Nelson, and S.~Labov,
  ``\BIBforeignlanguage{english}{Detection of radioactive sources in urban
  scenes using {Bayesian} {Aggregation} of data from mobile spectrometers},''
  \emph{\BIBforeignlanguage{english}{Information Systems}}, vol.~57, pp.
  195--206, Apr. 2016. [Online]. Available:
  \url{https://www.sciencedirect.com/science/article/pii/S0306437915001866}
\BIBentrySTDinterwordspacing

\bibitem{miller_gamma-ray_2018}
\BIBentryALTinterwordspacing
K.~Miller and A.~Dubrawski, ``Gamma-{Ray} {Source} {Detection} {With} {Small}
  {Sensors},'' \emph{IEEE Transactions on Nuclear Science}, vol.~65, no.~4, pp.
  1047--1058, Apr. 2018. [Online]. Available:
  \url{https://ieeexplore.ieee.org/document/8305516}
\BIBentrySTDinterwordspacing

\bibitem{bilton_non-negative_2019}
\BIBentryALTinterwordspacing
K.~J. Bilton, T.~H. Joshi, M.~S. Bandstra, J.~C. Curtis, B.~J. Quiter, R.~J.
  Cooper, and K.~Vetter, ``Non-negative {Matrix} {Factorization} of
  {Gamma}-{Ray} {Spectra} for {Background} {Modeling}, {Detection}, and
  {Source} {Identification},'' \emph{IEEE Transactions on Nuclear Science},
  vol.~66, no.~5, pp. 827--837, May 2019. [Online]. Available:
  \url{https://ieeexplore.ieee.org/abstract/document/8673769}
\BIBentrySTDinterwordspacing

\bibitem{lo_presti_baseline_2006}
\BIBentryALTinterwordspacing
C.~A. Lo~Presti, D.~R. Weier, R.~T. Kouzes, and J.~E. Schweppe,
  ``\BIBforeignlanguage{english}{Baseline suppression of vehicle portal monitor
  gamma count profiles: {A} characterization study},''
  \emph{\BIBforeignlanguage{english}{Nuclear Instruments and Methods in Physics
  Research Section A: Accelerators, Spectrometers, Detectors and Associated
  Equipment}}, vol. 562, no.~1, pp. 281--297, Jun. 2006. [Online]. Available:
  \url{http://www.sciencedirect.com/science/article/pii/S0168900206004402}
\BIBentrySTDinterwordspacing

\bibitem{burr_alarm_2007}
\BIBentryALTinterwordspacing
T.~Burr, J.~R. Gattiker, K.~Myers, and G.~Tompkins,
  ``\BIBforeignlanguage{english}{Alarm criteria in radiation portal
  monitoring},'' \emph{\BIBforeignlanguage{english}{Applied Radiation and
  Isotopes}}, vol.~65, no.~5, pp. 569--580, May 2007. [Online]. Available:
  \url{http://www.sciencedirect.com/science/article/pii/S0969804307000048}
\BIBentrySTDinterwordspacing

\bibitem{karnowski_design_2010}
\BIBentryALTinterwordspacing
T.~P. Karnowski, M.~F. Cunningham, J.~S. Goddard, A.~M. Cheriyadat, D.~E.
  Hornback, L.~Fabris, R.~A. Kerekes, K.-P. Ziock, E.~C. Bradley, J.~Chesser,
  and W.~Marchant, ``Design of dual-road transportable portal monitoring system
  for visible light and gamma-ray imaging,'' in \emph{Chemical, {Biological},
  {Radiological}, {Nuclear}, and {Explosives} ({CBRNE}) {Sensing} {XI}}, vol.
  7665.\hskip 1em plus 0.5em minus 0.4em\relax International Society for Optics
  and Photonics, May 2010, p. 76651J. [Online]. Available:
  \url{https://www.spiedigitallibrary.org/conference-proceedings-of-spie/7665/76651J/Design-of-dual-road-transportable-portal-monitoring-system-for-visible/10.1117/12.850191.short}
\BIBentrySTDinterwordspacing

\bibitem{ziock_performance_2013}
\BIBentryALTinterwordspacing
K.~Ziock, E.~Bradley, A.~Cheriyadat, M.~Cunningham, L.~Fabris, C.~Fitzgerald,
  J.~Goddard, D.~Hornback, R.~Kerekes, T.~Karnowski, W.~Marchant, and J.~Newby,
  ``Performance of the {Roadside} {Tracker} {Portal}-{Less} {Portal}
  {Monitor},'' \emph{IEEE Transactions on Nuclear Science}, vol.~60, no.~3, pp.
  2237--2246, Jun. 2013. [Online]. Available:
  \url{https://ieeexplore.ieee.org/document/6530685}
\BIBentrySTDinterwordspacing

\bibitem{livesay_rain-induced_2014}
\BIBentryALTinterwordspacing
R.~J. Livesay, C.~S. Blessinger, T.~F. Guzzardo, and P.~A. Hausladen,
  ``\BIBforeignlanguage{english}{Rain-induced increase in background radiation
  detected by {Radiation} {Portal} {Monitors}},''
  \emph{\BIBforeignlanguage{english}{Journal of Environmental Radioactivity}},
  vol. 137, pp. 137--141, Nov. 2014. [Online]. Available:
  \url{http://www.sciencedirect.com/science/article/pii/S0265931X1400215X}
\BIBentrySTDinterwordspacing

\bibitem{mitchell_gamma-ray_2015}
\BIBentryALTinterwordspacing
L.~J. Mitchell, B.~F. Phlips, E.~A. Wulf, A.~L. Hutcheson, C.~Gwon, R.~S.
  Woolf, and D.~Polaski, ``Gamma-ray and neutron background comparison of {US}
  metropolitan areas,'' \emph{Nuclear Instruments and Methods in Physics
  Research Section A: Accelerators, Spectrometers, Detectors and Associated
  Equipment}, vol. 784, pp. 311--318, Jun. 2015. [Online]. Available:
  \url{http://www.sciencedirect.com/science/article/pii/S0168900215000467}
\BIBentrySTDinterwordspacing

\bibitem{stewart_understanding_2018}
\BIBentryALTinterwordspacing
I.~R. Stewart, A.~D. Nicholson, D.~E. Archer, M.~J. Willis, M.~W. Swinney,
  I.~Garishvili, and W.~R. Ray, ``\BIBforeignlanguage{english}{Understanding
  and quantifying the systematic effects of clutter within a radiation
  detection scene},'' \emph{\BIBforeignlanguage{english}{Journal of
  Radioanalytical and Nuclear Chemistry}}, Aug. 2018. [Online]. Available:
  \url{https://doi.org/10.1007/s10967-018-6159-8}
\BIBentrySTDinterwordspacing

\bibitem{kaffine_background_2017}
C.~Kaffine, B.~Pires, A.~Laddha, D.~Bayani, K.~Miller, M.~Hebert, and
  A.~Dubrawski, ``\BIBforeignlanguage{english}{Background {Spectrum}
  {Estimation} from {Panoramic} {Images}},'' Atlanta, GA, Oct. 2017, 2017 IEEE
  Nuclear Science Symposium.

\bibitem{nicholson_multiagency_2017}
\BIBentryALTinterwordspacing
A.~D. Nicholson, I.~Garishvili, D.~E. Peplow, D.~E. Archer, W.~R. Ray, M.~W.
  Swinney, M.~J. Willis, G.~G. Davidson, S.~L. Cleveland, B.~W. Patton, D.~E.
  Hornback, J.~J. Peltz, M.~S.~L. McLean, A.~A. Plionis, B.~J. Quiter, and
  M.~S. Bandstra, ``Multiagency {Urban} {Search} {Experiment} {Detector} and
  {Algorithm} {Test} {Bed},'' \emph{IEEE Transactions on Nuclear Science},
  vol.~64, no.~7, pp. 1689--1695, Jul. 2017. [Online]. Available:
  \url{https://ieeexplore.ieee.org/document/7869400}
\BIBentrySTDinterwordspacing

\bibitem{archer_modeling_2017}
\BIBentryALTinterwordspacing
D.~E. Archer, M.~S. Bandstra, G.~G. Davidson, S.~L. Cleveland, I.~Garishvili,
  D.~E. Hornback, J.~O. Johnson, M.~S.~L. McLean, A.~D. Nicholson, B.~W.
  Patton, D.~E. Peplow, A.~A. Plionis, B.~J. Quiter, W.~R. Ray, A.~J. Rowe,
  M.~W. Swinney, and M.~J. Willis, ``\BIBforeignlanguage{english}{Modeling and
  {Urban} {Search} {Experiments}: {Fort} {Indiantown} {Gap} {Data}
  {Collections} {Summary} and {Analysis}},'' Oak Ridge National Lab. (ORNL),
  Oak Ridge, TN (United States), Tech. Rep. ORNL/LTR-2017/371, Oct. 2017.
  [Online]. Available: \url{https://www.osti.gov/biblio/1410932}
\BIBentrySTDinterwordspacing

\bibitem{bandstra_radmap:_2016}
\BIBentryALTinterwordspacing
M.~S. Bandstra, T.~J. Aucott, E.~Brubaker, D.~H. Chivers, R.~J. Cooper, J.~C.
  Curtis, J.~R. Davis, T.~H. Joshi, J.~Kua, R.~Meyer, V.~Negut, M.~Quinlan,
  B.~J. Quiter, S.~Srinivasan, A.~Zakhor, R.~Zhang, and K.~Vetter, ``{RadMAP}:
  {The} {Radiological} {Multi}-sensor {Analysis} {Platform},'' \emph{Nuclear
  Instruments and Methods in Physics Research Section A: Accelerators,
  Spectrometers, Detectors and Associated Equipment}, vol. 840, pp. 59--68,
  Dec. 2016. [Online]. Available:
  \url{http://www.sciencedirect.com/science/article/pii/S0168900216309780}
\BIBentrySTDinterwordspacing

\bibitem{bandstra_attribution_2020}
\BIBentryALTinterwordspacing
M.~S. Bandstra, B.~J. Quiter, J.~C. Curtis, K.~J. Bilton, T.~H.~Y. Joshi,
  R.~Meyer, V.~Negut, K.~Vetter, D.~E. Archer, D.~E. Hornback, D.~E. Peplow,
  C.~E. Romano, M.~W. Swinney, T.~L. McCullough, and M.~S.~L. McLean,
  ``\BIBforeignlanguage{english}{Attribution of gamma-ray background collected
  by a mobile detector system to its surroundings using panoramic video},''
  \emph{\BIBforeignlanguage{english}{Nuclear Instruments and Methods in Physics
  Research Section A: Accelerators, Spectrometers, Detectors and Associated
  Equipment}}, vol. 954, p. 161126, Feb. 2020. [Online]. Available:
  \url{http://www.sciencedirect.com/science/article/pii/S016890021831043X}
\BIBentrySTDinterwordspacing

\bibitem{swinney_methodology_2018}
\BIBentryALTinterwordspacing
M.~W. Swinney, D.~E. Peplow, B.~W. Patton, A.~D. Nicholson, D.~E. Archer, and
  M.~J. Willis, ``A {Methodology} for {Determining} the {Concentration} of
  {Naturally} {Occurring} {Radioactive} {Materials} in an {Urban}
  {Environment},'' \emph{Nuclear Technology}, vol.~0, no.~0, pp. 1--11, May
  2018. [Online]. Available:
  \url{https://doi.org/10.1080/00295450.2018.1458558}
\BIBentrySTDinterwordspacing

\bibitem{bandstra_correlations_2019}
M.~S. Bandstra, B.~J. Quiter, K.~J. Bilton, J.~C. Curtis, S.~Goldenberg,
  T.~H.~Y. Joshi, and M.~Salathe, ``Correlations between {Panoramic} {Imagery}
  and {Gamma}-{Ray} {Background} in an {Urban} {Area},'' in \emph{2019 {IEEE}
  {Nuclear} {Science} {Symposium} and {Medical} {Imaging} {Conference}
  ({NSS}/{MIC})}, Oct. 2019, pp. 1--5, iSSN: 2577-0829.

\bibitem{google_cartog_2016}
W.~Hess, D.~Kohler, H.~Rapp, and D.~Andor, ``{Real}-{Time} {Loop} {Closure} in
  {2D} {LIDAR} {SLAM},'' in \emph{2016 IEEE International Conference on
  Robotics and Automation (ICRA)}, 2016, pp. 1271--1278.

\bibitem{runkle_analysis_2006}
R.~C. Runkle, M.~F. Tardiff, K.~K. Anderson, D.~K. Carlson, and L.~E. Smith,
  ``Analysis of {Spectroscopic} {Radiation} {Portal} {Monitor} {Data} {Using}
  {Principal} {Components} {Analysis},'' \emph{IEEE Transactions on Nuclear
  Science}, vol.~53, no.~3, pp. 1418--1423, Jun. 2006.

\bibitem{stinnett_uncertainty_2017}
J.~Stinnett, C.~J. Sullivan, and H.~Xiong, ``Uncertainty {Analysis} of
  {Wavelet}-{Based} {Feature} {Extraction} for {Isotope} {Identification} on
  {NaI} {Gamma}-{Ray} {Spectra},'' \emph{IEEE Transactions on Nuclear Science},
  vol.~64, no.~7, pp. 1670--1676, Jul. 2017.

\bibitem{paatero_positive_1994}
\BIBentryALTinterwordspacing
P.~Paatero and U.~Tapper, ``\BIBforeignlanguage{english}{Positive matrix
  factorization: {A} non-negative factor model with optimal utilization of
  error estimates of data values},''
  \emph{\BIBforeignlanguage{english}{Environmetrics}}, vol.~5, no.~2, pp.
  111--126, Jun. 1994. [Online]. Available:
  \url{https://onlinelibrary.wiley.com/doi/abs/10.1002/env.3170050203}
\BIBentrySTDinterwordspacing

\bibitem{lee_learning_1999}
\BIBentryALTinterwordspacing
D.~D. Lee and H.~S. Seung, ``\BIBforeignlanguage{english}{Learning the parts of
  objects by non-negative matrix factorization},''
  \emph{\BIBforeignlanguage{english}{Nature}}, vol. 401, no. 6755, pp.
  788--791, Oct. 1999. [Online]. Available:
  \url{http://www.nature.com/nature/journal/v401/n6755/abs/401788a0.html}
\BIBentrySTDinterwordspacing

\bibitem{bandstra_modeling_2020}
\BIBentryALTinterwordspacing
M.~S. Bandstra, T.~H.~Y. Joshi, K.~J. Bilton, A.~Zoglauer, and B.~J. Quiter,
  ``Modeling {Aerial} {Gamma}-{Ray} {Backgrounds} {Using} {Non}-negative
  {Matrix} {Factorization},'' \emph{IEEE Transactions on Nuclear Science},
  vol.~67, no.~5, pp. 777--790, May 2020. [Online]. Available:
  \url{https://ieeexplore.ieee.org/abstract/document/9025259}
\BIBentrySTDinterwordspacing

\bibitem{pfund_examination_2007}
\BIBentryALTinterwordspacing
D.~Pfund, R.~Runkle, K.~Anderson, and K.~Jarman, ``Examination of
  {Count}-{Starved} {Gamma} {Spectra} {Using} the {Method} of {Spectral}
  {Comparison} {Ratios},'' \emph{IEEE Transactions on Nuclear Science},
  vol.~54, no.~4, pp. 1232--1238, Aug. 2007. [Online]. Available:
  \url{https://ieeexplore.ieee.org/document/4178949}
\BIBentrySTDinterwordspacing

\bibitem{lee_algorithms_2001}
\BIBentryALTinterwordspacing
D.~D. Lee and H.~S. Seung, ``Algorithms for {Non}-negative {Matrix}
  {Factorization},'' in \emph{Advances in {Neural} {Information} {Processing}
  {Systems} 13}, T.~K. Leen, T.~G. Dietterich, and V.~Tresp, Eds.\hskip 1em
  plus 0.5em minus 0.4em\relax MIT Press, 2001, pp. 556--562. [Online].
  Available:
  \url{http://papers.nips.cc/paper/1861-algorithms-for-non-negative-matrix-factorization.pdf}
\BIBentrySTDinterwordspacing

\bibitem{kullback_information_1951}
\BIBentryALTinterwordspacing
S.~Kullback and R.~A. Leibler, ``On {Information} and {Sufficiency},''
  \emph{The Annals of Mathematical Statistics}, vol.~22, no.~1, pp. 79--86,
  1951. [Online]. Available: \url{https://www.jstor.org/stable/2236703}
\BIBentrySTDinterwordspacing

\bibitem{pauca_nonnegative_2006}
\BIBentryALTinterwordspacing
V.~P. Pauca, J.~Piper, and R.~J. Plemmons, ``Nonnegative matrix factorization
  for spectral data analysis,'' \emph{Linear Algebra and its Applications},
  vol. 416, no.~1, pp. 29--47, Jul. 2006. [Online]. Available:
  \url{http://www.sciencedirect.com/science/article/pii/S002437950500340X}
\BIBentrySTDinterwordspacing

\bibitem{becker_adaptive_2015}
J.~M. Becker, M.~Rohbeck, and C.~Rohlfing, ``Adaptive weights for {NMF} with
  additional priors,'' in \emph{2015 {International} {Symposium} on
  {Intelligent} {Signal} {Processing} and {Communication} {Systems}
  ({ISPACS})}, Nov. 2015, pp. 89--94.

\bibitem{li_learning_2001}
S.~Z. Li, X.~W. Hou, H.~J. Zhang, and Q.~S. Cheng, ``Learning {Spatially}
  {Localized}, {Parts}-{Based} {Representation},'' in \emph{Proceedings of the
  2001 {IEEE} {Computer} {Society} {Conference} on {Computer} {Vision} and
  {Pattern} {Recognition}. {CVPR} 2001}, vol.~1, 2001, pp. I--207--I--212
  vol.1.

\bibitem{matrixcalculus}
\BIBentryALTinterwordspacing
S.~Laue, M.~Mitterreiter, J.~Giesen, and J.~Mueller. (2020) {Matrix Calculus}.
  [Online]. Available: \url{http://www.matrixcalculus.org}
\BIBentrySTDinterwordspacing

\bibitem{chen_encoder-decoder_2018}
\BIBentryALTinterwordspacing
L.-C. Chen, Y.~Zhu, G.~Papandreou, F.~Schroff, and H.~Adam, ``Encoder-{Decoder}
  with {Atrous} {Separable} {Convolution} for {Semantic} {Image}
  {Segmentation},'' in \emph{Proceedings of the {European} {Conference} on
  {Computer} {Vision} ({ECCV})}, 2018, pp. 801--818. [Online]. Available:
  \url{https://openaccess.thecvf.com/content_ECCV_2018/html/Liang-Chieh_Chen_Encoder-Decoder_with_Atrous_ECCV_2018_paper.html}
\BIBentrySTDinterwordspacing

\bibitem{cordts_cityscapes_2016}
M.~Cordts, M.~Omran, S.~Ramos, T.~Rehfeld, M.~Enzweiler, R.~Benenson,
  U.~Franke, S.~Roth, and B.~Schiele, ``{The} {Cityscapes} {Dataset} for
  {Semantic} {Urban} {Scene} {Understanding},'' in \emph{Proc. of the IEEE
  Conference on Computer Vision and Pattern Recognition (CVPR)}, 2016.

\bibitem{scikit-learn}
F.~Pedregosa, G.~Varoquaux, A.~Gramfort, V.~Michel, B.~Thirion, O.~Grisel,
  M.~Blondel, P.~Prettenhofer, R.~Weiss, V.~Dubourg, J.~Vanderplas, A.~Passos,
  D.~Cournapeau, M.~Brucher, M.~Perrot, and E.~Duchesnay, ``Scikit-learn:
  Machine learning in {P}ython,'' \emph{Journal of Machine Learning Research},
  vol.~12, pp. 2825--2830, 2011.

\bibitem{salathe_determining_2021}
\BIBentryALTinterwordspacing
M.~Salathe, B.~J. Quiter, M.~S. Bandstra, J.~C. Curtis, R.~Meyer, and C.~H.
  Chow, ``Determining urban material activities with a vehicle-based
  multi-sensor system,'' \emph{Physical Review Research}, vol.~3, no.~2, p.
  023070, Apr. 2021, publisher: American Physical Society. [Online]. Available:
  \url{https://link.aps.org/doi/10.1103/PhysRevResearch.3.023070}
\BIBentrySTDinterwordspacing

\bibitem{iyengar_systematic_2015}
\BIBentryALTinterwordspacing
A.~Iyengar, M.~Beach, R.~J. Newby, L.~Fabris, L.~H. Heilbronn, and J.~P.
  Hayward, ``Systematic measurement of fast neutron background fluctuations in
  an urban area using a mobile detection system,'' \emph{Nuclear Instruments
  and Methods in Physics Research Section A: Accelerators, Spectrometers,
  Detectors and Associated Equipment}, vol. 773, pp. 27--32, Feb. 2015.
  [Online]. Available:
  \url{http://www.sciencedirect.com/science/article/pii/S0168900214012029}
\BIBentrySTDinterwordspacing

\bibitem{davis_fast_2017}
\BIBentryALTinterwordspacing
J.~R. Davis, E.~Brubaker, and K.~Vetter, ``Fast neutron background
  characterization with the {Radiological} {Multi}-sensor {Analysis} {Platform}
  ({RadMAP}),'' \emph{Nuclear Instruments and Methods in Physics Research
  Section A: Accelerators, Spectrometers, Detectors and Associated Equipment},
  vol. 858, pp. 106--112, Jun. 2017. [Online]. Available:
  \url{http://www.sciencedirect.com/science/article/pii/S0168900217303820}
\BIBentrySTDinterwordspacing

\bibitem{glick_deployment_2021}
\BIBentryALTinterwordspacing
A.~Glick, E.~Brubaker, B.~Cabrera-Palmer, M.~Gerling, B.~J. Quiter, and
  K.~Vetter, ``\BIBforeignlanguage{english}{Deployment of a double scatter
  system for directional detection of background neutron radiation},''
  \emph{\BIBforeignlanguage{english}{Nuclear Instruments and Methods in Physics
  Research Section A: Accelerators, Spectrometers, Detectors and Associated
  Equipment}}, vol. 992, p. 165029, Mar. 2021. [Online]. Available:
  \url{https://www.sciencedirect.com/science/article/pii/S0168900221000139}
\BIBentrySTDinterwordspacing

\end{thebibliography}


\appendices
\section{Cosmic regularization functions and their gradients}\label{sec:append}
We only apply the penalty \(f_1\) to the components of the first column of \(\mathbf{W}\) that are above 1250\,keV, represented by spectrum index \(I_{1250}\).
To use the symmetric KL~divergence, we generate normalized vectors for the cosmic component and first row of \(\mathbf{W}\) above index \(I_{1250}\):
\begin{align}
    p_i &\equiv \frac{W_{i0}}{\sum_{i' > I_{1250}} W_{i'0}} \\
    q_i &\equiv \frac{w_{\mathrm{cos},i}}{\sum_{i' > I_{1250}} w_{\mathrm{cos},i'}} \\
    f_1(\mathbf{W}) &= \sum_{i > I_{1250}} (p_i - q_i) \log\left(\frac{p_i}{q_i}\right).
\end{align}

The gradient of \(f_1\) is
\begin{align}
    \begin{split}
    \frac{\partial f_1}{\partial W_{ij}} = {} & \frac{[j = 0] [i > I_{1250}]}{\sum_{i' > I_{1250}} W_{i'0}} \times \\
    & \ \Biggl\{ \Biggl(1 - \log q_i - \sum_{i' > I_{1250}} p_{i'} \log p_{i'} \Biggr) \\
    & \ \ - \Biggl( \frac{q_i}{p_i} - \log p_i - \sum_{i' > I_{1250}} p_{i'} \log q_{i'} \Biggr) \Biggr\}
    \end{split} \\
    \equiv {} & \left(\nabla_1^{+}\right)_{ij} - \left(\nabla_1^{-}\right)_{ij}
\end{align}
where \([]\) is the Iverson bracket (\(1\) if argument is true, \(0\) otherwise), and we have split the gradient into a sum of terms that are always positive (\(\boldsymbol\nabla_1^{+}\)) and terms that are always negative (\(-\boldsymbol\nabla_1^{-}\)).

The \(f_2\) function is the sum of \(\mathbf{W}\) for the spectral bins above 3\,MeV, represented by the spectral index \(I_{3000}\):
\begin{align}
    f_2(\mathbf{W}) &= \sum_{i > I_{3000}} \sum_{j > 0} W_{ij}.
\end{align}
The gradient of \(f_2\) is
\begin{align}
    \frac{\partial f_2}{\partial W_{ij}} &= [i > I_{3000}] [j > 0] \\
    &\equiv \left(\nabla_2^{+}\right)_{ij}.
\end{align}

\end{document}